%% file: crude.0.7_pc.tex
\documentclass[aps,twocolumn,showpacs,pra,citeautoscript,amsmath,amssymb,floatfix,superscriptaddress,10pt]{revtex4-1}
\pdfoutput=1

\usepackage{color}
\usepackage[normalem]{ulem}  
\usepackage{amsthm}
\usepackage{epsfig}
\usepackage[caption=false]{subfig}
\usepackage{hyperref,url}

\input{common/definitions}

\newcommand{\Wtran}[2]{W_{#1\rightarrow#2}}

\def\???{{\color{red}???}}

\def\final{\sigma}

\def\ITR{\textrm{PITR}}
\def\ITRs{\textrm{PITRs}}
\def\longITR{\textrm{Partial Isothermal Reversible Process}}
\def\longITRs{\textrm{Partial Isothermal Reversible Processes}}

\def\PLT{\textrm{PLT}}
\def\PLTs{\textrm{PLTs}}\def\longPLT{\textrm{Partial Level Thermalization}}
\def\longPLTs{\textrm{Partial Level Thermalizations}}

\def\SR{\textrm{SR}}

\def\longSR{\textrm{Subspace Rotation}}
\def\longSRs{\textrm{Subspace Rotations}}

\def\PF{\textrm{PF}}
\def\PFs{\textrm{PFs}}
\def\longPF{\textrm{Points Flow}}
\def\longPFs{\textrm{Points Flows}}

\def\LT{\textrm{LT}}
\def\LTs{\textrm{LTs}}
\def\longLT{\textrm{Level Transformation}}
\def\longLTs{\textrm{Level Transformations}}

\def\CO{\textrm{CO}}
\def\COs{\textrm{COs}}
\def\longCO{\textrm{Crude Operation}}
\def\longCOs{\textrm{Crude Operations}}

\begin{document}
\author{Christopher Perry}
\affiliation{Department of Physics and Astronomy, University College London, Gower Street, London WC1E 6BT, United Kingdom}
\author{Piotr \'Cwikli\'nski}
\affiliation{Faculty of Mathematics, Physics and Informatics, Institute of Theoretical Physics and Astrophysics, University of Gda\'nsk, 80-952 Gda\'nsk, Poland and  National Quantum Information Centre of Gda\'nsk, 81-824 Sopot, Poland}
\author{Janet Anders}
\affiliation{Department of Physics and Astronomy, University College London, Gower Street, London WC1E 6BT, United Kingdom}
\affiliation{Department of Physics and Astronomy, University of Exeter, Stocker Road, Exeter EX4 4QL, United Kingdom}
\author{Micha{\l} Horodecki}
\affiliation{Faculty of Mathematics, Physics and Informatics, Institute of Theoretical Physics and Astrophysics, University of Gda\'nsk, 80-952 Gda\'nsk, Poland and  National Quantum Information Centre of Gda\'nsk, 81-824 Sopot, Poland}
\author{Jonathan Oppenheim}
\affiliation{Department of Physics and Astronomy, University College London, Gower Street, London WC1E 6BT, United Kingdom}

\title{A sufficient set of experimentally implementable thermal operations}

\date{\today}

\begin{abstract}
Recent work using tools from quantum information theory has shown that at the nanoscale where quantum effects become prevalent, there is not one thermodynamical second law but many. Derivations of these laws assume that an experimenter has very precise control of the system and heat bath. Here we show that these multitude of laws can be saturated using two very simple operations: changing the energy levels of the system and thermalizing over any two system energy levels. Using these two operations, one can distill the optimal amount of work from a system, as well as perform the reverse formation process. Even more surprisingly, using only these two operations and one ancilla qubit in a thermal state, one can transform any state into any other state allowable by the second laws. We thus have the remarkable result that the second laws hold for fine-grained manipulation of system and bath, but can be achieved using very coarse control. This brings the full array of thermal operations into a regime accessible by experiment, and establishes the physical relevance of these second laws.

\end{abstract}

\maketitle

\section{Introduction}

Thermodynamics and statistical physics are one of the most successful areas of physics, owing to its broad applicability. One can make statements which do not depend on the particulars of the dynamics, and such laws govern much of the world around us. Thermodynamics puts limitations on the efficiency of our cars' engines, determines the weather, can be used to predict many phenomena in particle accelerators, and even plays a central role in areas of fundamental physics, providing one of the only clues we have to a quantum theory of gravity through the laws of black hole thermodynamics. However, traditional thermodynamics, as derived from statistical mechanics, generally concerns itself with the average behavior of large systems, composed of many particles. Here the experimenter is only able to manipulate macroscopic quantities of the material such as its pressure and volume, and does not have access to the microscopic degrees of freedom of the system, much less the heat bath. The basic operations are limited to very crude control of the system-bath -- isotherms, adiabats, isochors etc.

However, as our abilities to manipulate and control small thermodynamical systems improve, we are able to control the microscopic degrees of freedom of smaller and smaller systems \cite{Scovil1959masers,scully2002afterburner,rousselet1994directional,Faucheux1995ratchet,baugh2005experimental}. It thus seems natural to consider the thermodynamical behavior of small, finite sized systems or heat engines composed of just a few molecules.

\begin{figure}
\centering
\begin{minipage}{.5\columnwidth}
  \centering
	{\includegraphics[width=0.95\textwidth]{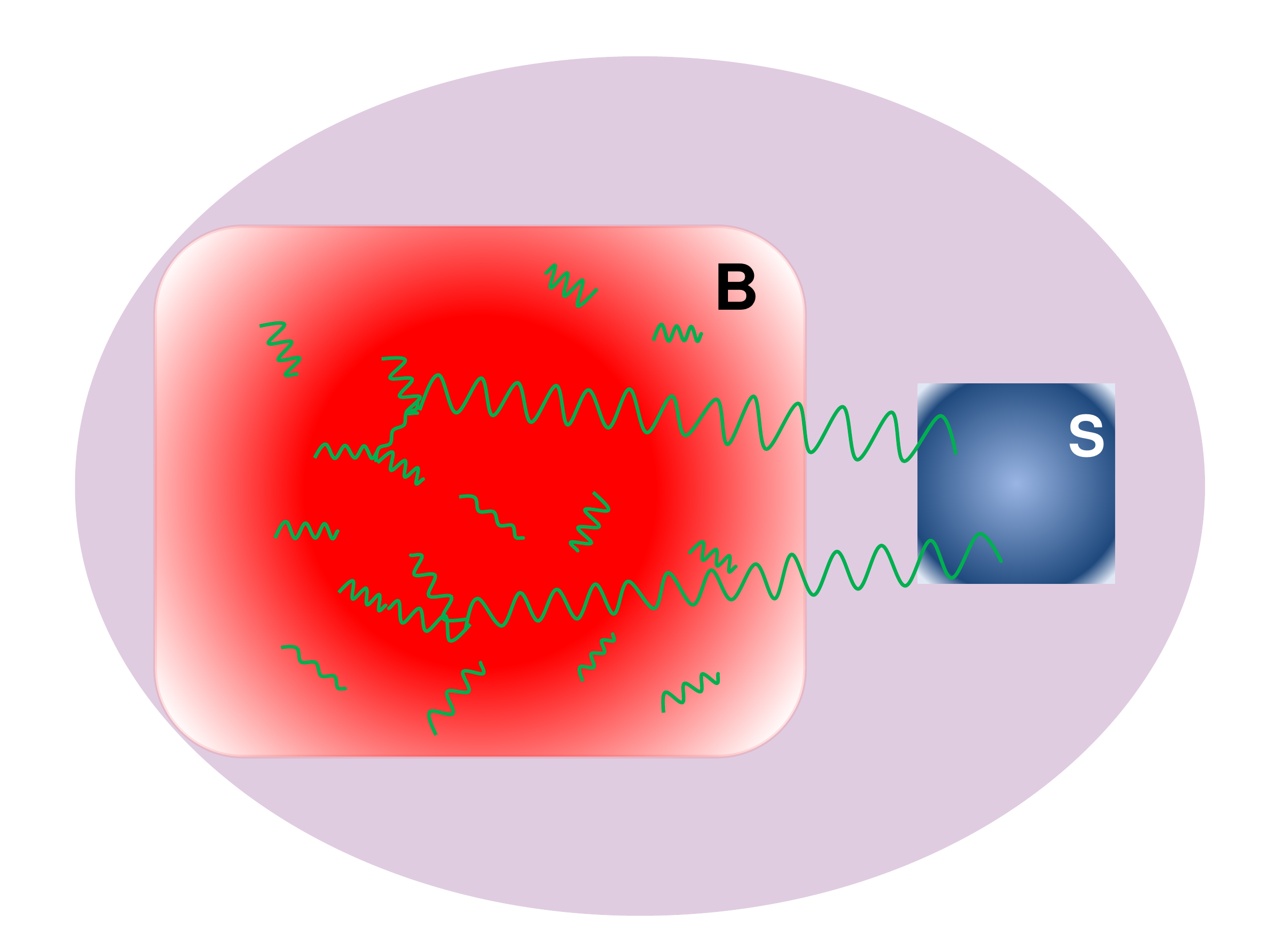}}
  \label{fig:comparison_1}
	\centerline{(a) Global interactions}
\end{minipage}%
\begin{minipage}{.5\columnwidth}
  \centering
  \includegraphics[width=0.95\textwidth]{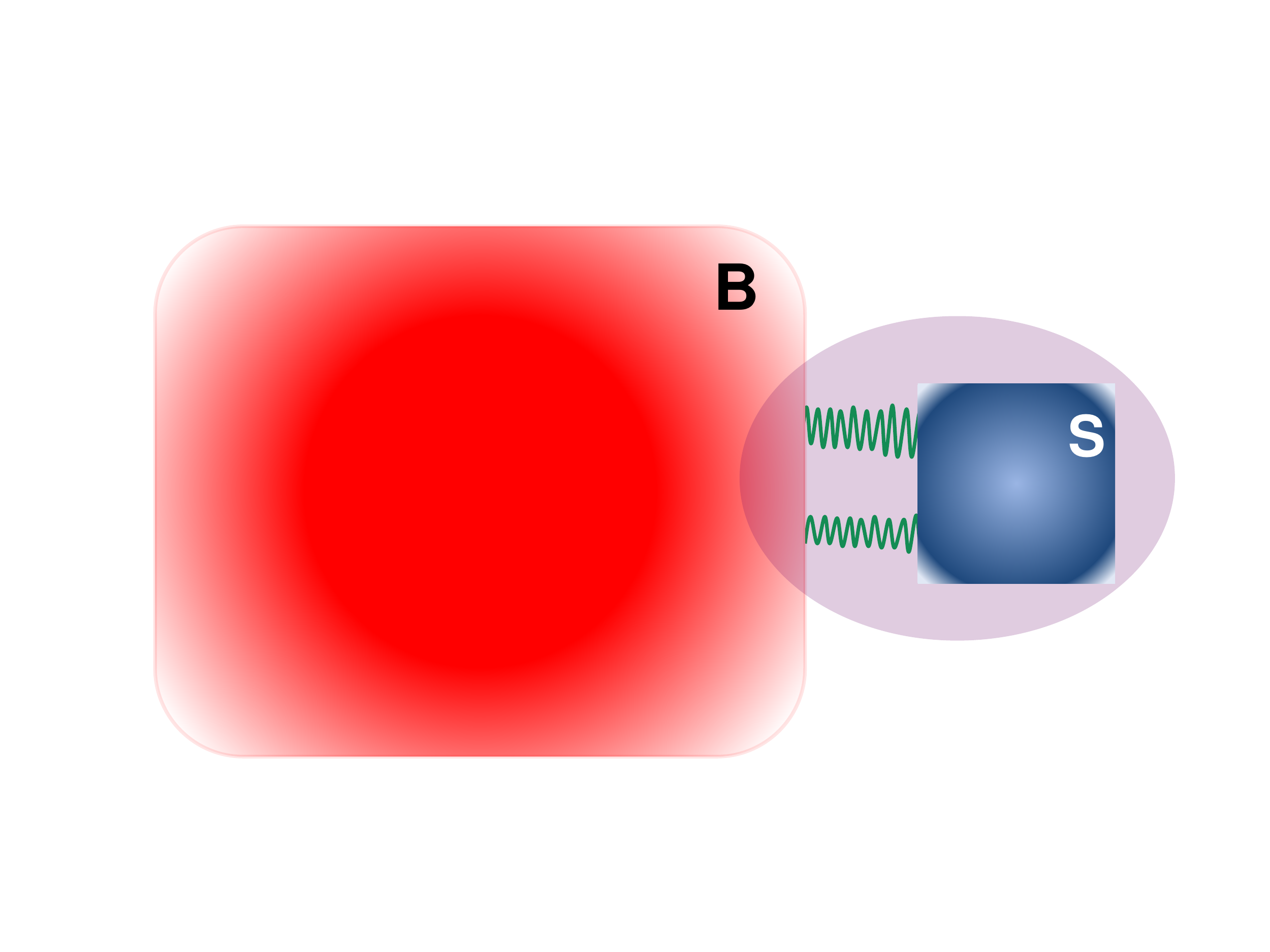}
  \label{fig:comparison_2}
	\centerline{(b) Simple interactions}
\end{minipage}\\
\vspace{.5cm}
\begin{minipage}{.5\columnwidth}
  \centering
  \includegraphics[width=0.95\textwidth]{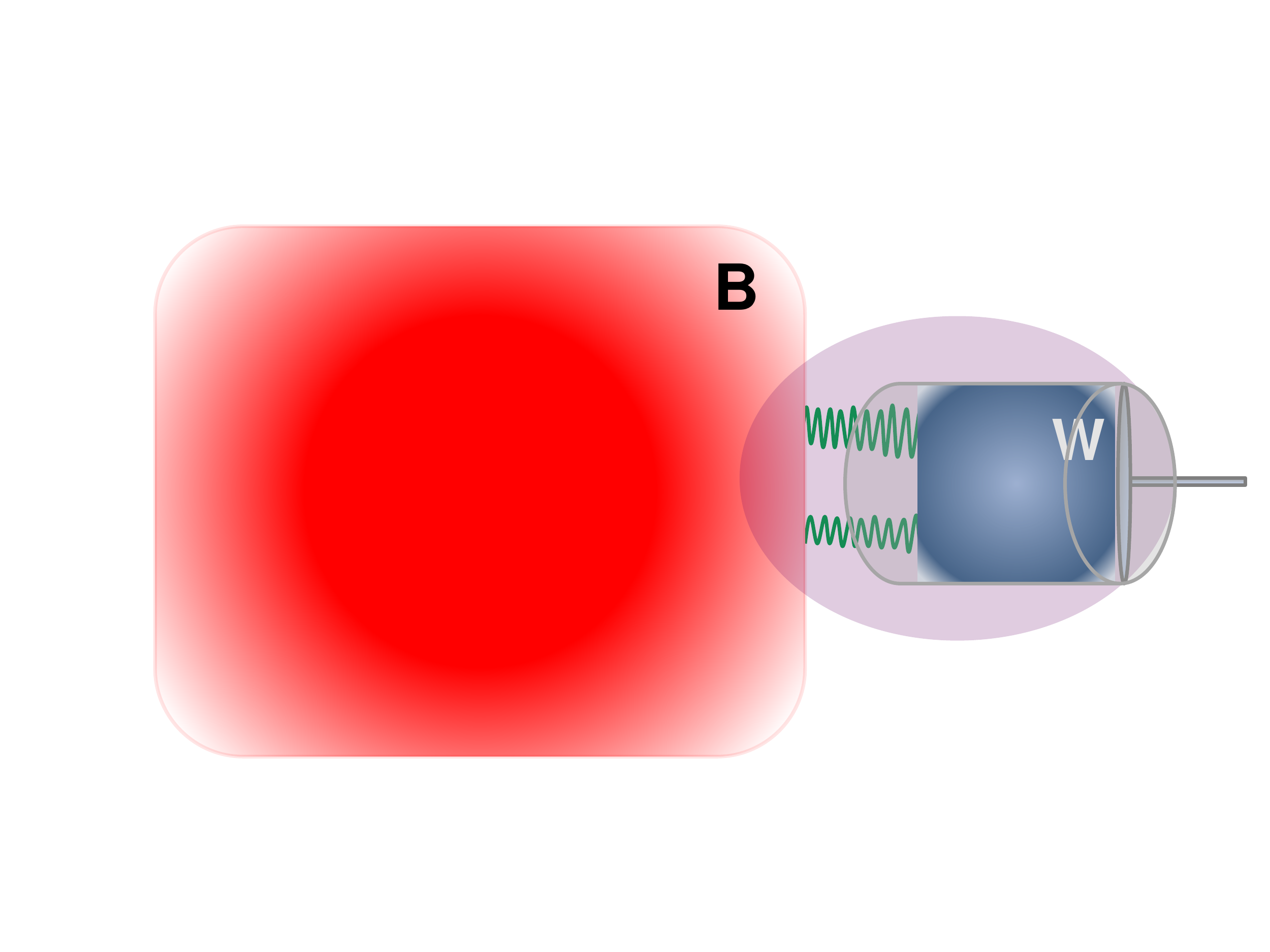}
  \label{fig:comparison_3}
	\centerline{(b') Classical thermodynamics - piston}
\end{minipage}%
\caption{\emph{Thermal operations vs. crude operations vs. classical operations.}
We consider a heat bath $B$ together with a system $S$ or working body $W$ and illustrate the different levels of control an experimenter can have on the setup and interactions. Figure (a): The most detailed, experimentally unfeasible control where the experimenter keeps a record of every microstate of the system and bath (the area contained within the purple oval) and controls interactions between the system and the entirety of the heat bath (illustrated by green stings).
Figure (b): The desired level of control where the experimenter keeps track of the system and a small portion of the bath (purple oval) and performs some simple interactions between these regions (green strings).
Figure (b'): The previous case can be regarded as analogous to the set up in traditional thermodynamics where one has a working body $W$ of which some parameters can be changed using simple processes such as moving a piston and weak couplings to the heat bath.}
\label{fig:comparison}
\end{figure}

For a $d$-level system interacting with a heat bath, one can imagine an experimenter manipulating the system, who has control over each of the levels and can interact the system in any way they want with the heat bath. From a practical point of view, needing to perform such arbitrary interactions is undesirable as they require very precise control over and be able to keep track of the entirety of the heat bath. Simpler interactions would be much more appealing. See Figure~\ref{fig:comparison} for a schematic of this comparison.


However, even if one allows for such fine-grained control, the most experimentally unfeasible scenario, the second law of thermodynamics still holds (provided one computes the entropy of the system in terms of its microstates rather than using a course grained entropy). In fact, not only does the traditional second law hold, but additional second laws emerge at the nano-scale such as the so-called \emph{thermo-majorization} criteria~\cite{ruch1976principle,horodecki2013fundamental}, and those given by a family of generalized free energies \cite{brandao2013second} \footnote{Note that these generalized free energy constraints can be obtained from the thermo-majorization criteria provided one is allowed to use an ancillary system which is returned in its initial state at the end of protocol. As such, we can restrict ourselves to considering the thermo-majorization criteria here.}. These constrain the set of states it is possible to transition to from a given starting state and converge to the familiar second law in the thermodynamic limit.

However, such precise control will be impossible to implement as it could require accurately manipulating all of the $10^{20}$ molecules contained in a typical heat bath. As such, it may seem what an experimenter can achieve without such incredibly fine-grained control must be very far from what is allowed by the second laws \cite{wilming2014weak}. This contrasts sharply with traditional, macroscopic thermodynamics. There, those transformations allowed by the standard second law can easily be achieved by controlling macroscopic, coarse-grained parameters such as a system's volume or an external field. If the same level of control was needed macroscopically as seems necessary at the nano-scale, then running a car efficiently would require control of all of the molecules in the exploding fuel and cooler. Clearly this would be an undesirable feature - must it exist at the nano-scale? The existence of a large gap between what is allowed by the most general class of operations, and what is achievable without detailed control of the heat bath, would make it hard to decide what the science of thermodynamics of microscopic systems should actually be about and how applicable the recently derived second laws are. 

Surprisingly, here we show that any state transformation permitted by the additional second laws can be achieved using three simple operations. These operations, which we term \emph{\longCOs}, are experimentally feasible and do not require fine control of bath degrees of freedom to implement, only weak coupling to the bath. All allowed transformations can be implemented by applying thermalizations, raising and lowering energy levels and rotations within energy subspaces to the system and a single thermal qubit taken from the heat bath. This serves to place the microscopic laws of thermodynamics on the same footing as their macroscopic counterpart: saturating them does not require unfeasibly precise levels of control.

As a by-product, our simple operations can be viewed analogously to a universal gate set in quantum computing: they provide building blocks for the construction of more elaborate protocols. 

\section{Thermal Operations and Thermo-majorization}

The thermo-majorization constraints were derived \cite{horodecki2013fundamental} under the largest class of operations one is allowed to implement under thermodynamics - \emph{Thermal Operations} \cite{janzing2000thermodynamic, Streater_dynamics, horodecki2013fundamental}. These are presented in full detail in Section \ref{ssec:thermal operations} of the \supl\  but in particular they allow an experimenter to perform any energy conserving unitary between system and bath. Energy conservation does not pose a constraint on what is allowed since it can be enforced by incorporating a work storage device into the system to account for any energy excess of deficit. Rather, imposing energy conservation allows us to account for all sources of energy as is necessary for thermodynamics in the micro-regime. Clearly needing to apply all such unitaries to realize all possible transformations would require an enormous amount of control. 

\begin{figure}
\centering
\includegraphics[width=0.9\columnwidth]{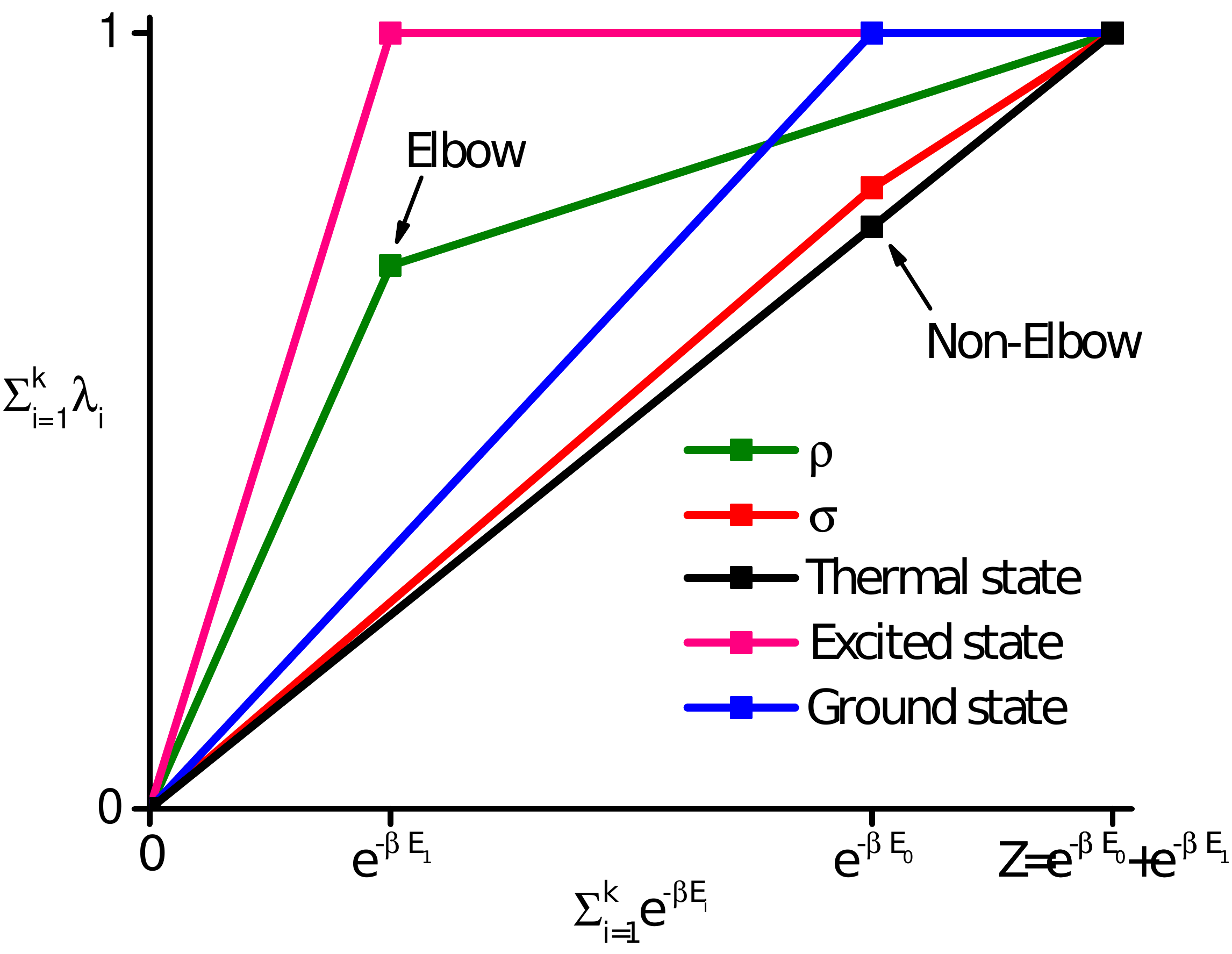}
\caption[Thermo-majorization curves.]{\emph{Thermo-majorization curves.} The thermo-majorization curve for $\rho$ is defined by plotting the points: $\left\{\sum_{i=1}^{k} e^{-\beta E_i^{\left(\rho\right)}},\sum_{i=1}^{k}p_i^{\left(\rho\right)}\right\}_{k=1}^{n}$. The superscript $\left(\rho\right)$ indicates that the occupation probabilities and their associated energy levels have been ordered such that $p_i^{\left(\rho\right)} e^{\beta E_i^{\left(\rho\right)}}$ is non-increasing in $i$. This is called $\beta${\it-ordering}. Note that states associated with the same Hamiltonian may have different $\beta$-orderings, as illustrated by $\rho$ and $\sigma$ here. We say that $\rho$ {\it thermo-majorizes} $\sigma$ as the thermo-majorization curve of $\rho$ is never below the thermo-majorization curve of $\sigma$. The thermo-majorization curve of a Gibbs state is given by a straight line between $\left(0,0\right)$ and $\left(Z,1\right)$. All other states thermo-majorize it. The pure state corresponding to the highest energy level of an $n$-level system thermo-majorizes all other states associated with that Hamiltonian. We call a point on a curve an elbow, if the gradient of the curve changes as it passes through the point. Otherwise, it is a non-elbow.}
\label{fig:thermomaj}
\end{figure}

For a system in state $\rho$ and associated Hamiltonian $H_S$ with energy eigenstates $\ket{i}$, thermo-majorization assigns to each state a curve determined by the state's energy levels $E_i$ and occupation probabilities $p_i=\tr\left[\rho\ketbra{i}{i}\right]$. The construction of these curves is detailed and illustrated in Figure \ref{fig:thermomaj} (see also Section \ref{ss:thermomaj} of the \supl).

A system in state $\rho$ can be transformed into state $\sigma$ using at heat bath at inverse temperature $\beta=\frac{1}{kT}$ (here $k$ is Boltzmann's constant) under thermal operations only if the thermo-majorization curve of $\rho$ is above that of $\sigma$. If $\sigma$ is block-diagonal in the energy eigenbasis or we have access to a source of coherence \cite{brandao2011resource, aberg2014catalytic, korzekwa2015extraction, cwiklinski2014limitations}, then they are sufficient too. In the thermodynamic limit, the thermo-majorization criteria collapses to the familiar second law: a transformation from $\rho$ to $\sigma$ is possible if and only if
\begin{equation*}
F\left(\rho,H_1\right)\geq F\left(\sigma,H_2\right)
\end{equation*}
where $F\left(\rho,H\right)=\tr\left[H\rho\right]-kTS\left(\rho\right)$ with $S\left(\rho\right)=-\tr\left[\rho\log\rho\right]$ is the free energy. For the thermal state of the system $\gibbs$, $F\left(\gibbs,H\right)=-kT\log Z$ where $Z=\sum_{i=1}^{d}e^{-\beta E_i}$ is the partition function. When the initial and final states are thermal states of their respective Hamiltonians, the difference in free energy equates to the optimal amount of work required (or gained) when transforming between the two states. For more general, nanoscopic states, thermo-majorization curves can be used to calculate the work required for a state transformation and we discuss this in Appendix~\ref{ssec:wits&switch}.

\section{Crude Operations}

The first of our three basic operations are \emph{\longPLTs}\ (\PLT). A thermalization essentially changes the state of the system into a thermal state and is usually achieved by putting the system in thermal contact with the reservoir until it equilibrates or by swapping the system with one from the reservoir. Thermalizations have no work cost or gain associated with it. A partial thermalization generalizes this, allowing one to thermalize with some probability $p$, implementing
\begin{equation*}
\rho\rightarrow p \rho +\left(1-p\right)\gibbs,
\end{equation*}
with $\gibbs$ being the thermal state at inverse temperature $\beta$. The probability $p$ can be determined by using the ambient heat bath as a source of noise or by putting the system in contact with it for a time shorter than the equilibration time.

\begin{figure}
\centering
\includegraphics[width=0.8\columnwidth]{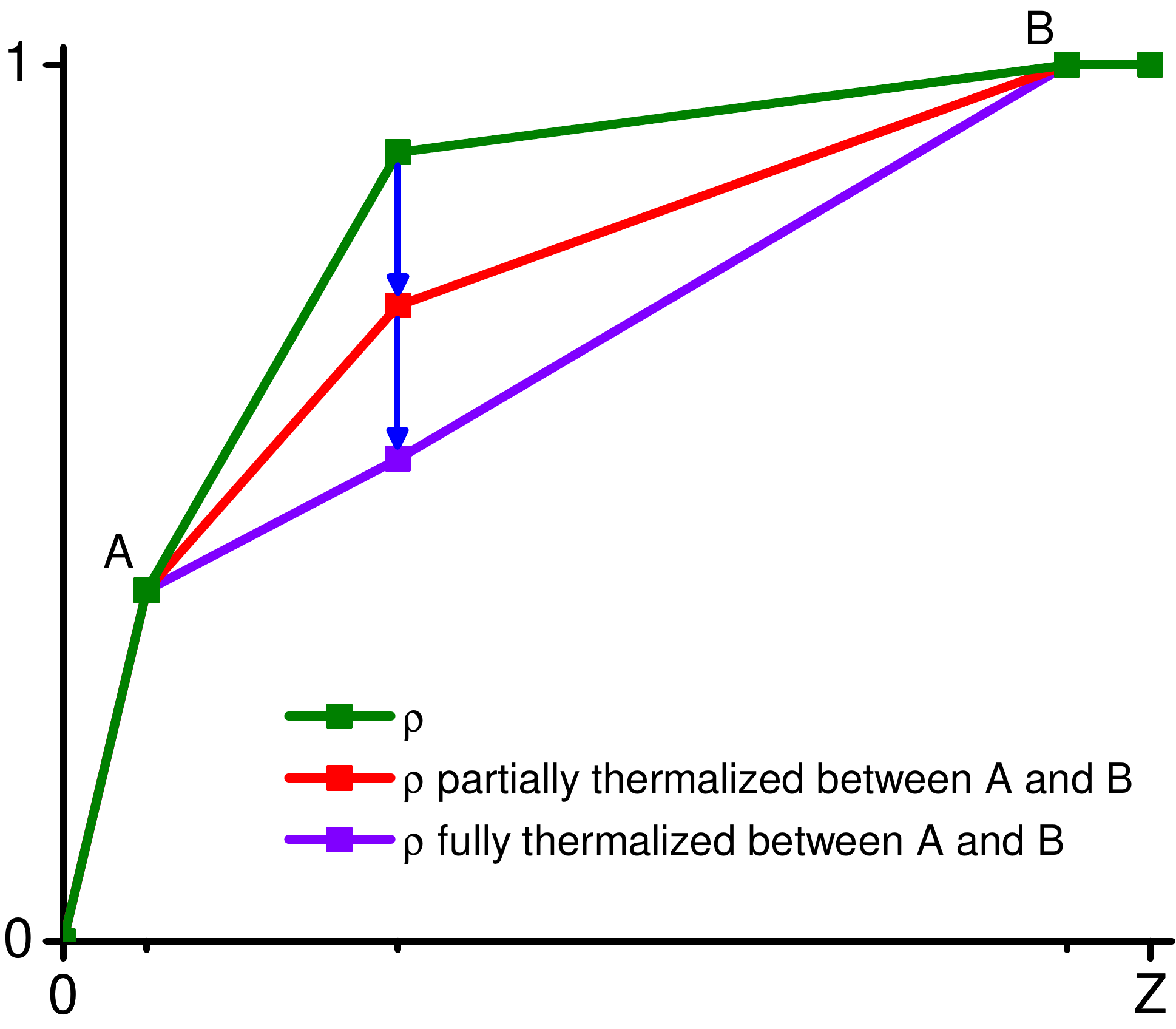}
\caption{\emph{Action of \longPLT.} Here we illustrate the action of \PLTs\ applied to a state $\rho$ over the two energy levels between points $A$ and $B$ for various degrees of thermalization $p$. Note, that as we apply the \PLTs\ to adjacent energy levels with respect to the $\beta$-ordering of $\rho$, the final states maintain this $\beta$-order.} 
\label{fig:PTmain}
\end{figure}

With \longPLTs\ we go one step further and allow for the partial thermalization to act on any pair of energy levels. In order to implement them, one needs to be able to either perform the SWAP gate between any two levels of the system and of the thermal bath or to selectively put system energy levels in contact with the reservoir, for example by making use of an optical cavity or intermediate system which acts as a filter to restrict what energy levels are being addressed by the thermal contact. The action of \PLTs\ on a state is illustrated in terms of thermo-majorization curves in Figure \ref{fig:PTmain}.

\begin{figure}
\centering
\includegraphics[width=0.8\columnwidth]{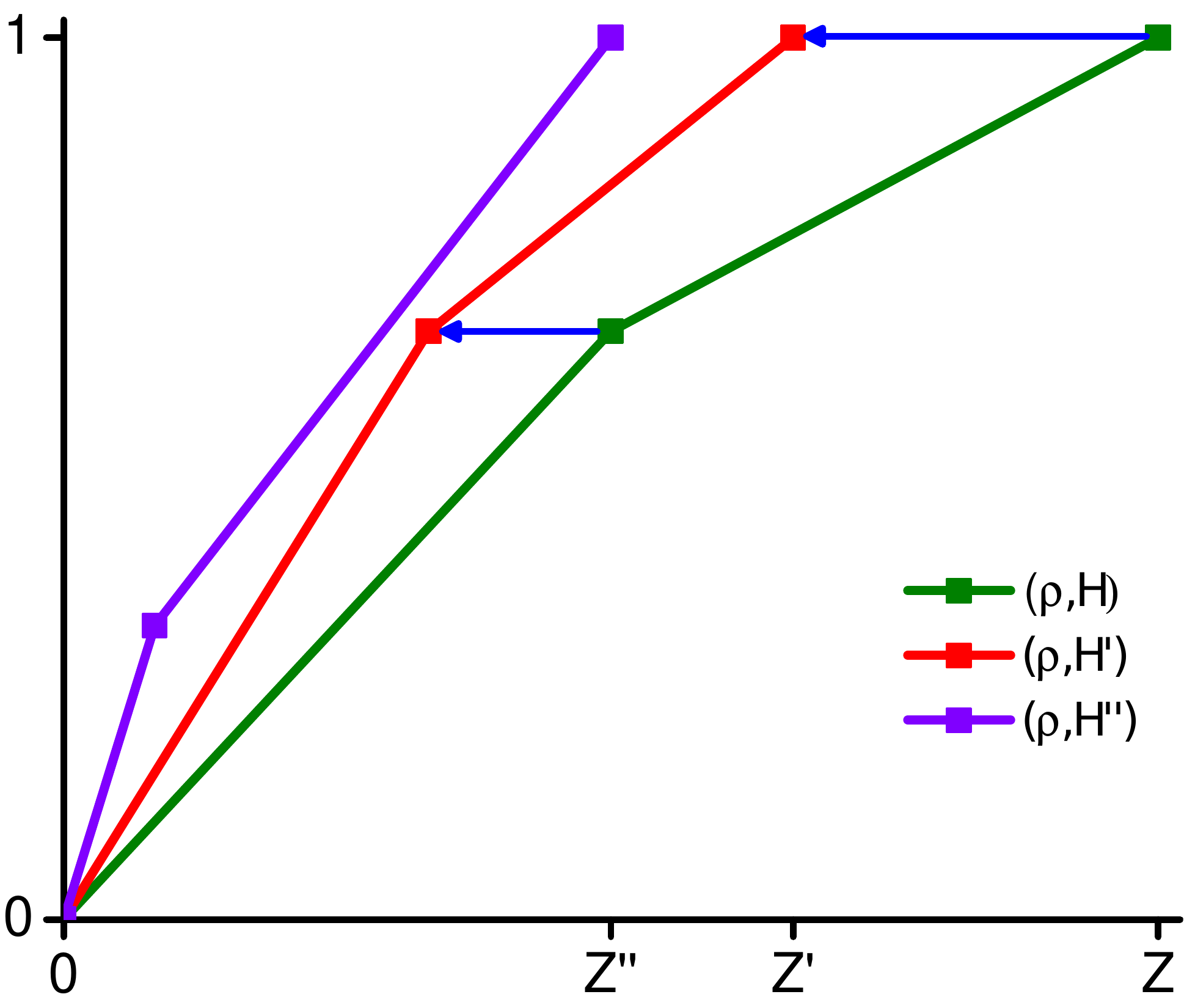}
\caption{\emph{Action of \longLTs.} Here we illustrate the action of \LTs\ applied to a system with a thermodynamical configuration $\left(\rho,H\right)$. Note that \LTs\ leave the occupation probabilities of $\rho$ unchanged but may alter the $\beta$ ordering as discussed in the \supl.} 
\label{fig:LTmain}
\end{figure}

Our second type of operation are \emph{\longLTs}\ (\LT), namely the raising and lowering of any subset of energy levels of the system's Hamiltonian. This type of transformation is common within thermodynamics and the work cost of implementing them is given by the change in energy of the level (when the level is populated). Their effect on thermo-majorization is shown in Figure~\ref{fig:LTmain}.

Finally, in the case where the system has degenerate energy levels, one may need to implement an energy conserving unitary acting within the degenerate subspace of the system. We will call this operation a \emph{\longSR}\ (\SR). It is only ever needed if the initial and final energy eigenstates are rotated with respect to each other and, in most examples we consider, it will not be needed and it does not effect the thermo-majorization curve of a state. 

These operations are detailed with greater specificity in Section \ref{sec:coarseops} of the \supl, where it is also shown that they are a subset of Thermal Operations. Here, we shall contrast them with operations that have appeared in other resource theoretic approaches to thermodynamics. In \cite{aaberg2013truly} it was shown that full thermalizations and \longLTs\ suffice for extracting the optimal amount of work from a given state under Thermal Operations as evaluated in \cite{horodecki2013fundamental}. Similarly they can be used to form a given state from the thermal state at the same work cost as under TO.

In \cite{egloff2015measure} it was shown, that for transitions between two diagonal states, perhaps with the expenditure or gain of work, it is enough to make thermal operations on the bath and the system, and instead of the  work system used in \cite{horodecki2013fundamental} just use level transformations. Still however, the system-bath coupling term used in \cite{egloff2015measure} requires being able to implement an arbitrary Thermal Operations (see \cite{renes2014work} and \cite{renes2015work}, where it is shown that the operations used in \cite{egloff2015measure} are the subset of Thermal Operations) and thus requires in principle unlimited control.

Finally, in \cite{Skrzypczyk13}, a subset of Thermal Operations were considered which involved interacting with a designer heat bath which contained an arbitrarily large number of systems in a series of states that interpolated between the input state and the target state. Again, this would require an unfeasible amount of control in preparing the states of the heat bath.  

As we shall see in the next Section, \longCOs\ allow all transformations allowed by Thermal Operations to be implemented without the need for unreasonable levels of control.

\section{Transformations using \longCOs}

In this paper, our goal is to consider the thermo-majorization curves of any two states in which the curve of $\initial$ lies above that of $\final$, and show that \longLTs\, \longPLTs\, and in some cases a \longSR, together with access to a single thermal qubit, are all that is needed to transform one curve into another. 

\begin{figure}
\centering
\includegraphics[width=0.8\columnwidth]{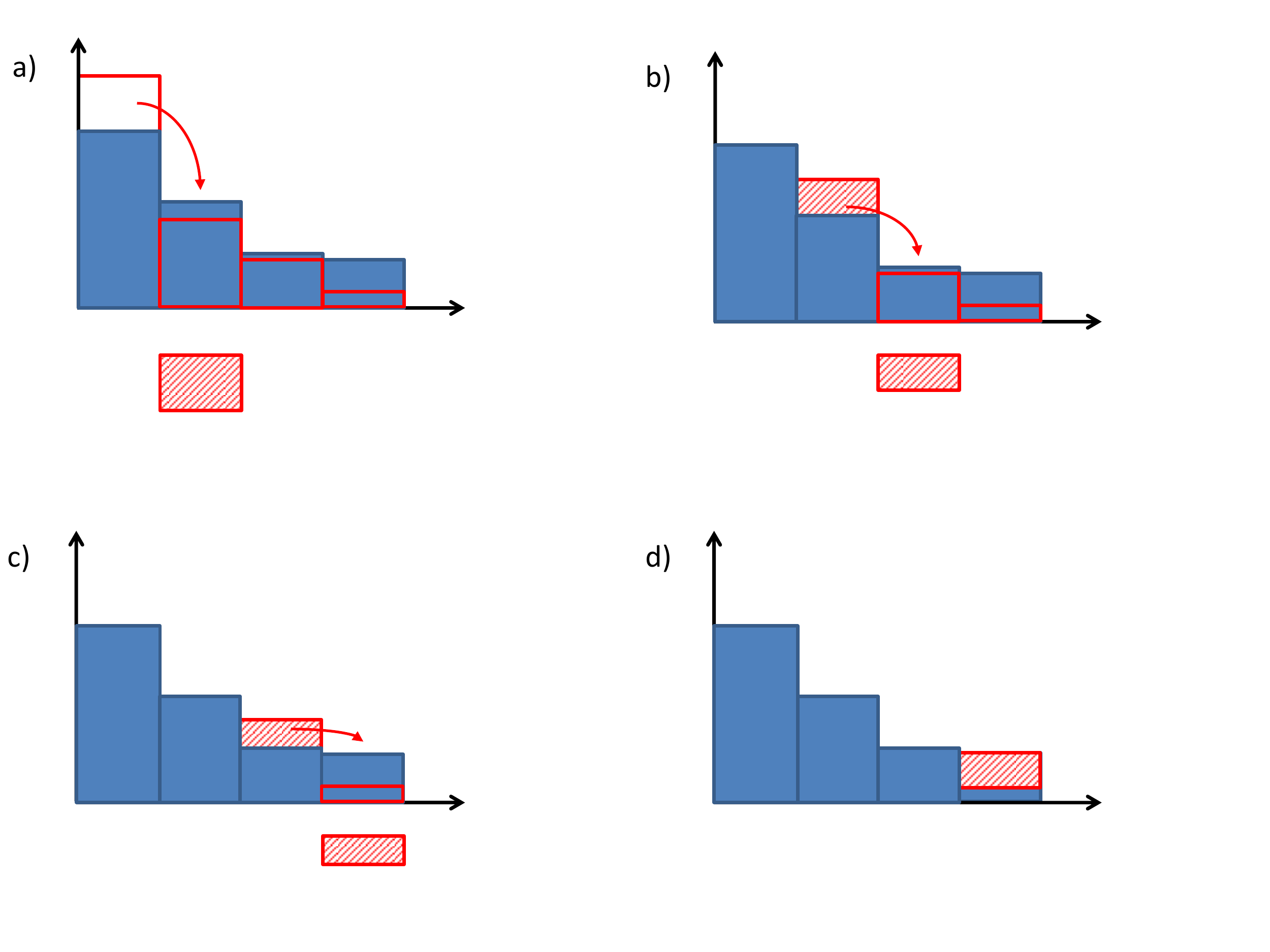}
\caption{The heights of each column above are given by the probability of being in a particular eigenstate. The action of {\it T-transforms} transforms an initial state represented by red columns into the state represented by blue columns. The probabilities (column heights) of the red state majorizes those of the blue state. We work our way from left to right, moving probability mass from the red histogram to the right until it matches the blue histogram. In each step we move some probability mass from a column of the histogram of the initial state and move it to the right until it either matches the probability required of the target state, or until the left column of the initial state matches the right column of the final state. Red dashed rectangles represent the part of the column that is added to the next one due to the action of the {\it T-transform}.} 
\label{fig:T-transforms}
\end{figure}

For the case where the Hamiltonian of the system is trivial, $H=0$, the thermo-majorization criteria is equivalent to the majorization criteria and any state $\initial$ can be transformed into any state $\final$ if and only if its eigenvalues majorize the eigenvalues of $\final$ \cite{uniqueinfo}. That is, for respective sets of eigenvalues $\left\{ p_1,\dots, p_d\right\}$ and $\left\{ q_1,\dots,q_d\right\}$ written in non-decreasing order, $\initial\rightarrow\final$ is possible if and only if $\sum_{i=1}^{k} p_i \geq \sum_{i=1}^k q_i$, $\forall k$. In such a case, Muirhead \cite{muirhead1902some} and Hardy, Littlewood, and Polya \cite{hardy1952inequalities} showed that for states of dimension $d$ one can perform the transformation, using at most $d-1$ {\it T-transforms}. For $H=0$, where the thermal state is a maximally mixes state, these are just \longPLTs. This is depicted in Figure \ref{fig:T-transforms} and discussed in the  context of thermodynamics in \cite{gour2013resource}.

However, in general such a protocol for $H=0$ does not extract any work during the state transformation. We show in Section \ref{sec:Szilard} of the \supl\ that by alternating $\longLTs$ and $\longPLTs$ one can make the transformation $\initial\rightarrow\final$ while extracting the optimal amount of work. We also show there how this can be accomplished in a physical set up involving a molecule in a box (a so called Szilard engine).

For an arbitrary Hamiltonian, the situation turns out to be much more complicated, as we will soon see. In fact, we will find that for an arbitrary state transformations, it is not enough to perform \COs\ on the system alone, but that we need to perform \COs\ on the system and an extra qubit in the thermal state.

In the case where we want to extract the maximum amount of work from some state $\rho$ (i.e. transform $\rho$ into $\gibbs$), this can be achieved using \longLTs\ and alternating \longPLTs\ and \longLTs. We term this alternating sequence a \longITR\ (\ITR) in reference to the isothermal reversible processes defined in \cite{aaberg2013truly} where \longLTs\ are interlaced with full thermalizations to distil work. We discuss this work extraction protocol in Section \ref{ss:extractwork} of the \supl\ and analyze the trade-off between the work extracted in the case where the protocol succeeds and the cost if it fails.

The reverse process to work extraction is that of formation. There one starts with the thermal state $\gibbs$ and uses work to form the state $\rho$. For the case where $\rho$ does not contain coherences, we construct a process to do this using \longCOs\ in Section \ref{ss:formwork} of the \supl. The work cost of this protocol is optimal, requiring the same amount of work to be expended as under Thermal Operations. 

\begin{figure}
\centering
\includegraphics[width=0.8\columnwidth]{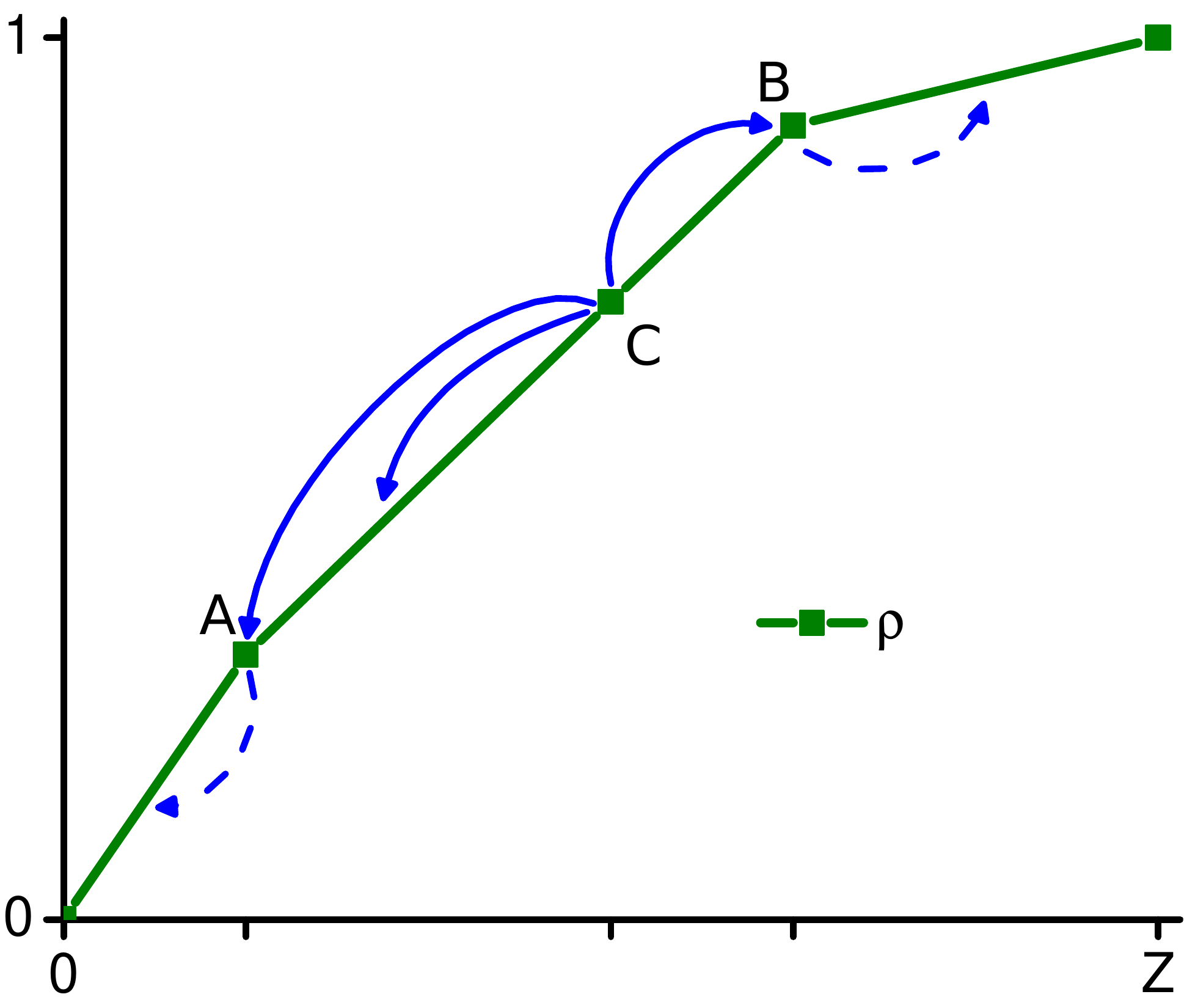}
\caption{\emph{Action of \longITRs.} Here we illustrate the action of alternating \longLTs\ and \longPLTs\ (so-called, \ITRs) applied to a system with state-Hamiltonian pair $\left(\rho,H\right)$. Using \ITRs, the point at $C$ can be moved such that it lies anywhere on the line-segment between $A$ and $B$ and without changing the shape of the overall thermo-majorization curve. By performing this process sufficiently slowly, this can be done with no deterministic work cost. If one moves the point $C$ to coincide with point $A$ (respectively $B$), one can then use a second \ITR\ to move point $A$ ($B$) as illustrated by the dashed arrows. Again, this does not alter the shape of the thermo-majorization curve.} 
\label{fig:ITRmain}
\end{figure}

Both work extraction and formation make use of \ITRs\ and these, and their associated work cost, are discussed more fully in Section \ref{ssec:points flow} of the \supl. There we show that they provide a method to move \emph{non-elbows} (as defined in Figure \ref{fig:thermomaj}) to different segments of a thermo-majorization curve, without altering the curves shape and without expending any work. This is illustrated in Figure \ref{fig:ITRmain}. Such a transformation will prove vital when we come to consider more general transformations.

\begin{figure}
\centering
\begin{minipage}{.5\columnwidth}
  \centering
	{\includegraphics[width=0.9\textwidth]{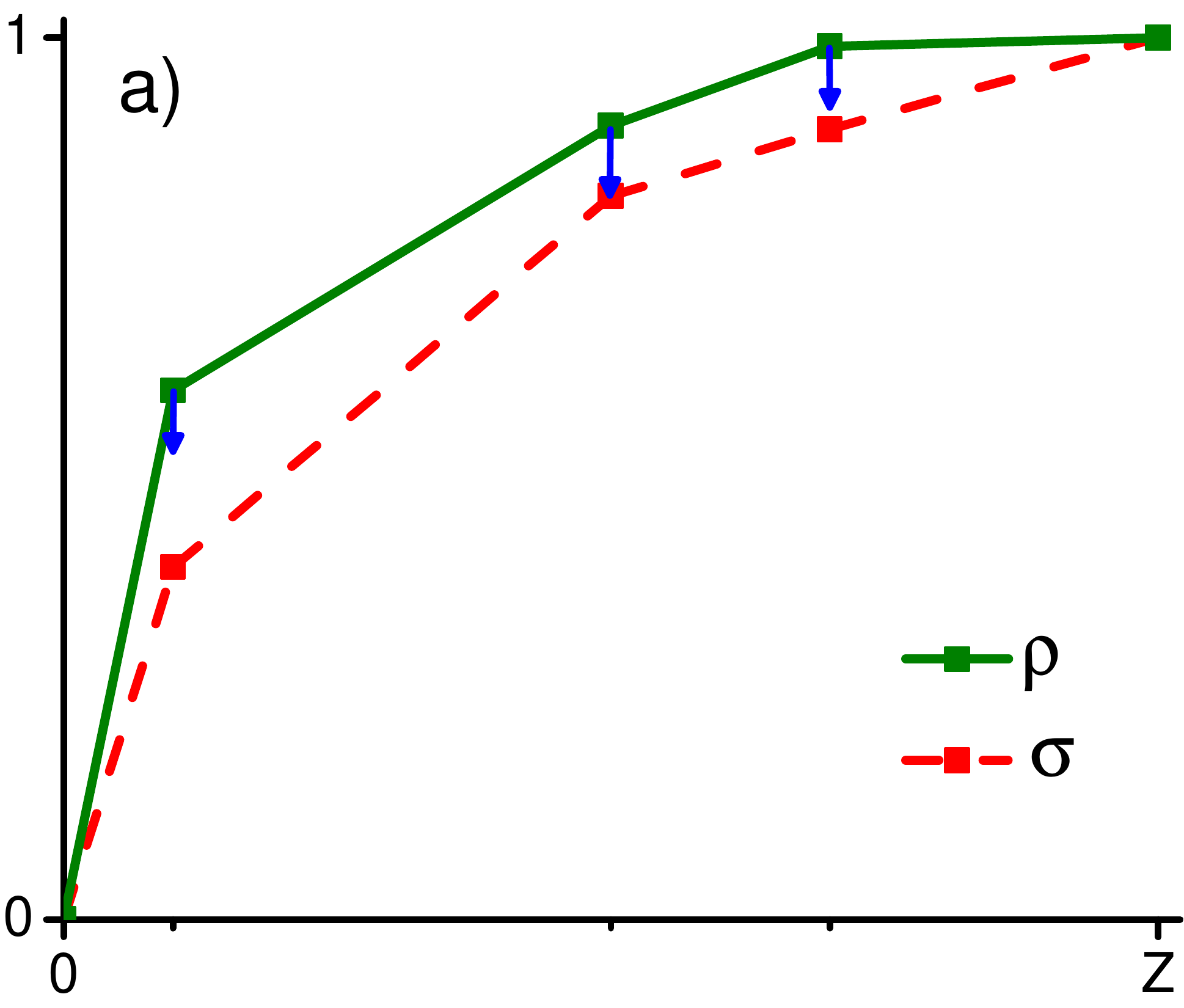}}
  \label{fig:SBO1main}
	\centerline{(a) First \PLT}
\end{minipage}%
\begin{minipage}{.5\columnwidth}
  \centering
  \includegraphics[width=0.9\textwidth]{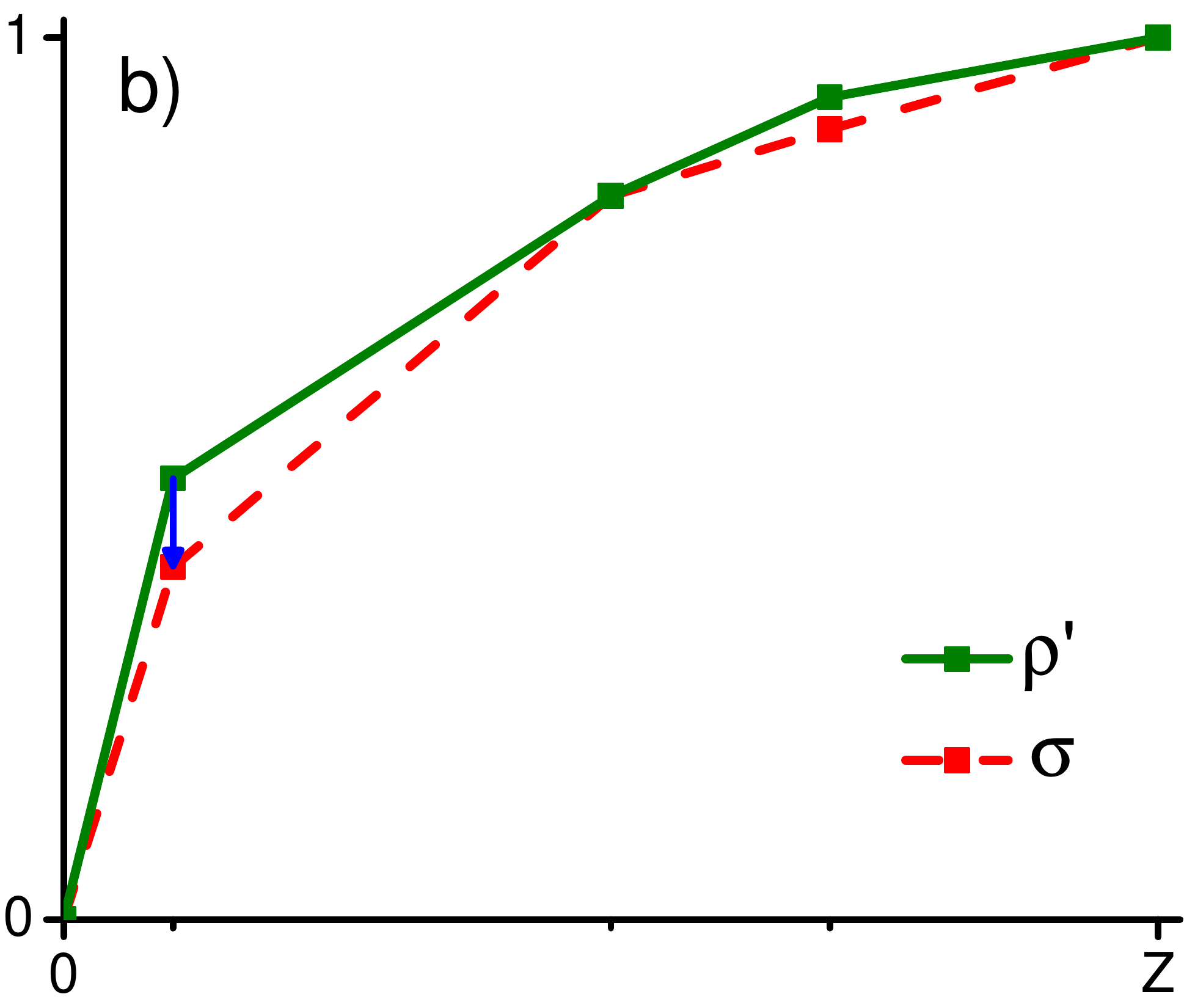}
  \label{fig:SBO2main}
	\centerline{(b) Second \PLT}
\end{minipage}\\
\vspace{.5cm}
\begin{minipage}{.5\columnwidth}
  \centering
  \includegraphics[width=0.9\textwidth]{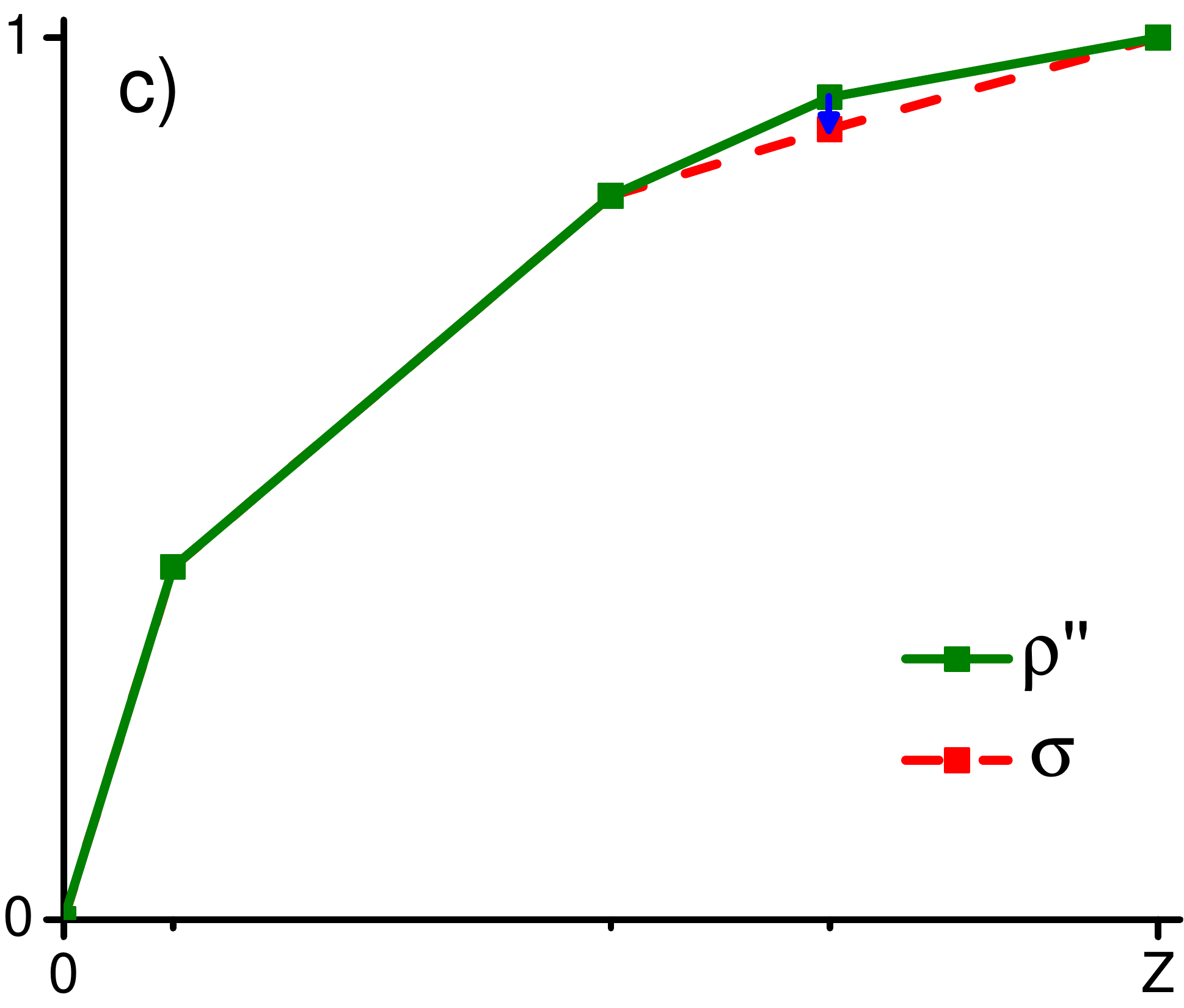}
  \label{fig:SBO3main}
	\centerline{(c) Final \PLT}
\end{minipage}%
\begin{minipage}{.5\columnwidth}
  \centering
  \includegraphics[width=.9\textwidth]{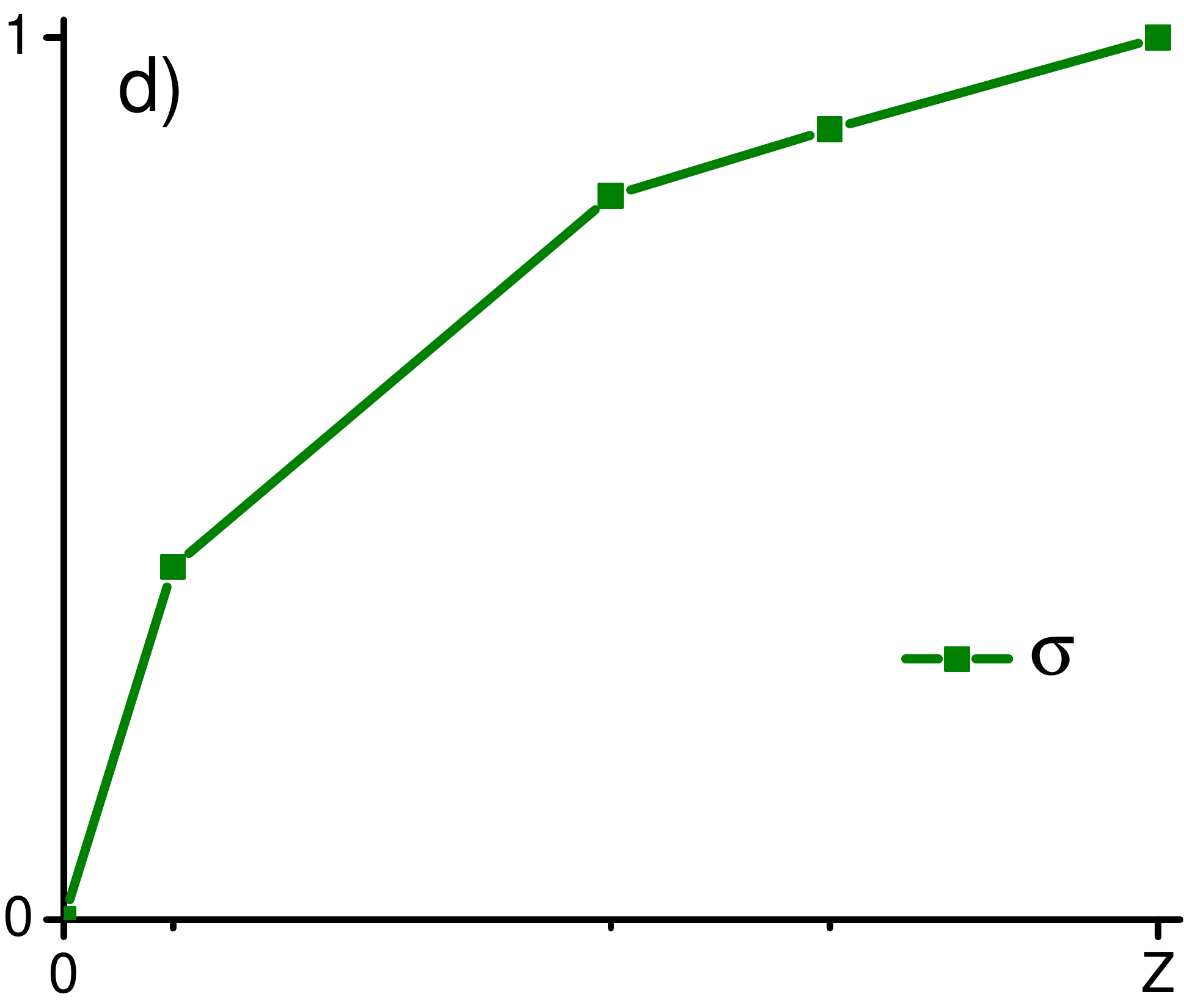}
  \label{fig:SBO4main}
	\centerline{(d) Final state, $\sigma$}
\end{minipage}
\caption{\emph{\longCOs\ protocol for transforming between states with the same $\beta$-order.} If two states, $\rho$ and $\sigma$, have the same $\beta$-order and are such that $\rho$ thermo-majorizes $\sigma$, then $\rho$ can be converted into $\sigma$ using \longPLTs. First a \PLT\ is applied to $\rho$ across the complete set of energy levels, Figure (a), lowering the thermo-majorization of $\rho$ until it meets that of $\sigma$, Figure (b). Next, a second \PLT\ is applied to those energy levels to the left of this meeting point, again lowering the curve until it meets that of $\sigma$ at a second point, Figure (c). By iterating this process, $\rho$ is transformed into $\sigma$, Figure (d).}
\label{fig:SBOmain}
\end{figure}

Transformations involving trivial Hamiltonian, and of distillation and formation in general, are straightforward because the initial and final states can be taken to be such that the elbows on their thermo-majorization curves are, respectively, vertically or horizontally aligned. For trivial Hamiltonians, all states have the same $\beta$-ordering and as such the points on the thermo-majorization diagrams for the initial and final states are vertically aligned. 
Building upon the results of \cite{muirhead1902some} and \cite{hardy1952inequalities}, transformations between states with the same $\beta$-order can be performed by moving points vertically downwards using \longPLTs. The protocol for achieving this is illustrated in Figure \ref{fig:SBOmain} and discussed in Section \ref{sec:same order} of the \supl. For work extraction and formation processes, the protocol is also simpler, since one of states is thermal and hence \ITRs\ can be used to horizontally align its points with those of the other state for free. Transformations then just involve shifting points horizontally through \longLTs.

\begin{figure}
\centering
\includegraphics[width=0.9\columnwidth]{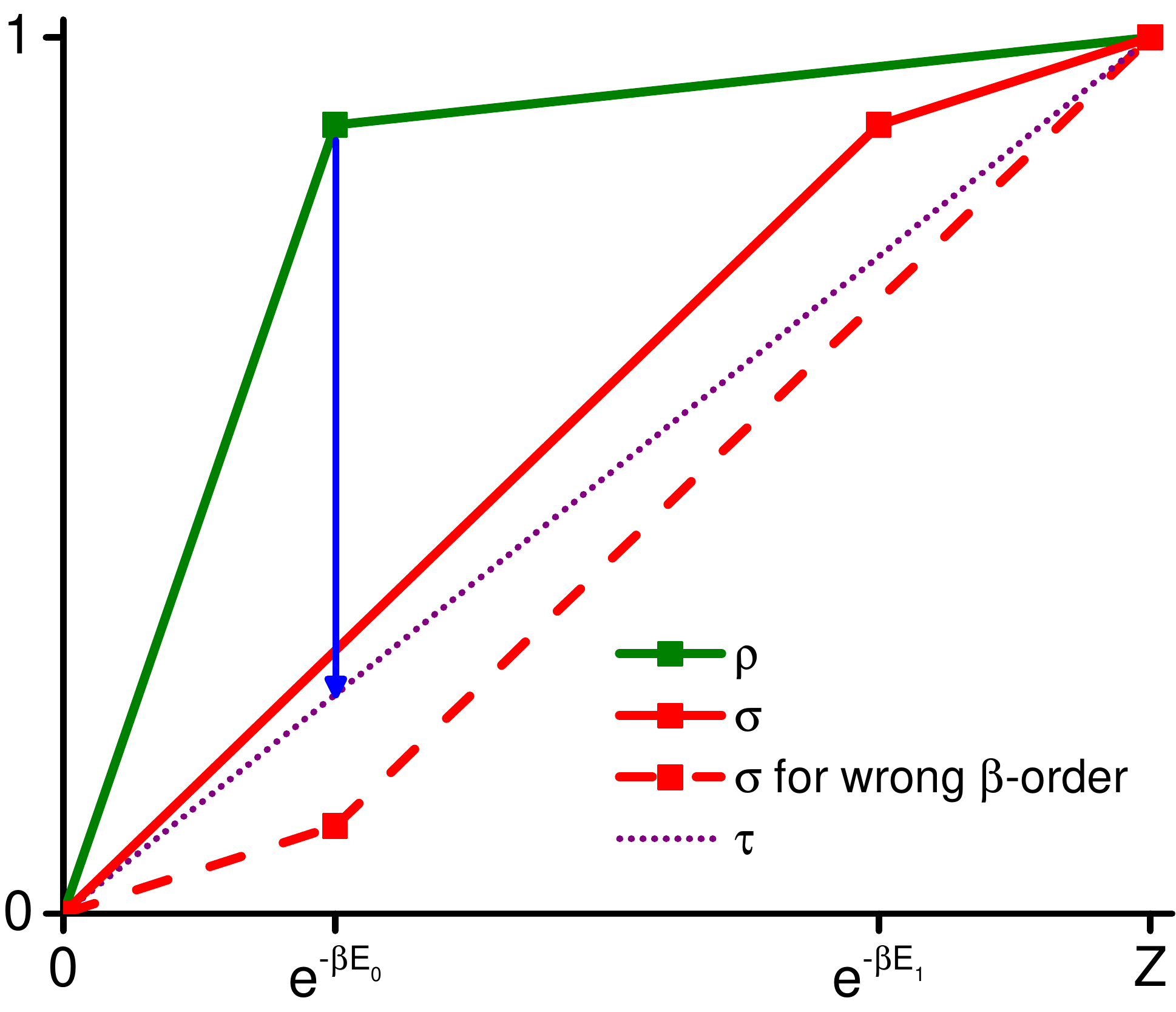}
\caption{\emph{\longPLTs\ are not enough.} Note that \PLTs\ alone cannot implement all transitions possible under thermal operations. More specifically, they cannot implement changes of $\beta$-order on qubits. In the above example, we see that a \PLT\ cannot take the initial state $\rho$ depicted in green to $\sigma$ depicted in red, because it first passes through the thermal state before it reaches $\sigma$ (which we have depicted in both its $\beta$-ordered, and non-$\beta$-ordered from). Because of this, work is required to go from $\rho$ to $\sigma$ even though the optimal process does not require work. } 
\label{fig:nogo}
\end{figure}

On the other hand, if the $\beta$-ordering between the initial and final state is different, then no method is known to transform $\initial$ into $\final$ using only \longCOs\ acting on the system alone. The essential reason for this can be seen by looking at a qubit transformation where the initial and final state have different $\beta$-ordering as depicted in Figure~\ref{fig:nogo}. Essentially, we cannot write $\final=p\initial + (1-p)\gibbs$, and therefore, a \longPLT\ cannot make $\initial\rightarrow\final$, nor can using any combination of \LTs\ and \PLTs.

\begin{figure}
\centering
\begin{minipage}{.5\columnwidth}
  \centering
  \includegraphics[width=0.9\textwidth]{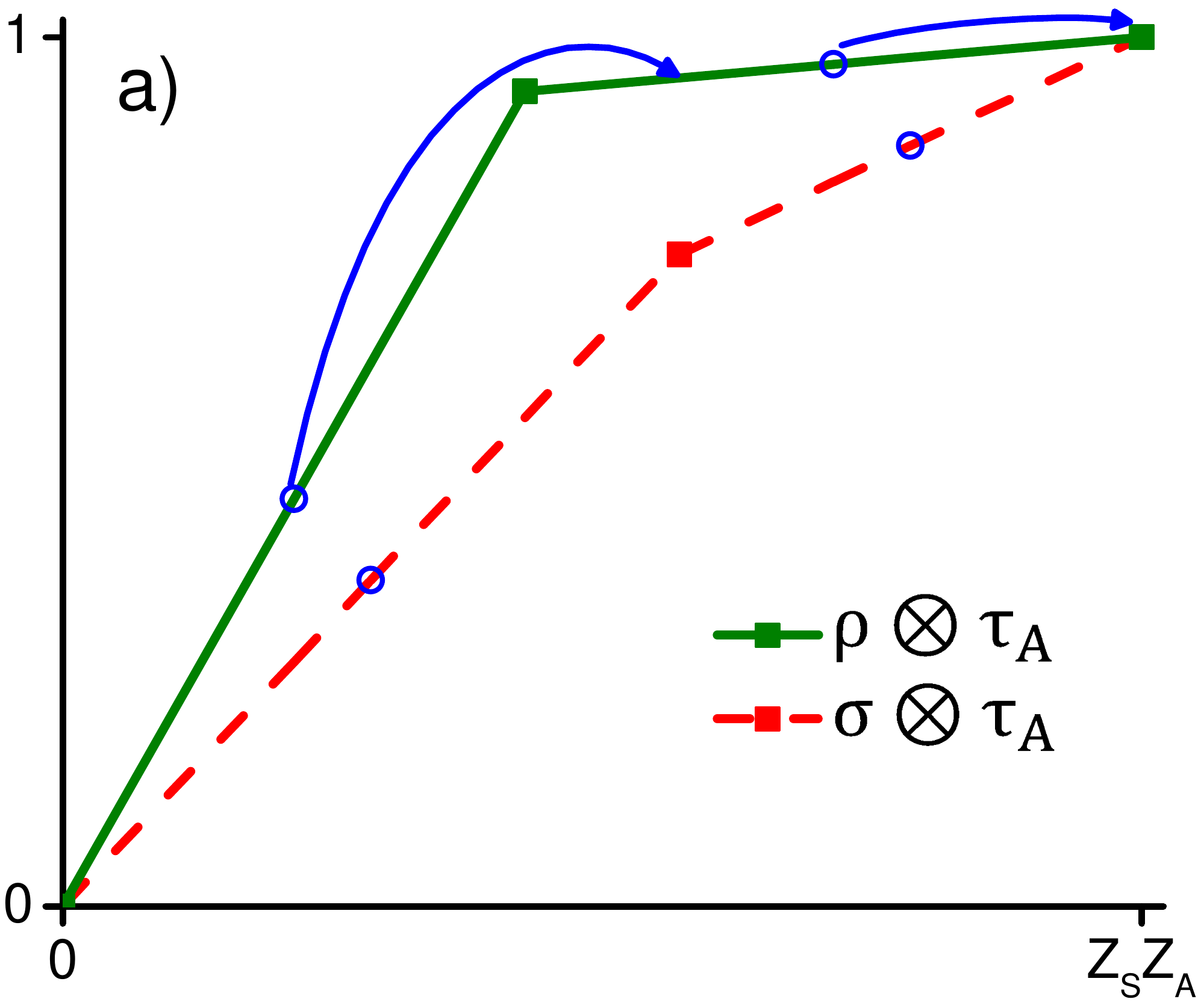}
  \label{fig:DBO1main}
	\centerline{(a) \ITRs\ and \PFs}
\end{minipage}%
\begin{minipage}{.5\columnwidth}
  \centering
  \includegraphics[width=0.9\textwidth]{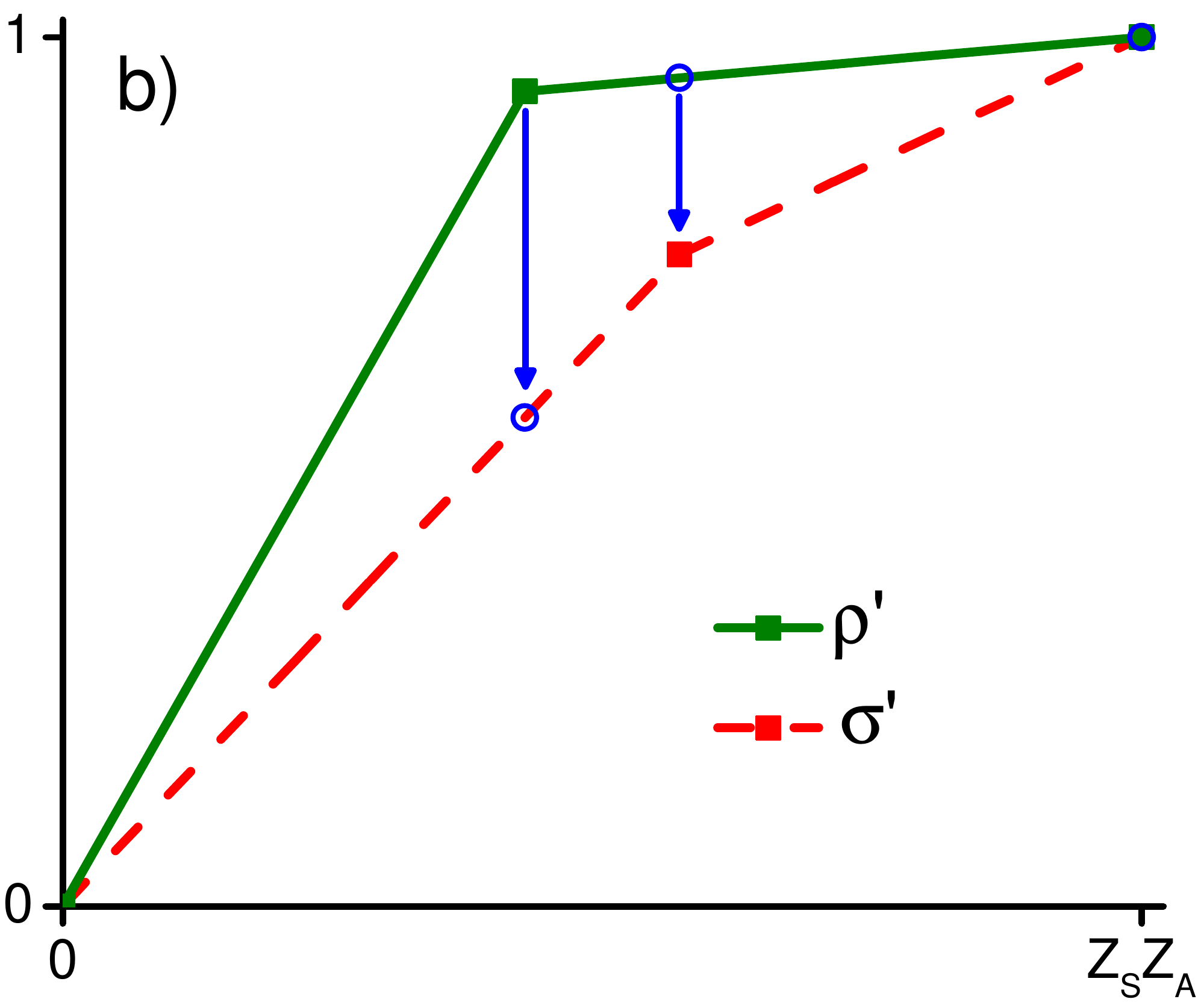}
  \label{fig:DBO2main}
	\centerline{(b) Same $\beta$-order protocol applied}
\end{minipage}\\
\vspace{.5cm}
\begin{minipage}{.5\columnwidth}
  \centering
  \includegraphics[width=0.9\textwidth]{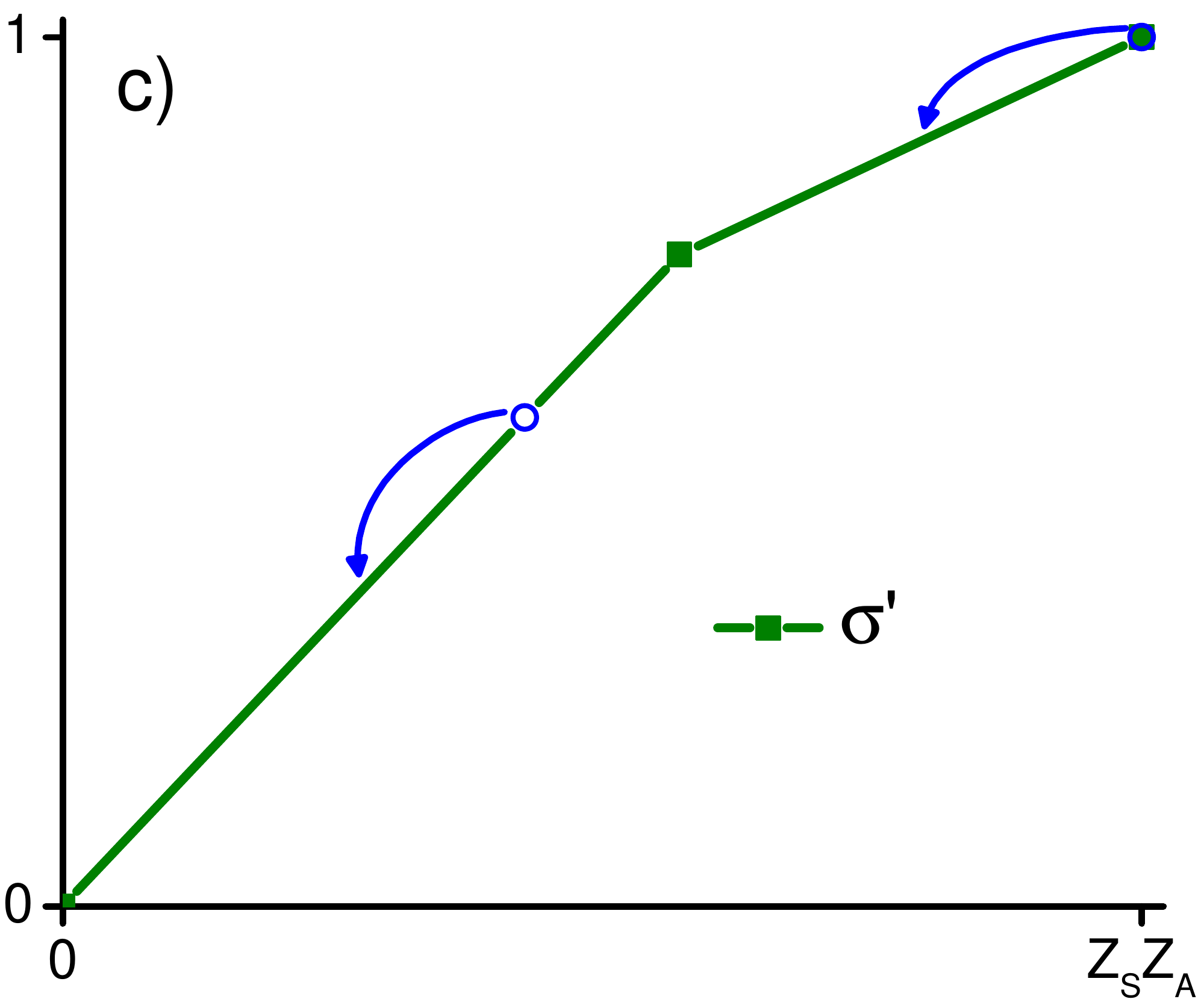}
  \label{fig:DBO3main}
	\centerline{(c) \ITRs\ and \PFs}
\end{minipage}%
\begin{minipage}{.5\columnwidth}
  \centering
  \includegraphics[width=.9\textwidth]{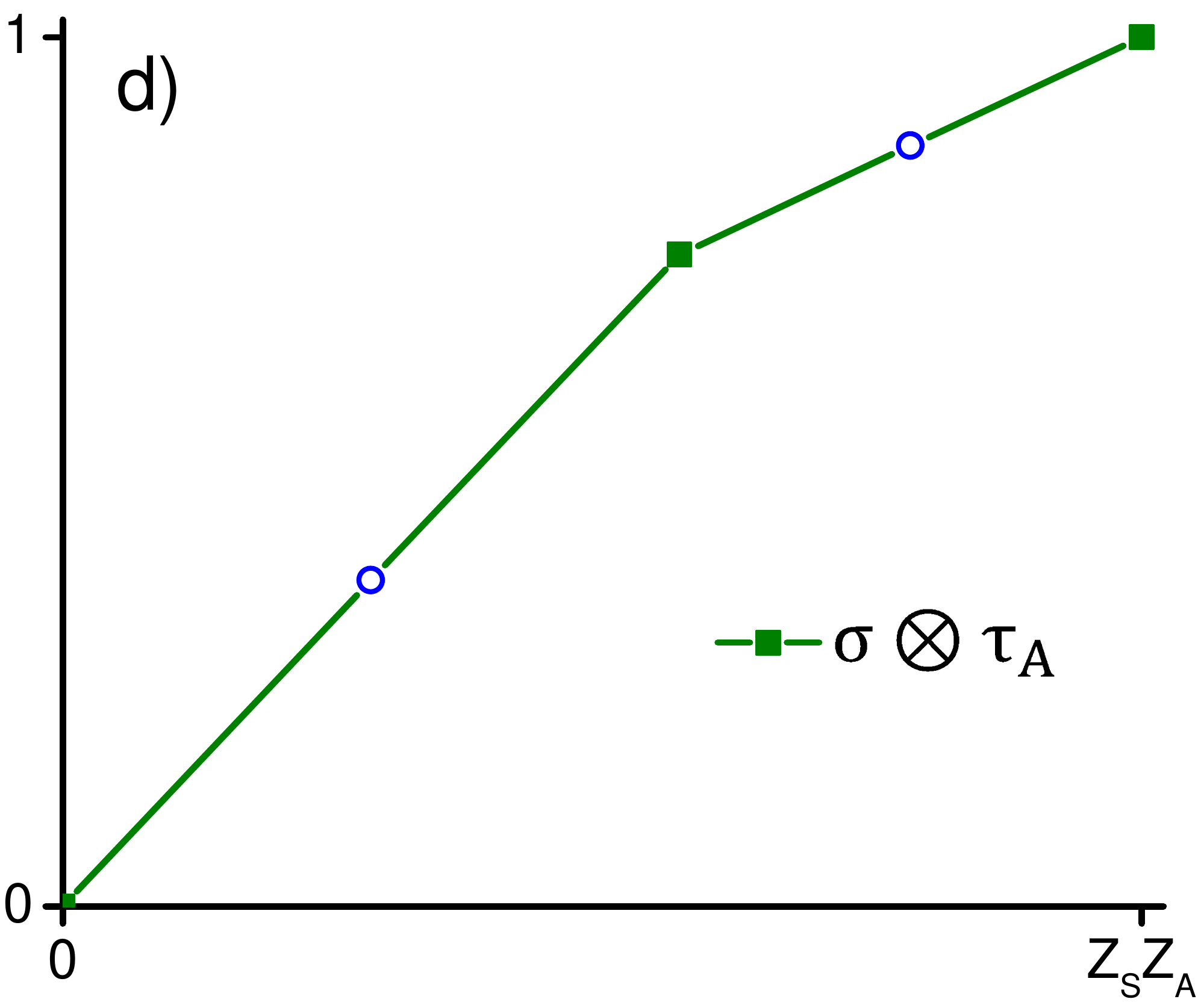}
  \label{fig:DBO4main}
	\centerline{(d) $\sigma\otimes\tau_A$}
\end{minipage}
\caption{\emph{\longCOs\ protocol for transforming between states with different $\beta$-orders.} If $\rho$ thermo-majorizes $\sigma$, then it is possible to transform $\rho$ into $\sigma$ using \longCOs. First, a thermal qubit, $\tau_A$, with known Hamiltonian is appended, Figure (a). Using \longITRs\, the blue circles on the thermo-majorization curve of $\rho\otimes\tau_A$ can be moved to be vertically aligned with the `elbows' on the curve for $\sigma\otimes\tau_A$. This forms the state $\rho'$, Figure (b), that has a thermo-majorization curve overlapping that of $\rho\otimes\tau_A$. Using the previously defined protocol for states with the same $\beta$-order, $\rho'$ can be converted into $\sigma'$, Figure (c), that has a thermo-majorization curve overlapping that of $\sigma\otimes\tau_A$. A final round of \ITRs\ converts $\sigma'$ into $\sigma\otimes\tau_A$, Figure (d), and upon discarding $\tau_A$ we obtain $\sigma$.}
\label{fig:DBOmain}
\end{figure}

However, if we are able to perform \longCOs, not just on the system, but on the system and a single qubit, $\tau_A$, from the thermal bath, then we show that one can transform $\initial$ into $\final$ even if the $\beta$-ordering is different. The protocol is given and explained in Section \ref{sec:DBO} and depicted in Figure \ref{fig:DBOmain}. The idea is to convert this scenario back into one where the states under consideration have the same $\beta$-order. Appending $\tau_A$ adds additional non-elbow points to the thermo-majorization curve associated with $\initial$ and, by using \ITRs, these can be moved so that they are vertically aligned with the elbows of the curve of $\final\otimes\tau_A$. Applying the protocol given in Figure \ref{fig:SBOmain} and a final round of \ITRs\ to properly align the resulting non-elbow points, produces $\final\otimes\tau_A$. To the best of our knowledge, such a thermodynamical state transformation has never been performed in the lab.

\section{Conclusion}

We have shown that Thermal Operations can be simulated by Crude Operations, a class of physical operations which can be implemented in the laboratory using current technology. This ought to bring thermodynamics of microscopic systems further into the experimental domain, and make the exploration of some of the results in the field~\cite{Geusic1967quatum, Alicki_1979heat,  Klimovsky_2013thermal, howard1997molecular, geva1992classical, Hanggi2009brownian, allahverdyan2000extraction, uniqueinfo, feldmann2006lubrication, linden2010small, dahlsten2011inadequacy, del2011thermodynamic, horodecki2013fundamental, aaberg2013truly, faist2012quantitative, Skrzypczyk13, brandao2013second, halpern2014beyond, anders2013thermodynamics, wilming2014weak, mueller2014work, huber2014thermodynamic, narasimhachar2014low, alhambra2015probability, woods2015maximum} more feasible. From a conceptual point of view, this shows that the paradigm of thermodynamics which allows for the maximum amount of control of the system and bath, is in some sense equivalent to one which allows only very crude control of the system and bath. The second laws of thermodynamics, since they are fundamental limitations on state transitions, need to be derived assuming the experimenter has as much control and technology as would be allowed by nature (i.e. Thermal Operations). Yet remarkably, the fundamental limitation of thermo-majorization and generalized free energies, which are derived assuming maximal control, can be achieved with very little control, namely by \COs. Control over bath degrees of freedom, with the exception of one qubit, is not needed.

There are additional second laws, which place constraints not only on the diagonal elements of the density matrix if written in the energy eigenbasis, but also place restrictions on the coherences over energy levels \cite{brandao2013second,lostaglio2015description,cwiklinski2014limitations,korzekwa2015extraction}. While we expect \longCOs\ are also sufficient for the control of quantum coherences, we do not yet know what the necessary conditions are for coherence manipulation, so verifying this is an important open question.
\\

\noindent {\bf Acknowledgments}
Part of this work was carried out during the program "Mathematical Challenges in Quantum Information" (2013) at the Isaac Newton Institute for Mathematical Sciences in the University of Cambridge and we thank them for their hospitality. J.O. is supported by an EPSRC Established Career Fellowship, the Royal Society, and FQXi. M.H. and P.\'C. thank EU grant RAQUEL. M.H. is also supported by  Foundation for Polish Science TEAM project co-financed by the EU European Regional Development Fund. P.\'C. also acknowledges the support from the grant PRELUDIUM 2015/17/N/ST2/04047 from National Science Centre.
\newpage
\widetext

\include{Supplementary_Material/Supplementary}
\newpage

\bibliography{References,common/refthermo,common/refjono2,common/thermo_IC}
\bibliographystyle{apsrev4-1}

\end{document}

%% file: common/definitions.tex
\def\<{\langle}
\def\>{\rangle}

\newcommand{\be}{\begin{eqnarray} \begin{aligned}}
\newcommand{\ee}{\end{aligned} \end{eqnarray} }
\newcommand{\benn}{\begin{eqnarray*} \begin{aligned}}
\newcommand{\eenn}{\end{aligned} \end{eqnarray*} }

\newcommand{\ben}{\begin{eqnarray} \begin{aligned}}
\newcommand{\een}{\end{aligned} \end{eqnarray} }

\newcommand{\bc}{\begin{center}}
\newcommand{\ec}{\end{center}}

\newcommand{\id}{\mathbb{I}}

\newcommand{\tr}{\mathop{\mathsf{tr}}\nolimits}

\newcommand{\beq}{\begin{eqnarray} \begin{aligned}}
\newcommand{\eeq}{\end{aligned} \end{eqnarray} }
\newcommand{\bea}{\begin{array}}
\newcommand{\eea}{\end{array}}

\newcommand{\bee}{\begin{enumerate}}
\newcommand{\eee}{\end{enumerate}}
\newcommand{\bei}{\begin{itemize}}
\newcommand{\eei}{\end{itemize}}
\newtheorem{theorem}{Theorem}

\newtheorem{lemma}[theorem]{Lemma}

\newtheorem{definition}[theorem]{Definition}

\usepackage{amsfonts}

\def\id{\mathbb{I}}

\def\01{\{0,1\}}

\newcommand{\ket}[1]{|#1\rangle}

\newcommand{\ketbra}[2]{|#1\rangle\langle#2|}

\newcommand{\supl}{Appendix}

\def\<{\langle}
\def\>{\rangle}

\def\gibbs{\tau_\beta}
\def\final{\rho'}

\def\initial{\rho}

\newcommand{\newreptheorem}[2]{%
\newenvironment{rep#1}[1]{%
 \def\rep@title{#2 \ref{##1} (restatement)}%
 \begin{rep@theorem}}%
 {\end{rep@theorem}}}
\makeatother

\newreptheorem{thm}{Theorem}
\newreptheorem{lem}{Lemma}

%% file: Supplementary_Material/Supplementary.tex
\appendix

In this \supl, we provide detailed protocols for simulating Thermal Operations with \longCOs. We begin in Section \ref{sec:prelim} with a review of Thermal Operations, and the criteria for state transformations -- thermo-majorization. Next, in \ref{sec:coarseops} we introduce \longCOs, describe each of the basic operations and show that they are a subset of Thermal Operations. 
We then consider the simpler case of transformations between states with the same $\beta$-ordering in Section \ref{sec:same order}, showing that they can be performed using \longCOs. In Section \ref{sec:DBO} we consider transitions between states where the $\beta$-ordering is different, and show that they can be achieved using \longCOs. Quantifying the amount of work used in these processes is done in \ref{sec:worstwork}. Finally, 
Section \ref{sec:Szilard}, contains some very simple examples of state transitions in the case $H=0$, described
using Szilard boxes. It serves an illustrative purpose, but also describes simple examples for extracting work while making a state transition. These were previously presented at \cite{talk_qip2013}.

\section{Preliminaries}
\label{sec:prelim}

In this appendix, we recall and define relevant concepts and results from the resource theory of thermal operations (TO).

\subsection{Thermal Operations} \label{ssec:thermal operations}

The resource theory of thermal operations \cite{ Streater_dynamics,janzing2000thermodynamic, horodecki2013fundamental} is a way of more rigorously defining thermodynamics, and allows us to account for all sources of work and heat, something which needs to be done precisely when investigating the thermodynamics of a small system in the presence of a large heat bath. It is equivalent to other well studied paradigms of thermodynamics as shown in \cite{brandao2011resource}.
Given a system in state $\rho$ with Hamiltonian $H$ and such a heat bath at temperature $T$, the following operations can be applied:
\begin{enumerate}
\item Since heat baths are a free resource, an ancillary system with any Hamiltonian, in the Gibbs state of that Hamiltonian at temperature $T$, can be appended as an ancilla. For the Hamiltonian $H_B=\sum_{i=1}^{n} E_i \ketbra{i}{i}$, the corresponding Gibbs state at temperature $T$ is defined as:
\begin{equation}
\tau_B=\frac{1}{Z_B}\sum_{i=1}^{n}e^{-\beta E_i} \ketbra{i}{i},
\end{equation}
where $\beta$ is the inverse temperature and $Z_B=\sum_{i=1}^{n} e^{-\beta E_i}$ is the partition function.
\item Any energy-conserving unitary, i.e. those unitaries that commute with the total Hamiltonian, can be applied to the global system. This is not a restriction, since any non-energy conserving unitary can be implemented using a work system and performing an energy conserving unitary on the work system and the original one. It does however allow us to properly account for work.
\item Any subsystem can be discarded through tracing out.
\end{enumerate}
Given these, the action of a thermal operation on the state $\rho$ of a state-Hamiltonian pair $\left(\rho,H_S\right)$ at temperature $T$ can be written: 
\begin{equation}
\rho\stackrel{\textrm{TO}}{\longrightarrow}\tr_A\left[U\left(\rho\otimes\tau_B\right)U^\dagger\right],
\end{equation}
where $\tau_B$ is the Gibbs state of a Hamiltonian $H_B$, $A$ is the subsystem to be discarded and $U$ is a unitary such that:
\begin{equation}
\left[U,H_S+H_B\right]=0,
\end{equation}
and we have used the shorthand notation $H_S+H_B=H_S\otimes\id_B+\id_S\otimes H_B$.

\subsection{Thermo-majorization} \label{ss:thermomaj}

Suppose we have a system $\left(\rho,H_S\right)$. Determining whether it can be transformed into the system $\left(\sigma,H_S\right)$ using thermal operations, can be formulated using thermo-majorization diagrams \cite{horodecki2013fundamental}. 
\begin{definition}[Thermo-majorization diagrams]
Given an $n$-level system, let $H_S=\sum_{i=1}^{n}E_i \ketbra{i}{i}$ be the system's Hamiltonian and let us consider the energy occupation
probabilities $\eta_i=\tr{\rho\ketbra{i}{i}}$ 
To define the thermo-majorization curve of $\rho$, we first $\beta$-order the probabilities $\eta_i$ and energy levels, listing them such that $\eta_i e^{\beta E_i}$ is in non-increasing order. 

The thermo-majorization curve of $\rho$ is then formed by plotting the $\beta$-ordered points:
\begin{equation} \label{eq:th-maj set}
\left\{\sum_{i=1}^{k} e^{-\beta E^{\left(\rho\right)}_i},\sum_{i=1}^{k} \eta_i^{\left(\rho\right)}\right\}_{k=1}^{n},
\end{equation}
together with $\left(0,0\right)$, and connecting them piecewise linearly to form a concave curve. Here, the superscript $\left(\rho\right)$ on $E_i$ and $\eta_i$ indicate that they have been $\beta$-ordered and that this ordering depends on $\rho$.
\end{definition}

\begin{figure}
\centering
\includegraphics[width=0.7\columnwidth]{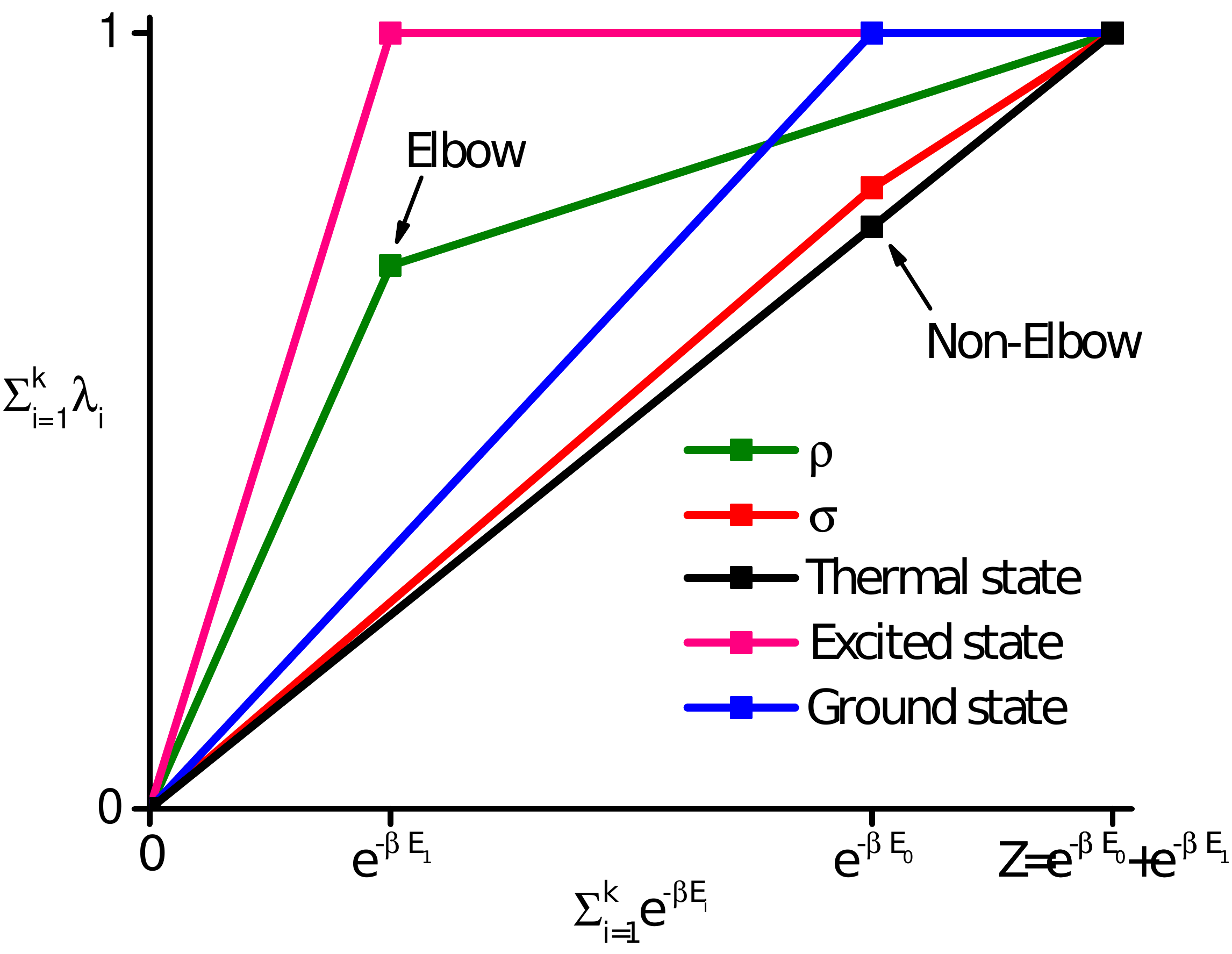}
\caption[Thermo-majorization curves.]{\emph{Thermo-majorization curves.} \emph{a)} The thermo-majorization curve for 
$\rho$, is defined by plotting the points: $\left\{\sum_{i=1}^{k} e^{-\beta E_i^{\left(\rho\right)}},\sum_{i=1}^{k}\eta_i^{\left(\rho\right)}\right\}_{k=1}^{n}$.
The superscript $\left(\rho\right)$ indicates that the occupation probabilities and their associated energy levels have been ordered such that $\eta_i^{\left(\rho\right)} e^{\beta E_i^{\left(\rho\right)}}$ is non-increasing in $i$. This is called $\beta${\it-ordering}. Note that states associated with the same Hamiltonian may have different $\beta$-orderings, as illustrated by $\rho$ and $\sigma$ here. \emph{b)} Here, we say that $\rho$ {\it thermo-majorizes} $\sigma$ as the thermo-majorization curve of $\rho$ is never below the thermo-majorization curve of $\sigma$. \emph{c)} The thermo-majorization curve of a Gibbs state is given by a straight line between $\left(0,0\right)$ and $\left(Z,1\right)$. All other states thermo-majorize it. \emph{d)} The pure state corresponding to the highest energy level of an $n$-level system thermo-majorizes all other states associated with that Hamiltonian. \emph{e)} We call a point on a curve an elbow, if the gradient of the curve changes as it passes through the point. Otherwise, it is a non-elbow.}
\label{fig:thermo diagrams}
\end{figure}

An example of thermo-majorization diagrams is shown in Figure \ref{fig:thermo diagrams}. We say that a state, thermo-majorizes another if its thermo-majorization curve never lies below that of the other.

With these definitions in place, we can give the result of \cite{horodecki2013fundamental}:
\begin{theorem}[Thermo-majorization] Given two states $\rho$ and $\sigma$ of an $n$-level system with Hamiltonian $H_S$:
\begin{enumerate}
\item If $\sigma$ is block-diagonal in the energy eigenbasis, then $\rho\stackrel{\textrm{TO}}{\longrightarrow}\sigma$, if and only if $\rho$ thermo-majorizes $\sigma$.
\item In general, $\rho\stackrel{\textrm{TO}}{\longrightarrow}\sigma$, only if $\rho$ thermo-majorizes $\sigma$.
\end{enumerate}
\end{theorem}

Thus, for states which are block diagonal in the energy basis, we have necessary and sufficient conditions to determine possible transitions between $\rho$ and $\sigma$. For quantum states with off-diagonal elements - coherences, the full set of constrains is only known in the case of a single 
qubit \cite{cwiklinski2014limitations}, although various necessary conditions are 
known \cite{brandao2013second,lostaglio2015description,cwiklinski2014limitations,korzekwa2015extraction}. On
the other hand, if the experimenter is allowed access to a reference frame, then thermo-majorization is a necessary and sufficient condition for all
state transformations \cite{brandao2011resource,aberg2014catalytic,korzekwa2015extraction}.

\subsection{Wits and switch qubits} \label{ssec:wits&switch}

\subsubsection{Work} \label{ssec:wits}

In general, if we want a transition from $\rho$ to $\sigma$ to be possible, work may have to be added. Alternatively, if a transition can be achieved with certainty, it can be possible to extract work. Within the thermal operations framework, the optimal amount of work that must be added or can be gained, the work of transition, can be quantified using a continuous system, or for example the energy gap $W$ of a 2-level system, a wit \cite{horodecki2013fundamental}, with zero energy state $\ket{0}$ and an additional state $\ket{1}$. The associated Hamiltonian is:
\begin{equation}
H_W= W\ketbra{1}{1}.
\end{equation}
The deterministic work of transition, denoted $\Wtran{\rho}{\sigma}$, is then defined to be the greatest value of $W$ such that the following holds:
\begin{equation}\label{eq:work def}
\left(\rho\otimes\ketbra{0}{0},H_S+H_W\right)\stackrel{\textrm{TO}}{\longrightarrow}\left(\sigma\otimes\ketbra{1}{1},H_S+H_W\right).
\end{equation}
If $\Wtran{\rho}{\sigma}$ is negative, to convert $\rho$ into $\sigma$ work has been taken from the work storage system to enable the transition to take place. On the other hand, if $\Wtran{\rho}{\sigma}$ is positive, in converting $\rho$ into $\sigma$ it has been possible to store some extracted work in the work system.

Defining work in such a way enables the quantification of the worst-case work of a process.  When $\Wtran{\rho}{\sigma}$ is negative, it can be interpreted as the smallest amount of work that must be supplied to guarantee the transition. If it is positive, it is the largest amount of work we are guaranteed to extract in the process. As the work system is both initially and finally in a pure state, no entropy is stored in it and its energy change must be completely due to work being exchanged with the system.  One can also quantify average work and consider fluctuations of the work system, however this does not change the considerations given here.

\subsubsection{Changes of Hamiltonian}

Thermodynamics is not just concerned with a system with Hamiltonian $H$ and whether it is possible to transform $\rho$ into $\sigma$ whilst keeping the $H$ fixed. One also wants to be able to consider transitions where the Hamiltonian changes, i.e.:
\begin{equation}
\left(\rho,H_1\right)\longrightarrow\left(\sigma,H_2\right),
\end{equation}
and to determine the work cost or yield of performing such a change. Following \cite{horodecki2013fundamental}, this scenario can be mapped to one with identical initial and final Hamiltonian using the \emph{switch qubit} construction. Here, we instead consider the transition between $\left(\rho\otimes\ketbra{0}{0}\right)$ and $\left(\sigma\otimes\ketbra{1}{1}\right)$ with Hamiltonian:
\begin{equation} \label{eq:Ham change}
H=H_1\otimes\ketbra{0}{0}+H_2\otimes\ketbra{1}{1}.
\end{equation}
Note that the partition function associated with $H$ is $Z=Z_1+Z_2$, where $Z_1$ and $Z_2$ are the partition functions of $H_1$ and $H_2$ respectively.

To model the work of transition in such a process, we combine this with Eq. \eqref{eq:work def} to give:
\begin{equation}
\left(\rho\otimes\ketbra{0}{0}\otimes\ketbra{0}{0},H+H_{W}\right)\stackrel{\textrm{TO}}{\longrightarrow}\left(\sigma\otimes\ketbra{1}{1}\otimes\ketbra{1}{1},H+H_{W}\right),
\end{equation}
and maximize over $W$.

\newpage

\section{\longCOs\ as thermal operations} \label{sec:coarseops}

In this appendix, we show how \longCOs\ (\CO) which we will now introduce are a subset of thermal operations. 

Under \longCOs\ the following operations can be applied:
\begin{enumerate}
\item A single 2-level system with known Hamiltonian, in the Gibbs state of that Hamiltonian at temperature $T$, can be appended as an ancilla.
\item \longPLT\ (\PLT) over any subset of energy levels.
\item \longLTs\ (\LT), provided the work cost is accounted for.
\item The ancilla system may be discarded.
\item \longSRs\ (\SR). Any energy conserving unitary $U$ such that $\left[U,H_S\right]=0$ may be applied to the system alone.
\end{enumerate}
Operation 1 can obviously be realized using Operation 1 of thermal operations as described in Appendix \ref{ssec:thermal operations} but as we restrict to one ancillary system, the converse does not hold. As we shall see, Operation 2 can be realized using thermal operations whilst in general, Operation 3 may cost work to perform. Operation 5 is only ever used in the case of a degenerate energy eigenbasis and when the eigenbasis of the initial state does not match the energy eigenbasis of the target state.  Let us now consider these operations in detail.

\subsection{\longPLTs}

\begin{figure}
\centering
\includegraphics[width=0.5\columnwidth]{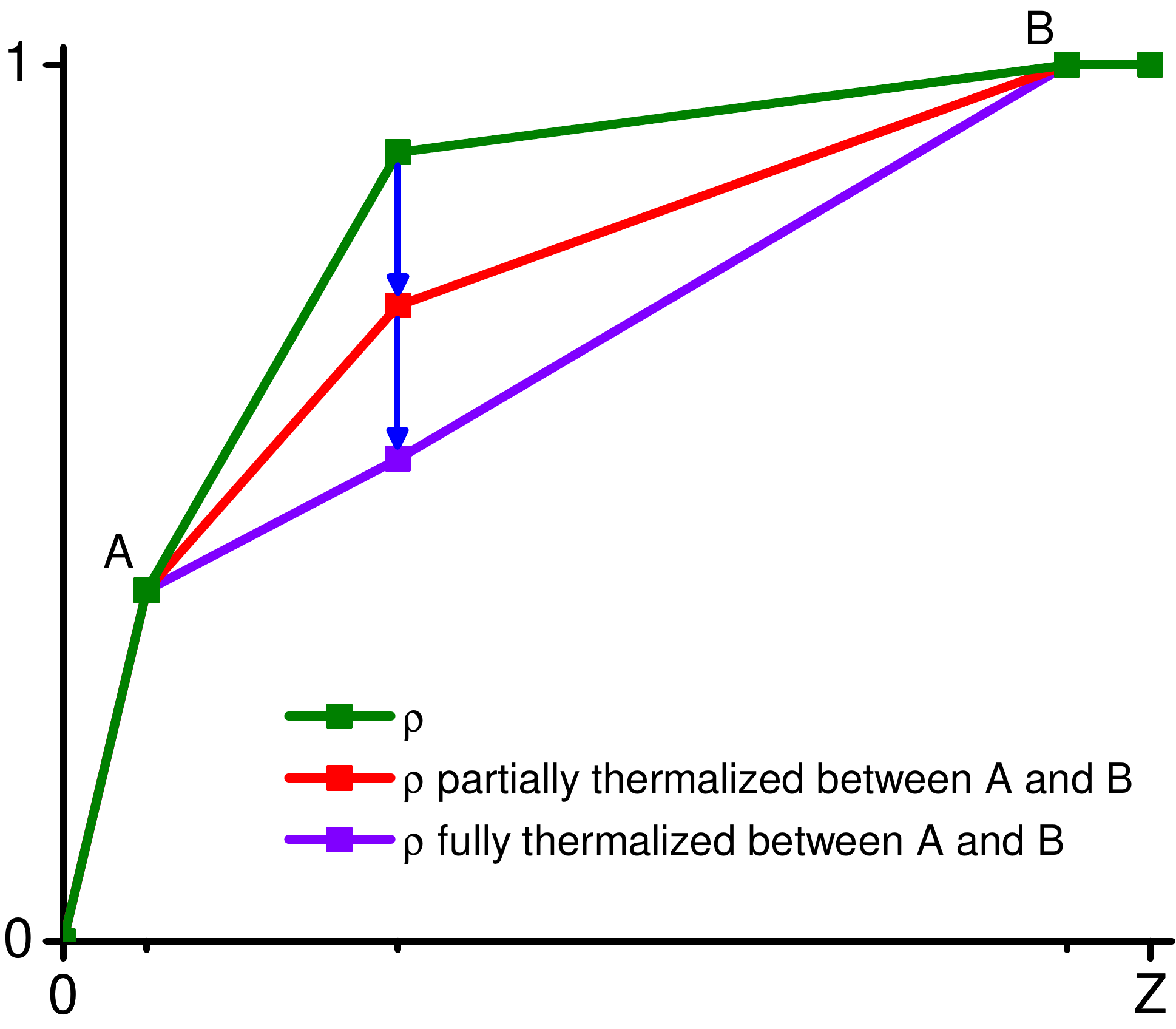}
\caption{\emph{Action of \longPLT.} Here we illustrate the action of \PLTs\ applied to a state $\rho$ over the two energy levels between points $A$ and $B$ for various degrees of thermalization. Note, that as we apply the \PLTs\ to adjacent energy levels with respect to the $\beta$-ordering of $\rho$, the final states maintain this $\beta$-order.} 
\label{fig:PT}
\end{figure}

\longPLTs\ act on a subset of a system's energy levels, partially thermalizing the state supported on those levels. More formally:
\begin{definition}[\longPLT]
Given an $n$-level system with Hamiltonian $H_S=\sum_{i=1}^{n} E_i \ketbra{i}{i}$, a \longPLT\ is parametrized by $\lambda\in\left[0,1\right]$ and acts on some subset of the system's energy levels. Denote this subset of energy levels by $\mathcal{P}$ and the \longPLT\ by $\PLT_{\mathcal{P}}\left(\lambda\right)$.

The action of $\textrm{\PLT}_{\mathcal{P}}\left(\lambda\right)$ on $\rho=\sum_{i=1}^{n}\eta_i\ketbra{i}{i}$, is defined by:
\begin{equation}
\rho\stackrel{\textrm{\PLT}_{\mathcal{P}}\left(\lambda\right)}{\longrightarrow} \rho',
\end{equation}
where $\rho'=\sum_{i=1}^{n} \eta'_i\ketbra{i}{i}$ and the $\eta'_i$ are such that for $i\in\mathcal{P}$:
\begin{equation} \label{eq:PLT components}
\eta'_i=\left(1-\lambda\right)\eta_i + \frac{\lambda e^{-\beta E_i}}{\sum_{i\in\mathcal{P}}e^{-\beta E_i}}\sum_{i\in\mathcal{P}}\eta_i,
\end{equation}
and $\eta'_i=\eta_i$ otherwise. The action of a \longPLT, is illustrated in terms of thermo-majorization curves in Figure \ref{fig:PT}.
\end{definition}


Note that such an operation preserves the $\beta$-ordering of the levels in $\mathcal{P}$ as, for $i,j\in\mathcal{P}$, if $\eta_i e^{\beta E_i}\geq \eta_j e^{\beta E_j}$, then  $\eta'_i e^{\beta E_i}\geq \eta'_j e^{\beta E_j}$. In particular, if for some $d$, $\mathcal{P}=\left\{i^{\left(\rho\right)}\right\}_{i=k}^{k+d-1}$ (with the superscript $\left(\rho\right)$ denoting that the energy levels are $\beta$-ordered with respect to $\rho$), then $\rho'$ will have the same $\beta$-ordering as $\rho$. 

That \longPLT\ form a subset of thermal operations, is captured in the following Lemma.
\begin{lemma} \label{le:PLT as TO}
The map $\textrm{\PLT}_{\mathcal{P}}\left(\lambda\right)$ can be implemented using thermal operations.
\end{lemma}
\begin{proof}
To show this, we give an explicit protocol implementing a \longPLT. For simplicity, we assume $\lambda$ is a positive rational of the form $\frac{a}{b}$ (if $\lambda$ is irrational, then $a$ and $b$ should be chosen such that the \longPLT\ is implemented to the desired accuracy). Let the state of the system be $\rho=\sum_{i=1}^{n}\eta_i\ketbra{i}{i}$ and the associated Hamiltonian, $H_S=\sum_{i=1}^{n}E_i\ketbra{i}{i}$. The protocol then runs as follows:
\begin{itemize}
\item Step 1. Using Operation 1 of thermal operations:
\begin{equation*}
\left(\rho,H_S\right)\rightarrow\left(\rho\otimes\tau_A\otimes\id_b,H_S+H_A+H_M\right),
\end{equation*}
where $\tau_A$ is the Gibbs state of a $|\mathcal{P}|$-level system with Hamiltonian:
\begin{equation*}
H_A=\sum_{j\in\mathcal{P}} E_j \ketbra{j}{j},
\end{equation*}
and $\id_b$ is the maximally mixed state of dimension $b$. This is a Gibbs state of the $b$-level system with Hamiltonian $H_M=0$.
\item Step 2. Let $\left\{\ket{r_S,s_A,t_M}\right\}$ be the set of orthonormal eigenvectors of $H_S+H_A+H_M$, each with associated energy level $E_r+E_s$. The eigenvalue of $\rho\otimes\tau_A\otimes\id_b$ associated with each energy level is $\frac{1}{bZ_A}\eta_r e^{-\beta E_s}$. Let $U$ be a unitary acting on the global system such that for $r \in \mathcal{P}$, $\forall s$, $t\in\left\{1,\dots,a\right\}$:
\begin{equation*}
U\ket{r_S,s_A,t_M}=\ket{s_S,r_A,t_M},
\end{equation*}
and $U\ket{r_S,s_A,t_M}=\ket{r_S,s_A,t_M}$ otherwise. By construction, $U$ is an energy conserving unitary that commutes with the total Hamiltonian.
\item Step 3. Discard the two ancilla systems.
\end{itemize}
After applying this protocol, the population of energy level $E_j$ for $j\in\mathcal{P}$ is:
\begin{align*}
&\eta_j - \frac{a}{bZ_A}\sum_{i\in\mathcal{P}}e^{-\beta E_i}\eta_j + \frac{a}{bZ_A}\sum_{i\in\mathcal{P}}e^{-\beta E_j}\eta_i,\\
=&\left(1-\frac{a}{b}\right)\eta_j+\frac{a}{b}\frac{e^{-\beta E_j}}{\sum_{i\in\mathcal{P}}e^{-\beta E_i}}\sum_{i\in\mathcal{P}}\eta_i,
\end{align*}
and $\eta_j$ otherwise. Comparing this with Eq.~\eqref{eq:PLT components}, we see that we have implemented the \longPLT\ as required.
\end{proof}

\subsection{\longLT}

\begin{figure}
\centering
\includegraphics[width=0.5\columnwidth]{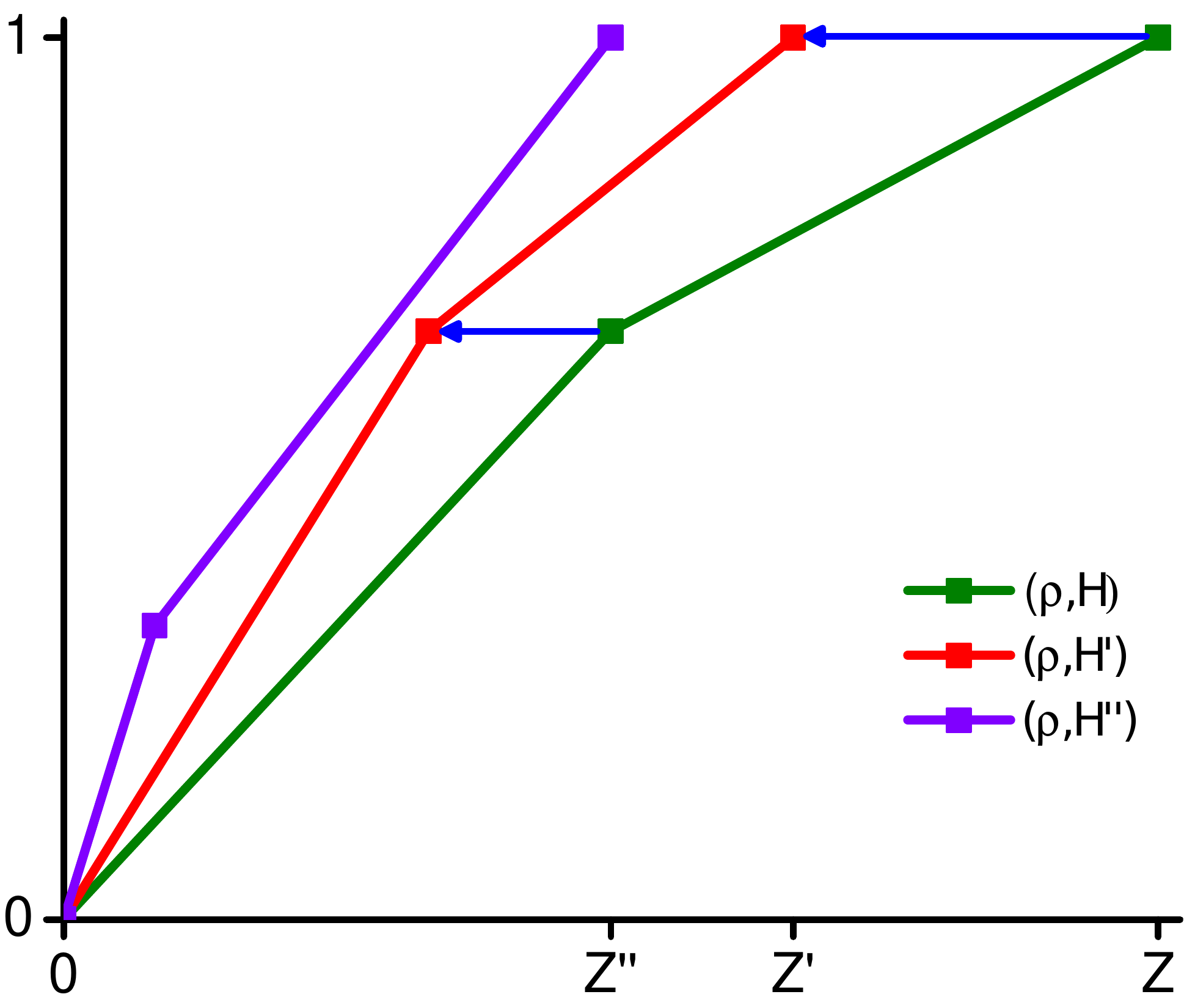}
\caption{\emph{Action of \longLTs.} Here we illustrate the action of \LTs\ applied to a system with state-Hamiltonian pair $\left(\rho,H\right)$. Note that \LTs\ leave the occupation probabilities of $\rho$ unchanged but may alter the $\beta$ ordering.} 
\label{fig:LT}
\end{figure}

\longLTs\ adjust the energy levels of a system's Hamiltonian while keeping the state of the system fixed.
\begin{definition}[\longLT]
Given an $n$-level system with Hamiltonian $H_S=\sum_{i=1}^{n} E_i \ketbra{i}{i}$ in the state $\rho=\sum_{i=1}^{n}\eta_i\ketbra{i}{i}$, a \longLT\ is parametrized by a set of real numbers $\mathcal{E}=\left\{h_i\right\}_{i=1}^{n}$ and denoted by $\LT_{\mathcal{E}}$.

The action of $\LT_{\mathcal{E}}$ on $\left(\rho,H_S\right)$ is:
\begin{equation}
\left(\rho,H_S\right)\stackrel{\LT_{\mathcal{E}}}{\longrightarrow}\left(\rho,H'_S\right),
\end{equation}
where:
\begin{equation}
H'_S=\sum_{i=1}^{n}\left(E_i+h_i\right)\ketbra{i}{i}.
\end{equation}

The single-shot, worst-case, work cost/yield of $\LT_{\mathcal{E}}$ is defined by:
\begin{equation} \label{eq:LT cost}
W_{\LT_{\mathcal{E}}}=-\max_{i:\eta_i>0} h_i.
\end{equation}
If $W_{\LT_{\mathcal{E}}}$ is negative, work must be added for the transformation to happen deterministically while if it is positive, it may be possible to extract some work.
\end{definition}
The action of a \longLT, is illustrated in terms of thermo-majorization curves in Figure \ref{fig:LT}.

\longLT\ can be modeled within thermal operations using the wit and switch qubit discussed in Appendix \ref{ssec:wits&switch}.
\begin{lemma}
The map $\LT_{\mathcal{E}}$ can be implemented using thermal operations with work cost at most $W_{\LT_{\mathcal{E}}}$.
\end{lemma}
\begin{proof}
Let $H=\sum_{i=1}^{n} E_i \ketbra{i}{i}$ be the initial Hamiltonian and $H'=\sum_{i=1}^{n} E'_i\ketbra{i}{i}$ be the final Hamiltonian after the application of $\LT_{\mathcal{E}}$. Let $\mathcal{E}=\left\{h_i\right\}_{i=1}^{n}$ so $E'_i=E_i+h_i$.

Consider modeling this transformation using the switch qubit construction:
\begin{equation*}
H_T=H\otimes\ketbra{0}{0}+H'\otimes\ketbra{1}{1}.
\end{equation*}
Let $H_W=W\ketbra{1}{1}$ be the Hamiltonian for the wit. The work require to implement the \longLT\, and convert the system $\left(\rho,H\right)$ into $\left(\rho,H'\right)$ under thermal operations $W_{H\rightarrow H'}$ is then given by the largest value of $W$ such that:
\begin{equation*}
\left(\rho\otimes\ketbra{0}{0}\otimes\ketbra{0}{0},H_T+H_W\right)\stackrel{\textrm{TO}}{\longrightarrow}\left(\rho\otimes\ketbra{1}{1}\otimes\ketbra{1}{1},H_T+H_W\right).
\end{equation*}

To see that $W_{H\rightarrow H'} \geq W_{\LT_{\mathcal{E}}}$, consider the \longLT\ parametrized by $\tilde{\mathcal{E}}=\left\{\tilde{h}_i\right\}_{i=1}^{n}$ where $\tilde{h}_i=W_{\LT_{\mathcal{E}}}$, $\forall i$. Let $\tilde{H}$ denote the Hamiltonian obtained by applying $\LT_{\tilde{\mathcal{E}}}$ to $H$ and note that the \longLT\ is such that $W_{\LT_{\mathcal{E}}}=W_{\LT_{\tilde{\mathcal{E}}}}$. To model this \longLT\ using thermal operations let:
\begin{equation*}
\tilde{H}_T=H\otimes\ketbra{0}{0}+\tilde{H}\otimes\ketbra{1}{1},
\end{equation*}
and its work cost under thermal operations $W_{H\rightarrow \tilde{H}}$ is given by the largest value of $W$ such that:
\begin{equation*}
\left(\rho\otimes\ketbra{0}{0}\otimes\ketbra{0}{0},\tilde{H}_T+H_W\right)\stackrel{\textrm{TO}}{\longrightarrow}\left(\rho\otimes\ketbra{1}{1}\otimes\ketbra{1}{1},\tilde{H}_T+H_W\right).
\end{equation*}
It can easily be seen that $W_{H\rightarrow \tilde{H}}=W_{\LT_{\tilde{\mathcal{E}}}}$ and $W_{H\rightarrow \tilde{H}}\leq W_{H\rightarrow H'}$. Hence the result follow.
\end{proof}

Note that this result implies that implementing the effect of a \longLT\ using a switch qubit and thermal operations, can be more cost-effective (in terms of work required to make the transformation deterministically) then performing a \longLT\ itself.

\subsection{\longITRs\ and \longPFs} \label{ssec:points flow}

\begin{figure}
\centering
\includegraphics[width=0.5\columnwidth]{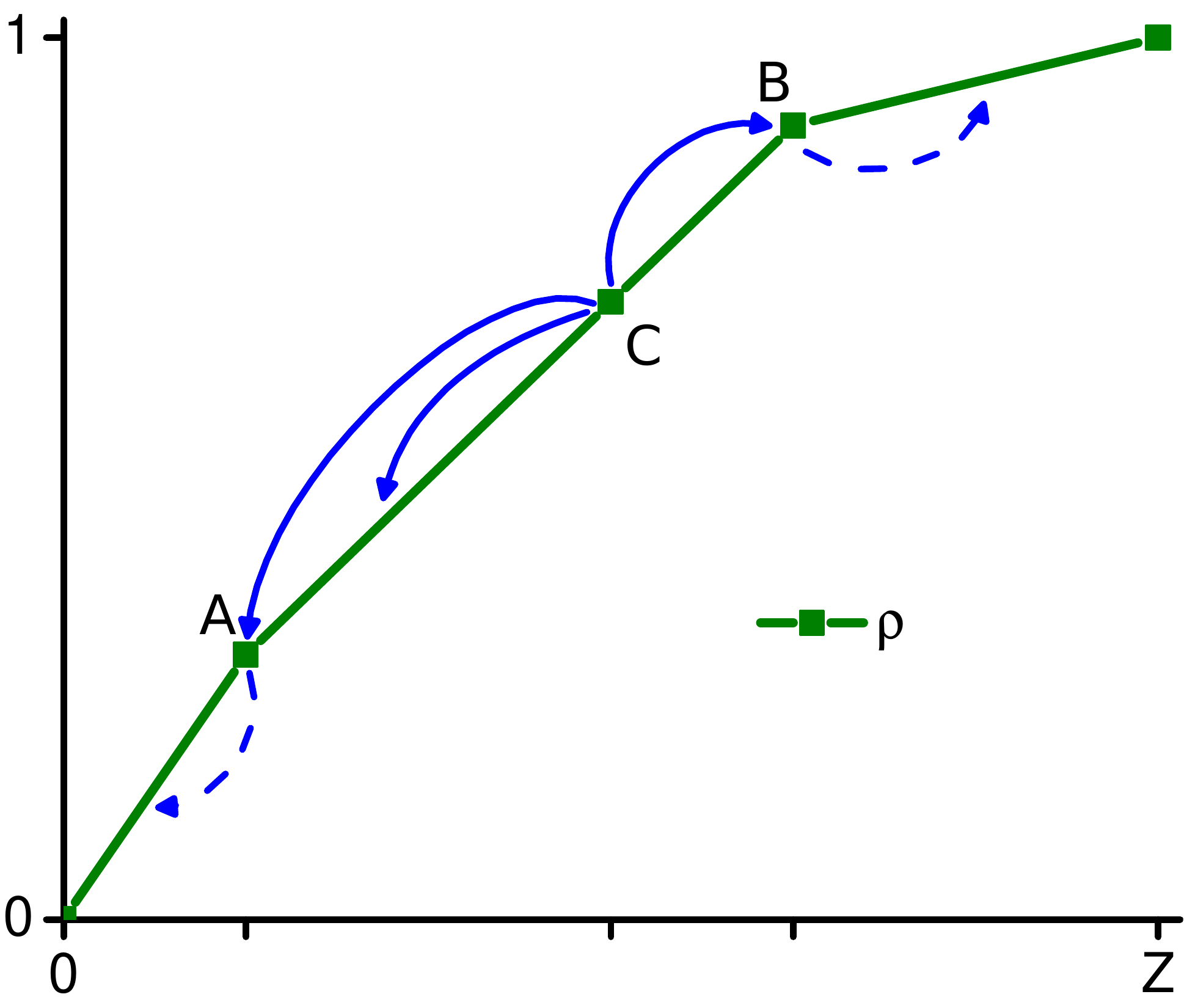}
\caption{\emph{Action of \longITRs.} Here we illustrate the action of \ITRs\ applied to a system with state-Hamiltonian pair $\left(\rho,H\right)$. Using \ITRs, the point at $C$ can be moved such that it lies anywhere on the line-segment between $A$ and $B$ and without changing the shape of the overall thermo-majorization curve. By performing this process sufficiently slowly, this can be done with no deterministic work cost. If one moves the point $C$ to coincide with point $A$ (respectively $B$), one can then use a second \ITR\ to move point $A$ ($B$) as illustrated by the dashed arrows. Again, this does not alter the shape of the thermo-majorization curve.} 
\label{fig:ITR}
\end{figure}

We can combine sequences of \longLTs\ and (full) \longPLTs\ (ie, with $\lambda = 1$) in such a way to form a useful protocol, termed an \emph{\longITR}\ as they are similar in construction to the \emph{Isothermal Reversible Processes} considered in \cite{aaberg2013truly} but require \longPLTs\ rather than full thermalizations. Note, that a similar protocol to that of \longITRs\ was developed in \cite{egloff2015measure} and termed an \emph{Isothermal shift of boundary}. In terms of thermo-majorization curves, \longITRs\ will enable us to move non-elbow points along the segments on which they exist, without changing the shape and structure of the rest of the curve. More formally:
\begin{definition}[\longITR] \label{def:ITR}
Given an $n$-level system with Hamiltonian $H_S=\sum_{i=1}^{n} E_i \ketbra{i}{i}$, an \longITR\ is parametrized by a positive constant $\kappa$ and acts on some pair of the system's energy levels, indexed by $j$ and $k$. Denote the \longITR\ by $\ITR_{j,k}\left(\kappa\right)$.

The action of $\ITR_{j,k}\left(\kappa\right)$ on $\left(\rho,H_S\right)$ where $\rho=\sum_{i=1}^{n}\eta_i\ketbra{i}{i}$, is defined by:
\begin{equation}
\left(\rho,H_S\right)\stackrel{\ITR_{j,k}\left(\kappa\right)}{\longrightarrow} \left(\rho',H'_S\right),
\end{equation}
where $\rho'=\sum_{i=1}^{n}\eta'_i$ and $H'_S=\sum_{i=1}^{n} E_i' \ketbra{i}{i}$. Defining $\tilde{\eta}_j=\frac{e^{-\beta E_j}}{e^{-\beta E_j}+e^{-\beta E_k}}\left(\eta_j+\eta_k\right)$ and $\tilde{\eta}_k=\frac{e^{-\beta E_k}}{e^{-\beta E_j}+e^{-\beta E_k}}\left(\eta_j+\eta_k\right)$, the components of $\left(\rho',H'_S\right)$ in terms of $\kappa$ are then:
\begin{align} 
\begin{split} \label{eq:ITR components}
\eta'_j&=\tilde{\eta}_j e^{-\beta\kappa},\\
\eta'_k&=\tilde{\eta}_k +\left(1-e^{-\beta \kappa}\right)\tilde{\eta}_j,  \\
E'_j&=E_j+\kappa,       \\
E'_k&=-\frac{1}{\beta} \ln\left[e^{-\beta E_k}+e^{-\beta E_j} \left(1-e^{-\beta \kappa}\right)\right],
\end{split}
\end{align}
with $\eta'_i=\eta_i$ and $E_i=E'_i$ for $i\notin\left\{j,k\right\}$. 
\end{definition}
The action of an \longITR\ in terms of therm-majorization diagrams is illustrated in Figure \ref{fig:ITR}.

Note that in such a process, $Z_S=Z'_S$ and that for all $\kappa$:
\begin{equation} \label{eq:ITR relation}
\eta'_j e^{\beta E'_j}=C=\eta'_k e^{\beta E'_k},
\end{equation}
where $C$ is some constant.

The work cost of an \longITR\ is captured in the following Lemma which is similar in construction to that of \cite[Lemma 7.]{egloff2015measure}:
\begin{lemma} \label{le:ITR cost}
The operation $\ITR_{j,k}\left(\kappa\right)$ does not cost any work.
\end{lemma}
\begin{proof}
To show this, we define a $t$-step procedure that implements $\ITR_{j,k}\left(\kappa\right)$ with each step consisting of a \longLT\ and a \longPLT. Let $\left(\rho,H_S\right)$ and $\left(\rho',H'_S\right)$ be defined as per Definition \ref{def:ITR}. Without loss of generality, assume that:
\begin{equation*}
\eta_j e^{\beta E_j}=C=\eta_k e^{\beta E_k}.
\end{equation*}
for some constant $C$ (if not, we can always perform $\textrm{\PLT}_{\left\{j,k\right\}}\left(\lambda=1\right)$ to make it so and as by Lemma \ref{le:PLT as TO} this is a thermal operation, this costs no work). Hence in the language of Definition \ref{def:ITR}, $\eta_j=\tilde{\eta}_j$ and $\eta_k=\tilde{\eta}_k$.

Define  $\epsilon=\frac{E'_j-E_j}{t}$. Let the Hamiltonian after step $r$ be $H^{\left(r\right)}_S=\sum_{i=1}^{n}E^{\left(r\right)}_i\ketbra{i}{i}$ and the state of the system be $\rho^{\left(r\right)}=\sum_{i=1}^{n}\eta^{\left(r\right)}\ketbra{i}{i}$. In step $r$, we perform the level transformation such that:
\begin{align*}
E^{\left(r\right)}_j&=E_j+r\epsilon,\\
E^{\left(r\right)}_k&=-\frac{1}{\beta}\ln\left[e^{-\beta E_k}+e^{-\beta E_j}\left(1-e^{-\beta r\epsilon}\right)\right]
\end{align*}
and fully thermalize over energy levels $j$ and $k$ so that:
\begin{align*}
\eta^{\left(r\right)}_j&=\eta_j e^{-\beta r\epsilon},\\
\eta^{\left(r\right)}_k&=\eta_k+\left(1-e^{-\beta r\epsilon}\right)\eta_j.
\end{align*}
All other energy levels and occupation probabilities remain unchanged. It can readily be verified that $Z^{\left(r\right)}_S=Z_S$ and that:
\begin{equation*}
\eta^{\left(r\right)}_j e^{\beta E^{\left(r\right)}_j}=C=\eta^{\left(r\right)}_k e^{\beta E^{\left(r\right)}_k}.
\end{equation*}
Hence, this protocol produces the desired $\left(\rho',H_{S'}\right)$ after $t$ steps. 

Such a protocol alters the state and Hamiltonian of the system but does not change the shape of the system's thermo-majorization curve. Following the proof of \cite[Supplementary Lemma 1.]{aaberg2013truly} regarding Isothermal Reversible Processes, we shall now show that the work cost of this protocol becomes increasingly peaked around zero as the number of steps taken tends to infinity.

Let $W^{\left(r\right)}$ denote the random variable for the work distribution in step $r$ of the \ITR. The work distribution for the whole $t$-step \ITR\ process is then:
\begin{equation*}
W^{\ITR}=\sum_{r=1}^{t}W^{\left(r\right)}.
\end{equation*}
Using Eqs. \eqref{eq:ITR components} and \eqref{eq:ITR relation}, $W^{\left(r\right)}$ is such that with probability:
\begin{align*}
C e^{-\beta E_j^{\left(r\right)}}, &\quad W^{\left(r\right)}=\epsilon,\\
C e^{-\beta E_k^{\left(r\right)}}, &\quad W^{\left(r\right)}=-\frac{1}{\beta}\ln\left[e^{-\beta E^{\left(r\right)}_k}+e^{-\beta E^{\left(r\right)}_j}\left(1-e^{-\beta\epsilon}\right)\right]-E^{\left(r\right)}_k,\\
\textrm{otherwise}, &\quad W^{\left(r\right)}=0.
\end{align*}
Now, for large $t$, small $\epsilon$, this becomes such that with probability:
\begin{align*}
C e^{-\beta E_j^{\left(r\right)}}, &\quad W^{\left(r\right)}=\epsilon,\\
C e^{-\beta E_k^{\left(r\right)}}, &\quad W^{\left(r\right)}=-\epsilon e^{\beta\left(E^{\left(r\right)}_k-E^{\left(r\right)}_j\right)}+O(\epsilon^2),\\
\textrm{otherwise}, &\quad W^{\left(r\right)}=0.
\end{align*}
Hence as $t\rightarrow\infty$, $\left\langle W^{\left(r\right)}\right\rangle\rightarrow0$, for all $r$. As $t=\frac{E^{(t)}_j-E^{(0)}_j}{\epsilon}=\frac{E'_j-E_j}{\epsilon}$, we have that:
\begin{equation*}
\left\langle W^{\ITR}\right\rangle=\sum_{r=1}^{t}\left\langle W^{\left(r\right)}\right\rangle\rightarrow 0, \quad \textrm{as } t\rightarrow\infty.
\end{equation*}

Now consider the variance of $W^{\ITR}$. For large $t$, hence small $\epsilon$:
\begin{align*}
\left\langle {W^{\left(r\right)}}^2\right\rangle &= C e^{-\beta E^{\left(r\right)}_j}\epsilon^2+C e^{-\beta E^{\left(r\right)}_k}\epsilon^2 e^{2\beta\left(E^{\left(r\right)}_k-E^{\left(r\right)}_j\right)}+O(\epsilon^4)\\
&=C e^{-\beta E^{\left(r\right)}_j}\epsilon^2 \left(1 + e^{\beta\left(E^{\left(r\right)}_k-E^{\left(r\right)}_j\right)}\right)+O(\epsilon^4)\\
&\rightarrow 0 \quad \textrm{as } \epsilon\rightarrow 0.
\end{align*}
Hence, $\textrm{Var}\left(W^{\left(r\right)}\right)\rightarrow 0$ as $t\rightarrow \infty$. As the $W^{\left(r\right)}$ are independent:
\begin{align*}
\textrm{Var}\left(W^{\ITR}\right)&=\sum_{r=1}^{t}\textrm{Var}\left(W^{\left(r\right)}\right)\\
&\rightarrow 0 \quad \textrm{as } \epsilon\rightarrow 0.
\end{align*}

Note that this analysis extends to the case where $E'_j\rightarrow\infty$. If we parametrize $E'_j$ in terms of the number of steps taken in the \ITR\ protocol so that $E'_j=\ln t$, then in the limit $t\rightarrow\infty$, $E'_j\rightarrow\infty$, $\left\langle W^{\ITR}\right\rangle\rightarrow 0$  and $\textrm{Var}\left(W^{\ITR}\right)\rightarrow0$.

Now, Chebyshev's inequality gives us that:
\begin{align*}
P\left(\left|W^\ITR\right|\geq k\sqrt{\textrm{Var}\left(W^{\ITR}\right)}\right)\leq \frac{1}{k^2},
\end{align*}
so by taking $t$ and $k$ to be large, we obtain that the work distribution for the \ITR\ becomes increasingly peaked around 0.
\end{proof}

\begin{figure}
\centering
\begin{minipage}{.5\textwidth}
  \centering
  \includegraphics[width=.9\linewidth]{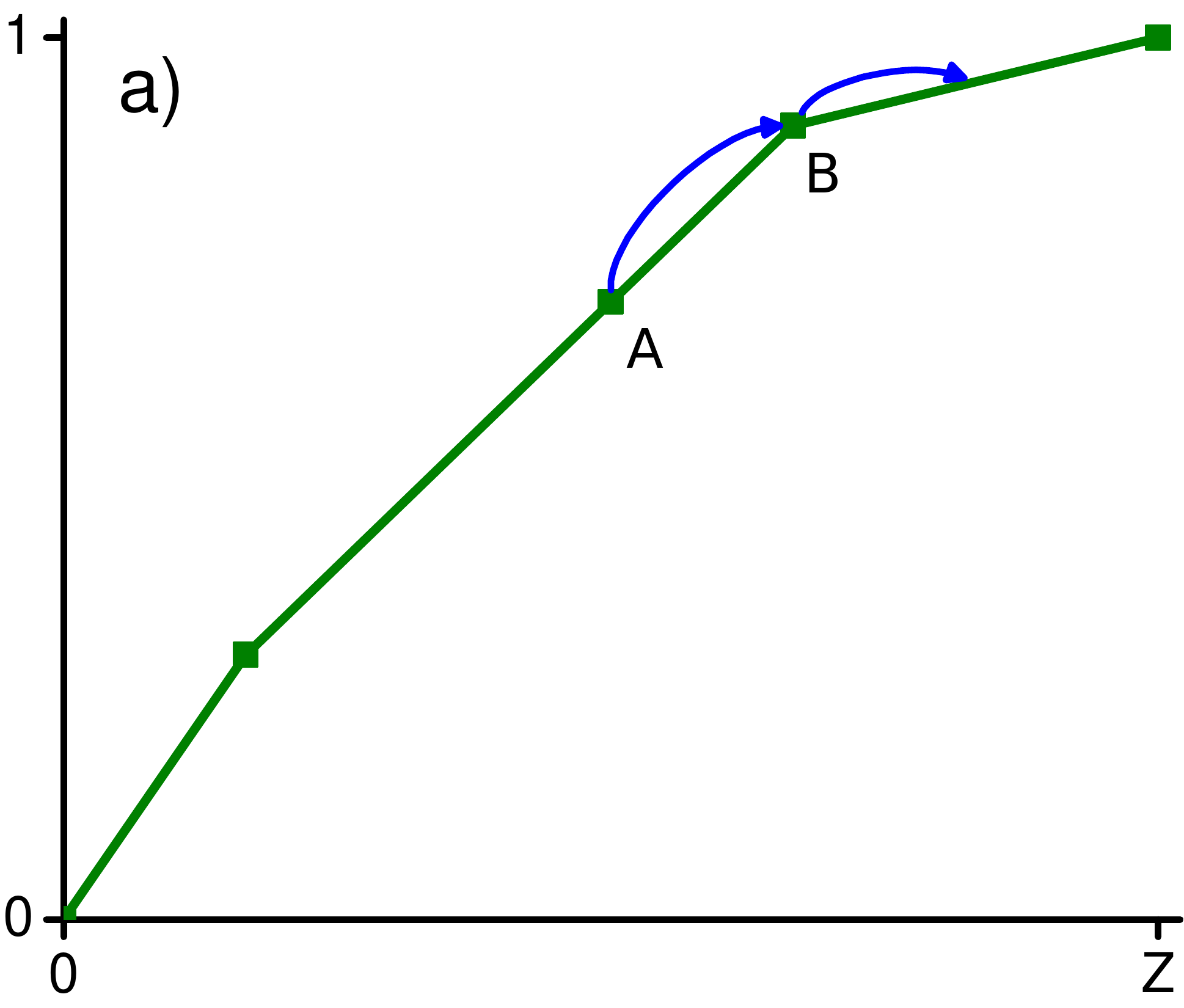}
  \centerline{(a) Initial state.}
  \label{fig:PF initial}
\end{minipage}%
\begin{minipage}{.5\textwidth}
  \centering
  \includegraphics[width=.9\linewidth]{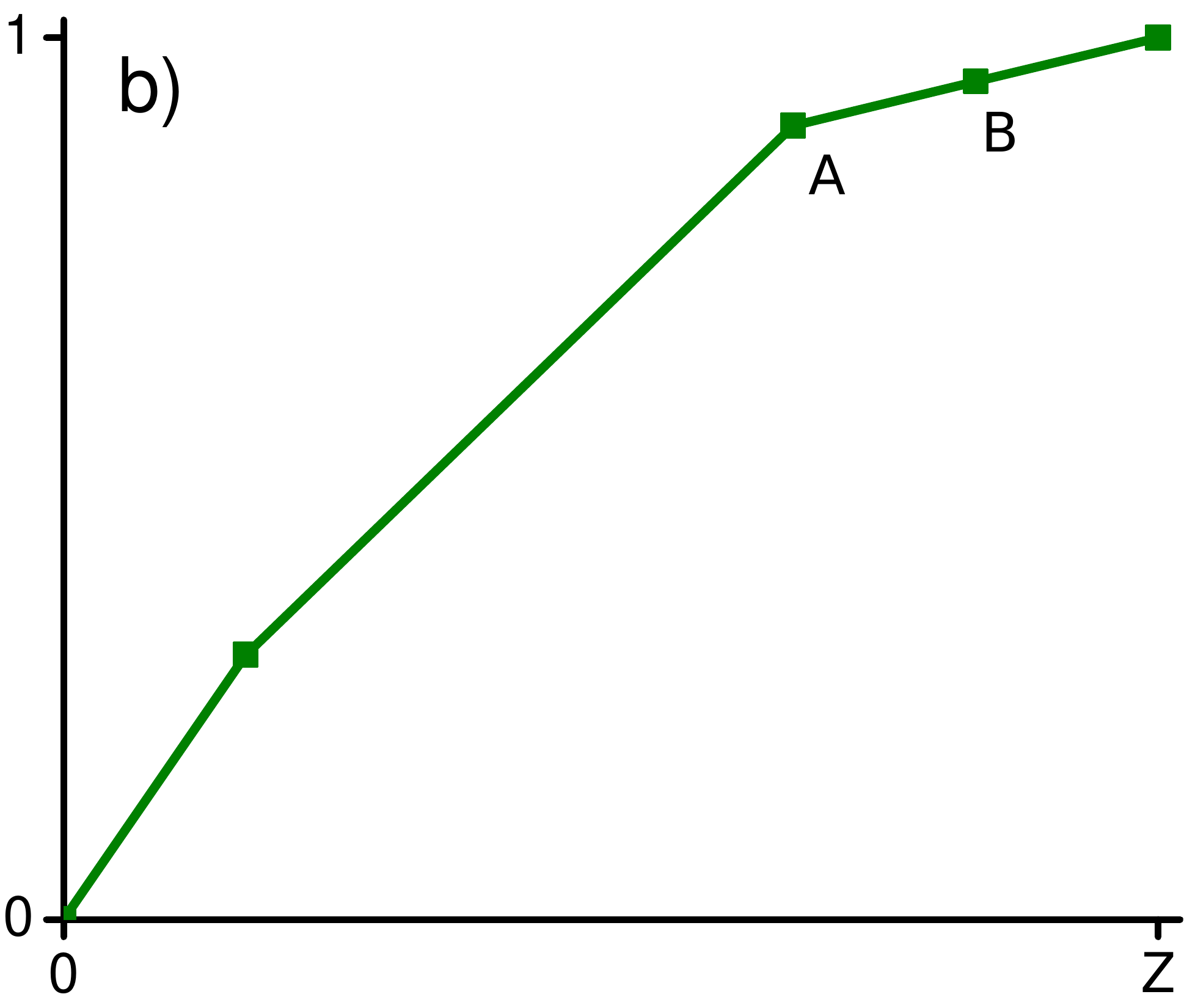}
  \centerline{(b) Final state.}
  \label{fig:PF final}
\end{minipage}
\caption{\emph{Exact \longPF.} Here we illustrate how to move a non-elbow point through an elbow. First, we perform a \ITR\ that sends the non-elbow, $A$, towards the elbow, $B$. As the appropriate energy level is raised to infinity during the protocol, $A$ tends towards $B$ until they coincide. Next, a second \ITR\ lowers the energy level from infinity, keeping it in partial thermal equilibrium with respect to another line-segment. This moves $B$, now a non-elbow, to this new line-segment.}
\label{fig:exact PF}
\end{figure}

Hence, \longITRs\ can be used to move non-elbow points along straight-line segments of a thermo-majorization curve, using \longCOs\ and without expending any work. By combining two \ITRs, it is possible to commute non-elbow points with elbows, meaning that non-elbows can be moved to any point of the thermo-majorization curve for free. We term this operation \emph{Exact \longPF}.
\begin{definition}[Exact \longPF\ (E\PF)] \label{def:EPF}
Exact \longPF\ is illustrated in Figure \ref{fig:exact PF}. Given an $n$-level system with Hamiltonian $H_S=\sum_{i=1}^{n} E_i \ketbra{i}{i}$ and state $\rho=\sum_{i=1}^{n} \eta_i\ketbra{i}{i}$ such that:
\begin{equation} \label{eq:exact PF elbow}
\eta_j e^{\beta E_j}=\eta_k e^{\beta E_k},
\end{equation}
for some $j,k$ (i.e. there is a non-elbow point on the thermo-majorization curve), an Exact \longPF\ moves this non-elbow to another part of the thermo-majorization curve whilst keeping the shape of the curve fixed.  
\end{definition}
To implement E\PF, we first apply a \ITR\ that sends $E_j\rightarrow\infty$. This lowers the energy of $E_k$ and does not alter the shape of the thermo-majorization curve. Next, to move the non-elbow to another part of the curve, we apply another \ITR\ to $j$ and a third level labeled by $l$, bringing the energy level associated with $j$ back down from infinity to some $E'_j$. This leaves us with a system $\left(\rho',H'_S\right)$ with thermo-majorization curve identical to that of $\left(\rho,H_S\right)$ and such that:
\begin{equation}
\eta'_j e^{\beta E'_j}=\eta'_l e^{\beta E'_l},
\end{equation}
(i.e. the elbow defined in Eq. \eqref{eq:exact PF elbow} has moved to another part of the curve).


\begin{figure}[!htbp]
\begin{center}
$
\begin{array}{ccccc}
\includegraphics[width=0.4\textwidth, height=0.311\textheight]{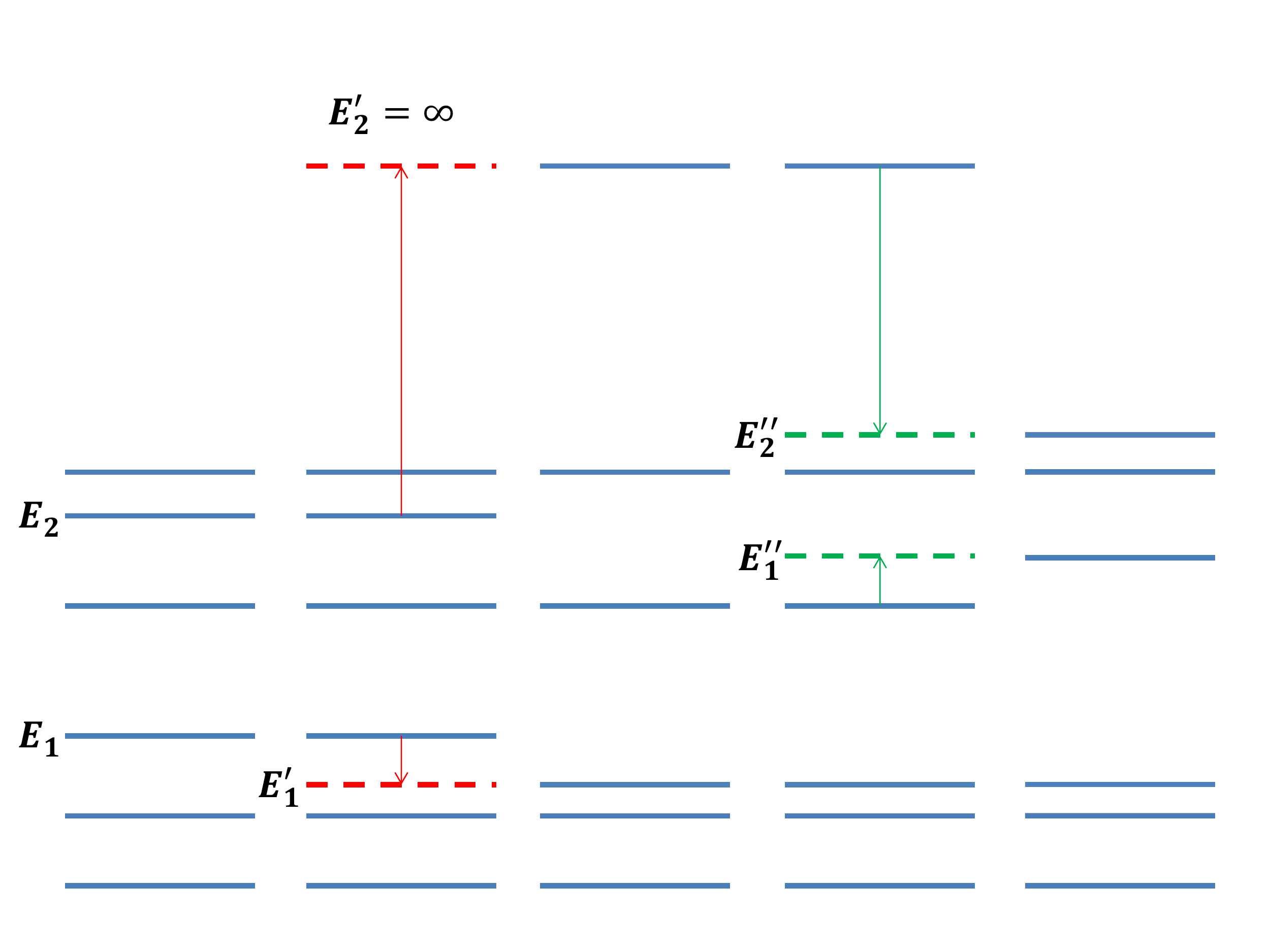} & \qquad
\includegraphics[width=0.4\textwidth, height=0.311\textheight]{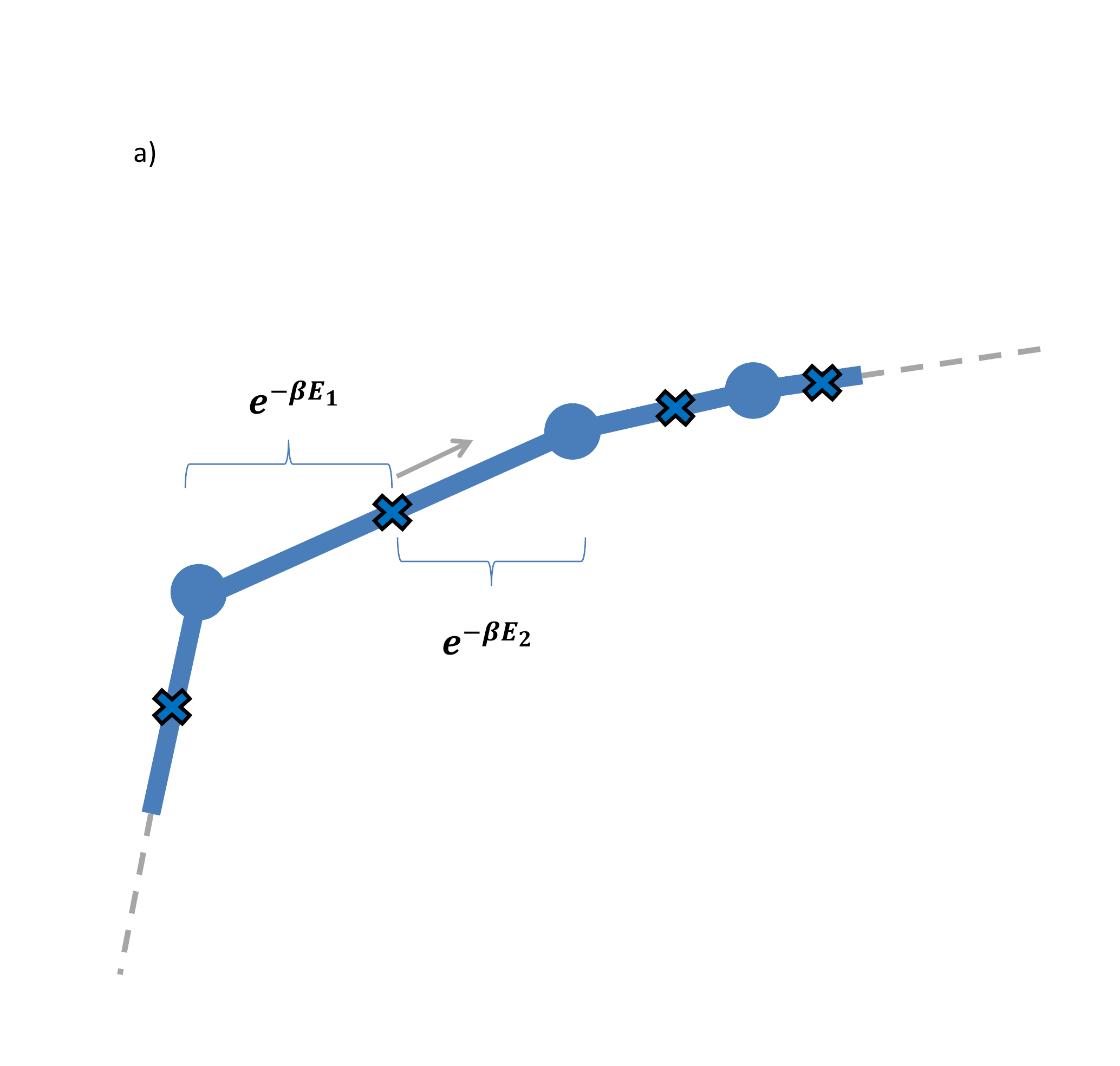} \\
\includegraphics[width=0.4\textwidth, height=0.311\textheight]{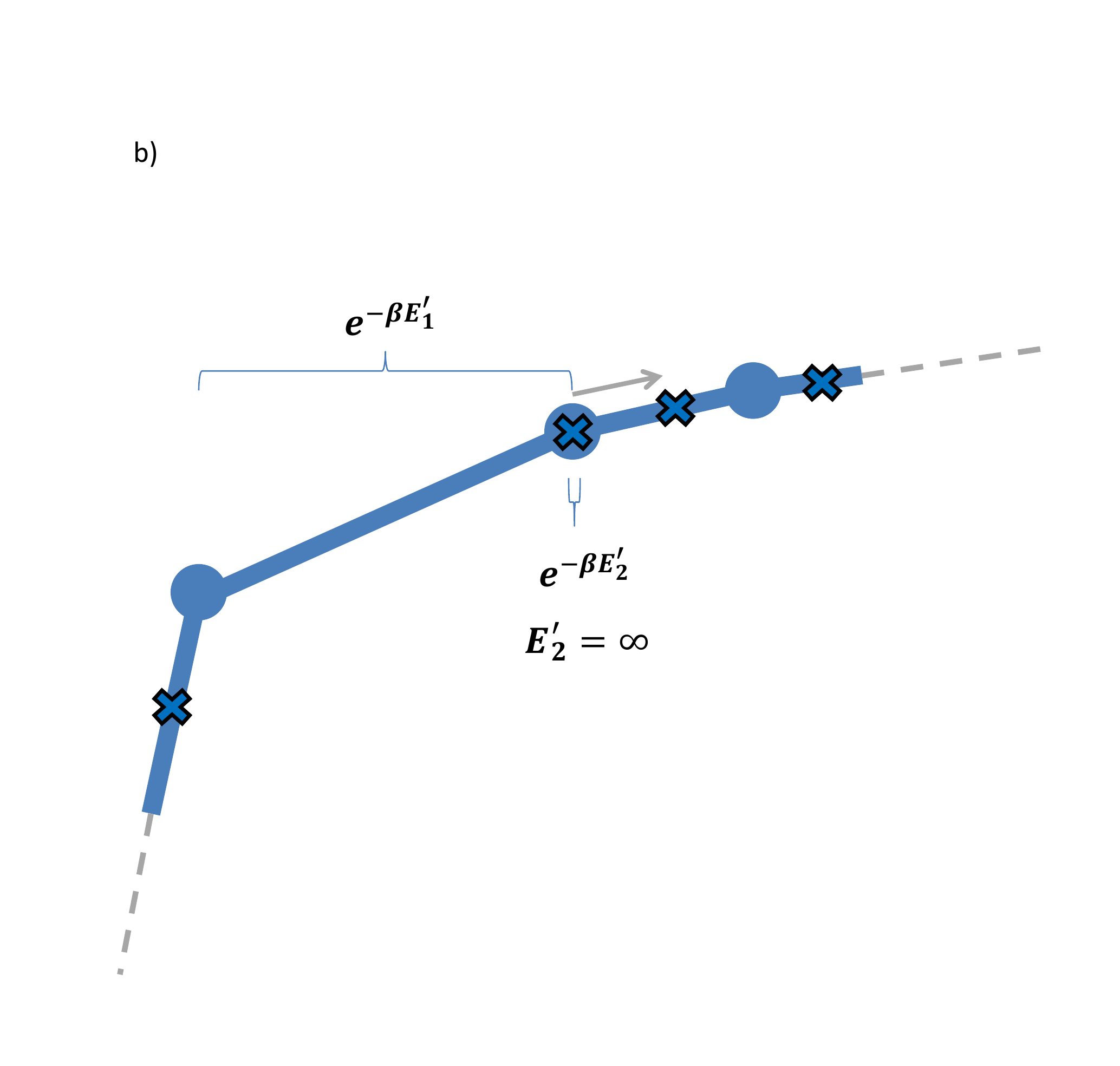}
\end{array}
$
\end{center}
\caption{\longPF\ - example with infinity}
\label{fig:PF_ex}
\end{figure}

We also present an example of Exact \longPF \ showing what happens with energy levels and how the points are moved with respect to Gibbs weights in Figure~\ref{fig:PF_ex}.

\begin{figure}
\centering
\begin{minipage}{.5\columnwidth}
  \centering
  \includegraphics[width=0.9\textwidth]{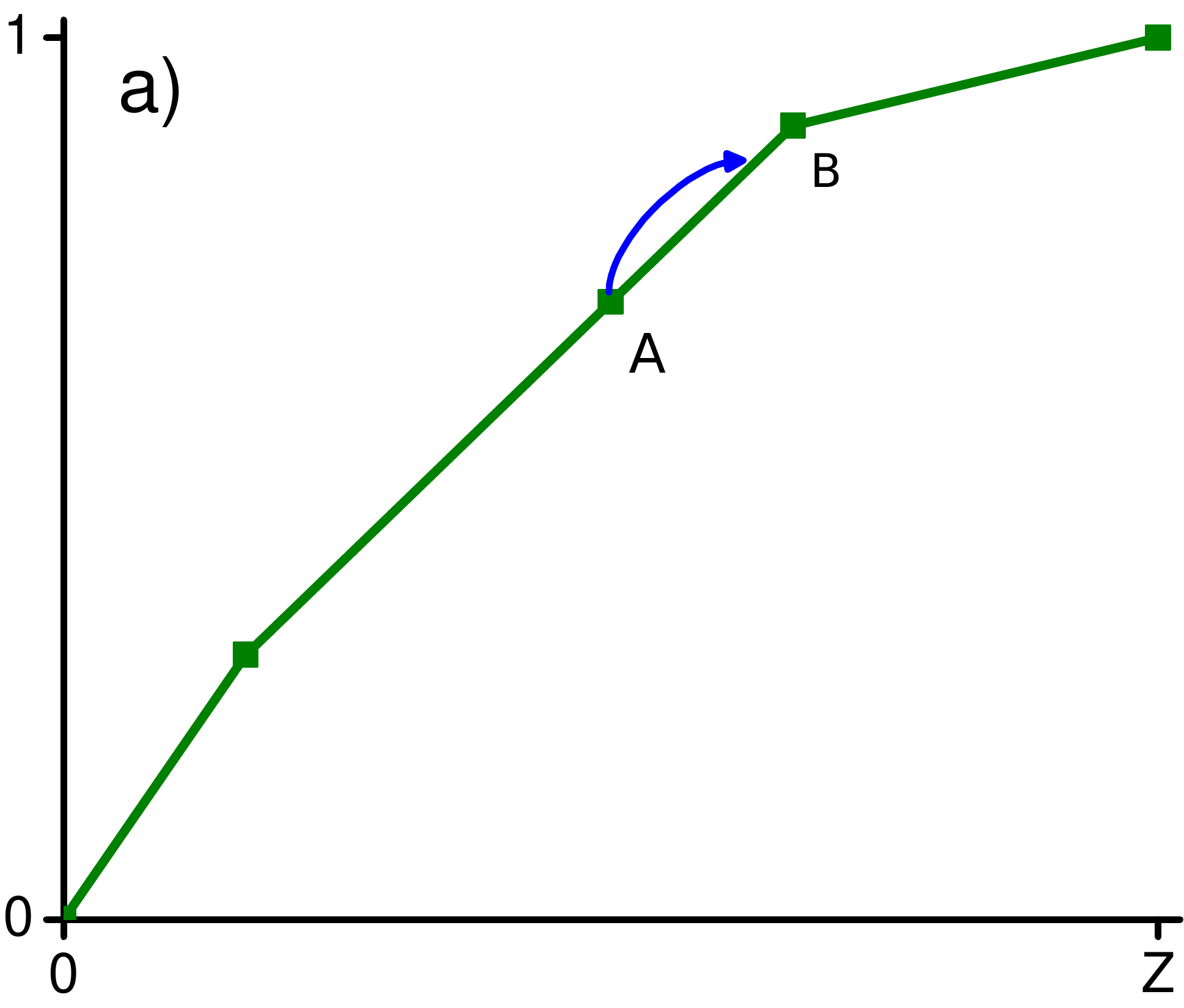}
  \label{fig:APF1}
	\centerline{(a) Initial State}
\end{minipage}%
\begin{minipage}{.5\columnwidth}
  \centering
  \includegraphics[width=0.9\textwidth]{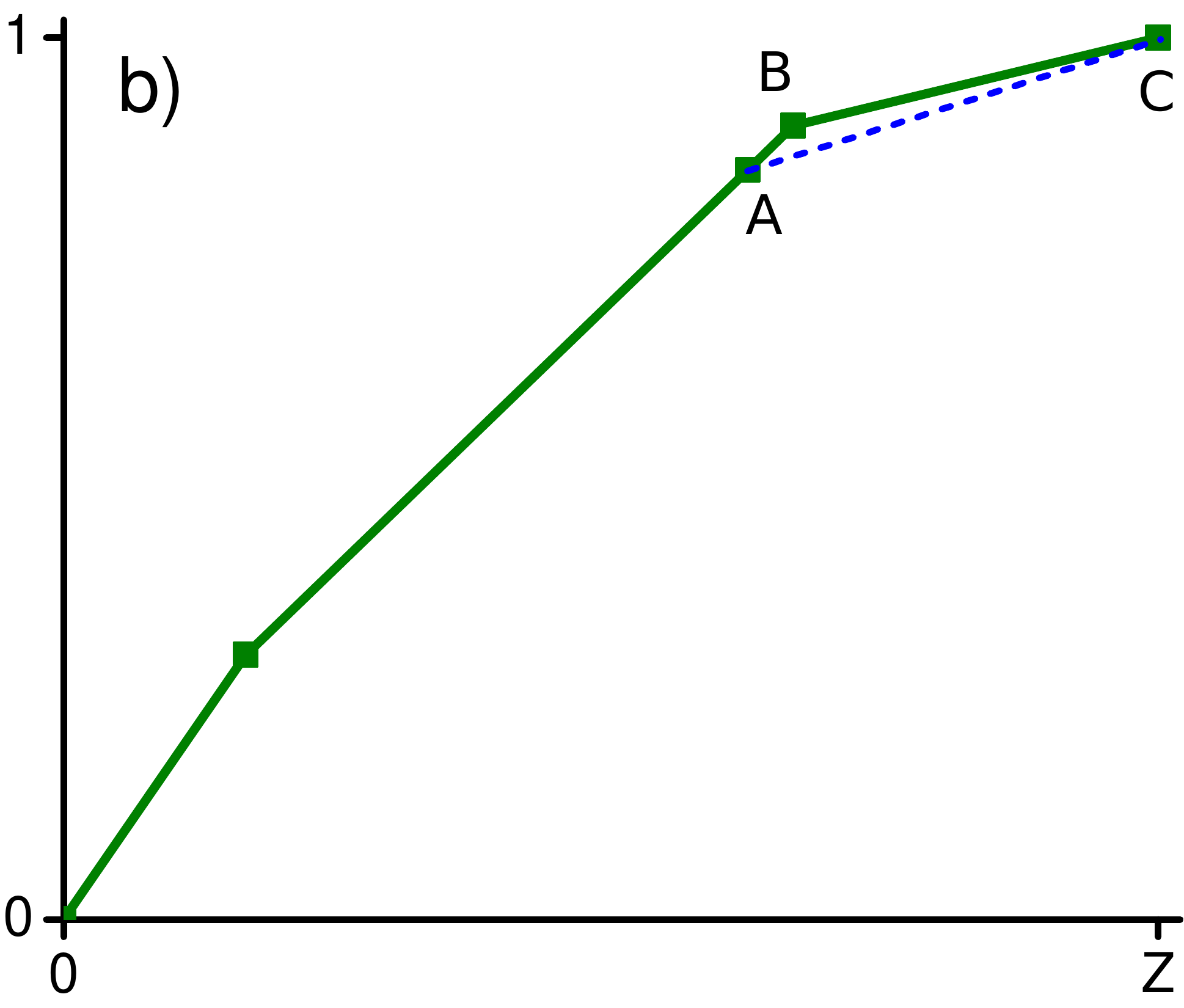}
  \label{fig:APF2}
	\centerline{(b) After first \ITR.}
\end{minipage}\\
\vspace{.5cm}
\begin{minipage}{.5\columnwidth}
  \centering
  \includegraphics[width=0.9\textwidth]{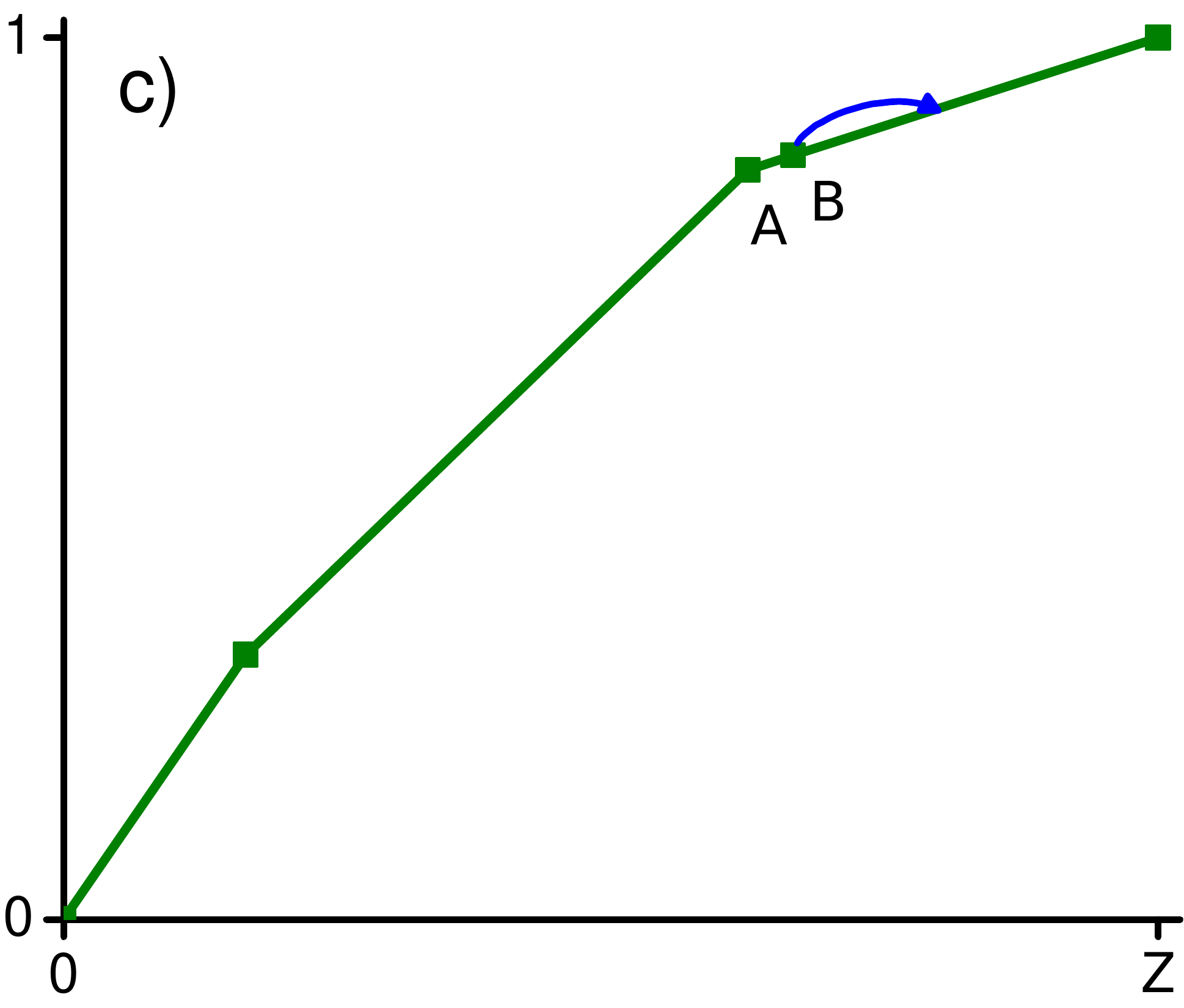}
  \label{fig:APF3}
	\centerline{(c) After \PLT.}
\end{minipage}%
\begin{minipage}{.5\columnwidth}
  \centering
  \includegraphics[width=.9\textwidth]{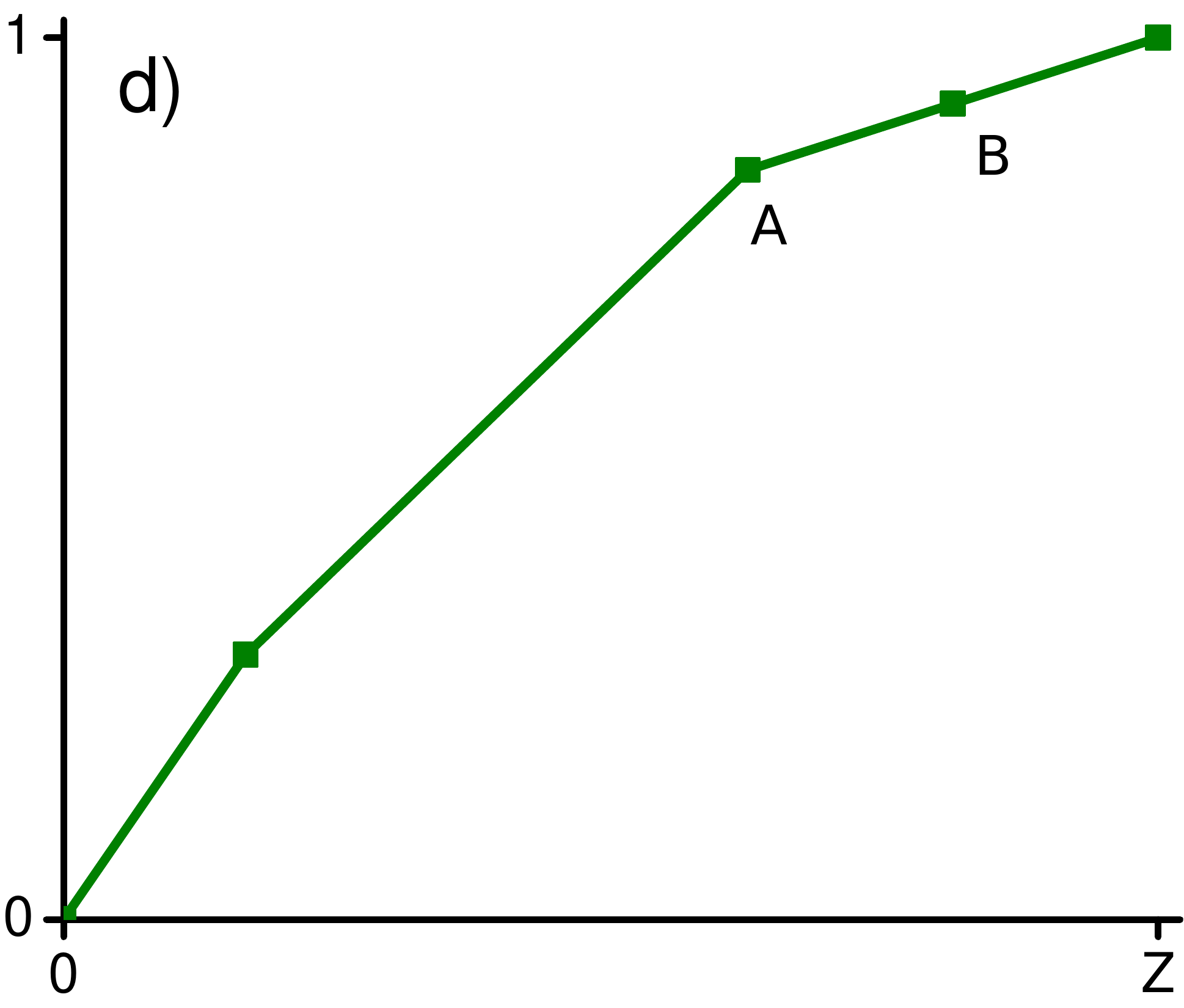}
  \label{fig:APF4}
	\centerline{(d) After second \ITR.}
\end{minipage}
\caption{\emph{Approximate \longPF.} Here we illustrate the protocol of Approximate \longPF\ using thermo-majorization diagrams. Initially the system is as per Figure (a). Using a \ITR, the non-elbow point, $A$, is moved towards the elbow at point $B$. This results in Figure (b). Next, a \PLT\ is applied between points $A$ and $C$, leading to Figure (c). Point $B$ is now a non-elbow and can be moved using a \ITR, giving Figure (d).}
\label{fig:approximate PF}
\end{figure}

Exact \longPF\ requires that an energy level is raised to infinity during a \longITR. If it is not possible, or undesirable, to raise an energy level to infinity, a similar effect to an Exact \longPF\ can be achieved while altering the shape of the thermo-majorization curve slightly in a process we call \emph{Approximate \longPF}.
\begin{definition}[Approximate \longPF\ (A\PF)]
Approximate \longPF\ is illustrated in Figure \ref{fig:approximate PF}. Given an $n$-level system with Hamiltonian $H_S=\sum_{i=1}^{n} E_i \ketbra{i}{i}$ and state $\rho=\sum_{i=1}^{n} \eta_i\ketbra{i}{i}$ such that:
\begin{equation} \label{eq:approx PF elbow}
\eta_j e^{\beta E_j}=\eta_k e^{\beta E_k},
\end{equation}
for some $j,k$ (i.e. there is a non-elbow point on the thermo-majorization curve), an Approximate \longPF\ moves this non-elbow to an adjacent segment of the thermo-majorization curve whilst modifying the shape of the thermo-majorization curve by an arbitrarily small amount and without sending an energy level to infinity.
\end{definition}

To implement it, without loss of generality, assume that the set $\left\{i\right\}_{i=1}^{n}$ has been $\beta$-ordered, we wish to move the non-elbow point to the right and take $j=k+1$. To do this, we:
\begin{enumerate}
\item Apply a \ITR\ that raises the energy level of $E_j$ to some fixed, large but finite amount. This lowers the energy of $E_k$ and does not alter the shape of the thermo-majorization curve.
\item We now apply:
\begin{equation}
\textrm{\PLT}_{\left\{j,j+1\right\}}\left(\lambda=1\right),
\end{equation}
to the system. This turns the non-elbow associated with $j$ and $k$ into an elbow and the elbow associated with $j$ and $j+1$ into a non-elbow.
\item Using a \ITR, we can now move the new non-elbow point without altering the shape of the thermo-majorization curve.
\end{enumerate}
(i.e. the elbow defined in Eq. \eqref{eq:exact PF elbow} has moved to another part of the curve). By adjusting the height to which $E_j$ is raised in Step 1, we can tune the extent to which the thermo-majorization curve is altered.

\subsection{\longSR}

\longSRs\ are only needed in the case when the eigenbasis of the initial state or target state are not diagonal in the energy eigenbasis. Since the thermo-majorization criteria is about transitions between energy eigenstates, we want to consider transitions to and from energy eigenstates. Since we are only considering such transitions, we can, as described in \cite{horodecki2013fundamental}, take the target state to be block-diagonal in the energy eigenbasis $\sigma=\sum\eta_i^{(\sigma)}\ketbra{i}{i}$, which one can also show allows one to consider only initial states which are block-diagonal. However, the eigenbasis of the initial and final state, even if block diagonal in the energy eigenbasis, may not be the same. In such a case we use a \longSR\ to rotate the state, either initially, or at the end of the protocol. Note also that if an \SR\ is applied at random, then it also serves to decohere in the energy eigenbasis of the system if that is desired. As the unitary is applied only to the system under consideration, it is much easier to implement experimentally than a more general thermal operation.
 We will henceforth assume that if needed, an \SR\ has been applied to the initial state, or will be applied at the end of the protocol, such that we need only consider transitions between states which are both diagonal in the energy eigenbasis.

\section{Transformations for states with same $\beta$-order} \label{sec:same order}

In this appendix, we show that for two states with the same $\beta$-ordering, if one thermo-majorizes the other, then the transition can be made using a finite number of \longPLTs. Furthermore, these \longPLTs\ need only act on two energy levels at a time.

We say that two states $\rho$ and $\sigma$, associated with the same Hamiltonian $H_S=\sum_{i=1}^{n} E_i \ketbra{i}{i}$, have the same $\beta$-ordering if:
\begin{equation}
\sum_{i=1}^{k} e^{-\beta E^{\left(\rho\right)}_i} = \sum_{i=1}^{k} e^{-\beta E^{\left(\sigma\right)}_i}, \quad\forall k.
\end{equation}

\begin{figure}
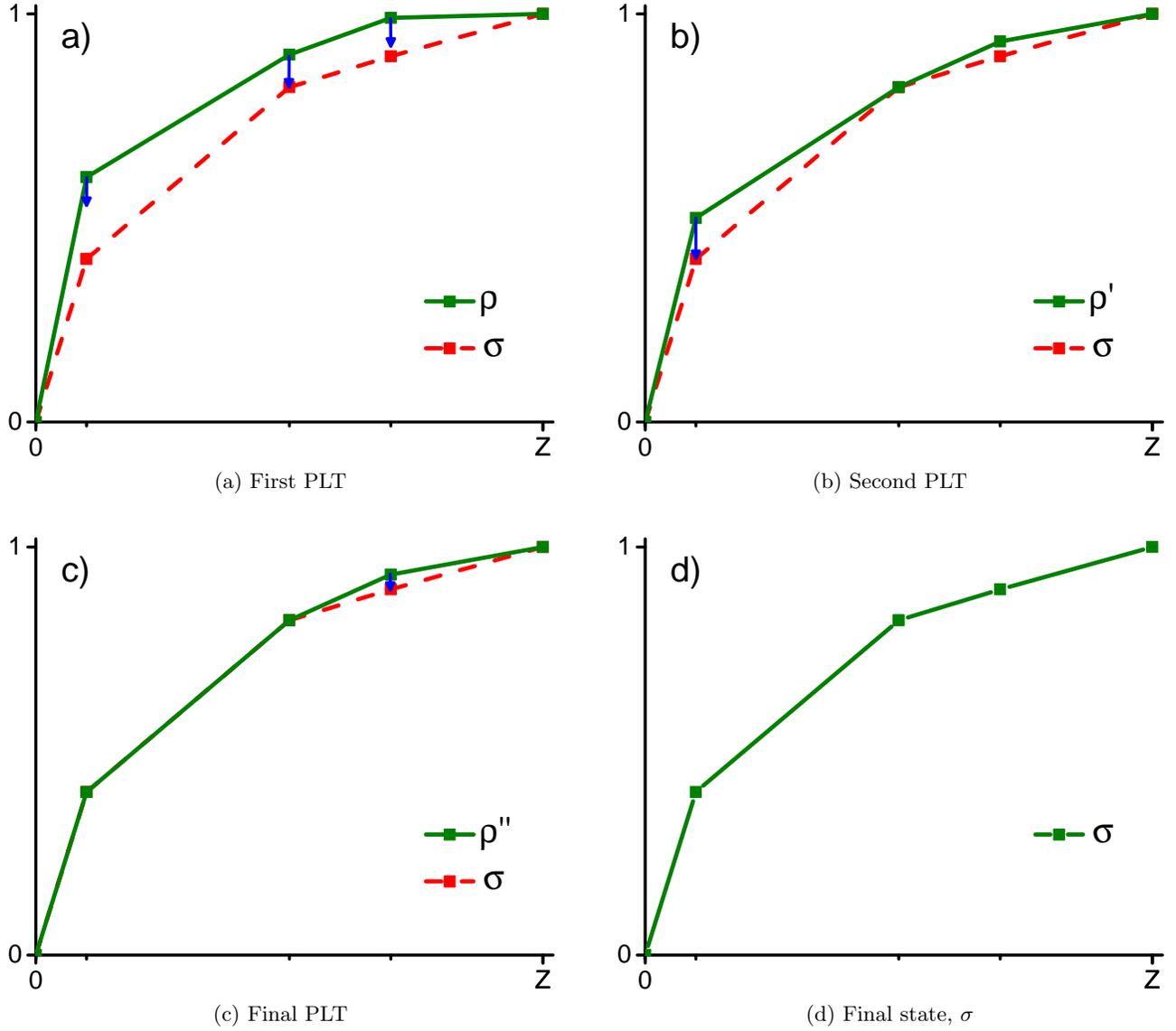

\centering
\begin{minipage}{.5\columnwidth}
  \centering
  \includegraphics[width=0.9\textwidth]{article_figures/SBO1Graph.pdf}
  \label{fig:SBO1}
	\centerline{(a) First \PLT}
\end{minipage}%
\begin{minipage}{.5\columnwidth}
  \centering
  \includegraphics[width=0.9\textwidth]{article_figures/SBO2Graph.pdf}
  \label{fig:SBO2}
	\centerline{(b) Second \PLT}
\end{minipage}\\
\vspace{.5cm}
\begin{minipage}{.5\columnwidth}
  \centering
  \includegraphics[width=0.9\textwidth]{article_figures/SBO3Graph.pdf}
  \label{fig:SBO3}
	\centerline{(c) Final PLT}
\end{minipage}%
\begin{minipage}{.5\columnwidth}
  \centering
  \includegraphics[width=.9\textwidth]{article_figures/SBO4Graph.pdf}
  \label{fig:SBO4}
	\centerline{(d) Final state, $\sigma$}
\end{minipage}
\caption{\emph{\longCOs\ protocol for transforming between states with the same $\beta$-order.} If two states, $\rho$ and $\sigma$, have the same $\beta$-order and are such that $\rho$ thermo-majorizes $\sigma$, then $\rho$ can be converted into $\sigma$ using \longPLTs. First a \PLT\ is applied to $\rho$ across the complete set of the energy levels, Figure (a), lowering the thermo-majorization of $\rho$ until it meets that of $\sigma$, Figure (b). Next, a second \PLT\ is applied to those energy levels to the left of this meeting point, again lowering the curve until it meets that of $\sigma$ at a second point, Figure (c). By iterating this process, $\rho$ is transformed into $\sigma$, Figure (d).}
\label{fig:SBO}
\end{figure}

\begin{theorem} \label{th:same beta order}
Suppose that $\rho$ and $\sigma$ are states of an $n$-level system with Hamiltonian $H_S=\sum_{i=1}^{n} E_i \ketbra{i}{i}$ such that:
\begin{enumerate}
\item $\sigma$ is block-diagonal in the energy eigenbasis.
\item $\rho$ and $\sigma$ have the same $\beta$-order.
\item $\rho\stackrel{\textrm{TO}}{\longrightarrow}\sigma$.
\end{enumerate}
Then $\rho$ can be converted into $\sigma$ using at most $n-1$ \longPLTs.
\end{theorem}
\begin{proof}
To prove this, we give a protocol consisting only of \longPLTs\ that converts $\rho$ into $\sigma$. An illustrative outline for the protocol is given in Figure \ref{fig:SBO}. As we can perform any energy conserving unitary on the system, we can assume that both $\rho$ and $\sigma$ are diagonal in the energy eigenbasis (by decohering $\rho$ if necessary).

Let $\left\{\eta_i\right\}_{i=1}^{n}$ be the $\beta$-ordered eigenvalues of $\rho$, $\left\{\zeta_i\right\}_{i=1}^{n}$ be the $\beta$-ordered eigenvalues of $\sigma$ and $\left\{E_i\right\}_{i=1}^{n}$ be the $\beta$-ordered energy-eigenvalues of $H_S$.

Given that $\rho$ and $\sigma$ have the same $\beta$-order, $\rho$ majorizes $\sigma$ if and only if:
\begin{equation} \label{eq:majorization1}
\sum_{i=1}^{m} \eta_i \geq \sum_{i=1}^{m} \zeta_i, \quad \forall m.
\end{equation}
Let $\mathcal{M}$ be the set of $m$ for which equality occurs in Eq.~\eqref{eq:majorization1}. Suppose that are $r$ of them, and label them in ascending order by $\left\{t_s\right\}_{s=1}^{r}$. Note that $t_r=n$ and for convenience, define $t_0=0$.

We now proceed to perform a \PLT\ on each set of energy levels $\mathcal{P}^s\equiv\left\{t_{s-1}+1,\dots,t_{s}-1\right\}$ for $1\leq s\leq r$. For each $m\in\mathcal{P}^s$, let $\lambda^s_m$ be defined as the solution to:
\begin{equation}
\sum_{j=t_{s-1}+1}^{m}\left[ \left(1-\lambda^s_m\right)\eta_j+\frac{\lambda^s_m e^{-\beta E_j}}{\sum_{l\in\mathcal{P}^s}e^{-\beta E_l}}\sum_{l\in\mathcal{P}^s}\eta_l\right]=\sum_{j=t_{s-1}+1}^{m} \zeta_j,
\end{equation}
i.e. $\lambda^s_m$ is such that if one were to apply $\textrm{PT}_{\mathcal{P}^s}\left(\lambda^s_m\right)$ to $\rho$, the thermo-majorization curve of the resultant state would touch or cross the thermo-majorization curve of $\sigma$ at $m$. To ensure the resultant state thermo-majorizes $\sigma$, we actually apply the transformation $\textrm{PT}_{\mathcal{P}^s}\left(\lambda^s_{\textrm{min}}\right)$ for each $s$ where:
\begin{equation}
\lambda^s_{\textrm{min}}=\min_{m\in\left\{t_{s-1}+1,\dots,t_{s}-1\right\}} \lambda^s_m.
\end{equation}
This leaves us with a state $\rho'$ that has the same $\beta$-ordering as $\rho$ and $\sigma$ and thermo-majorizes $\sigma$.

Let $\left\{\eta'_i\right\}_{i=1}^{n}$ be the $\beta$-ordered eigenvalues of $\rho'$. Then $r'$, the number of $m$ such that:
\begin{equation}
\sum_{i=1}^{m} \eta'_i = \sum_{i=1}^{m} \zeta_i,
\end{equation}
is greater than or equal to $r$ with equality if and only if $r=n$. Hence by iterating the above procedure at most $n-1$ times, we transform $\rho$ into $\sigma$ using \longPLTs.
\end{proof}

\longPLTs\ that act on 2 energy levels at a time are analogous to the concept of $T$-transforms (see, for example, \cite[Chapter 2, Section B]{marshall2010inequalities}) in majorization theory. Within majorization theory, it is know that if the vector $\vec{v}$ majorizes $\vec{u}$, then $\vec{v}$ can be converted into $\vec{u}$ using a finite number of $T$-transforms \cite{muirhead1902some,hardy1952inequalities}. The equivalent result for thermo-majorization is captured in the following theorem:
\begin{theorem}
Suppose that $\rho$ and $\sigma$ are states of an $n$-level system with Hamiltonian $H_S=\sum_{i=1}^{n} E_i \ketbra{i}{i}$ such that:
\begin{enumerate}
\item $\sigma$ is block-diagonal in the energy eigenbasis.
\item $\rho$ and $\sigma$ have the same $\beta$-order.
\item $\rho\stackrel{\textrm{TO}}{\longrightarrow}\sigma$.
\end{enumerate}
Then $\rho$ can be converted into $\sigma$ using at most $n-1$ \longPLTs\ that each act on 2 energy levels.
\end{theorem}
\begin{proof}
The aim is to construct a protocol consisting of such \PLTs\ that converts $\rho$ into $\sigma$. To do this, we perform a sequence of \PLTs. Each \PLT\ adjusts the gradients of two line-segments of the thermo-majorization curve of $\rho$ until one of them matches the gradient of the corresponding segment on $\sigma$. By picking the segments of $\rho$ such that one has gradient strictly greater than the corresponding segment on $\sigma$ and one has gradient strictly less than the corresponding segment on $\sigma$ this can always be done. Once all of the gradients have been matched, $\rho$ has been converted into $\sigma$. The full details of the protocol are below.
Again, by first imagining that we have decohered $\rho$ in the energy eigenbasis if necessary, we can assume that $\rho$ and $\sigma$ are diagonal in the energy eigenbasis.

Let $\left\{\eta_i\right\}_{i=1}^{n}$ be the $\beta$-ordered eigenvalues of $\rho$, $\left\{\zeta_i\right\}_{i=1}^{n}$ be the $\beta$-ordered eigenvalues of $\sigma$ and $\left\{E_i\right\}_{i=1}^{n}$ be the $\beta$-ordered energy-eigenvalues of $H_S$. Hence we have:
\begin{align}
\eta_1 e^{\beta E_1}\geq\dots\geq\eta_n e^{\beta E_n},
\end{align}
and
\begin{align}
\zeta_1 e^{\beta E_1}\geq\dots\geq\zeta_n e^{\beta E_n}.
\end{align}
Given that $\rho$ and $\sigma$ have the same $\beta$-order, $\rho$ majorizes $\sigma$ if and only if:
\begin{equation} \label{eq:majorization}
\sum_{i=1}^{m} \eta_i \geq \sum_{i=1}^{m} \zeta_i, \quad \forall m.
\end{equation}

Let $j$ be the largest index such that $\eta_j e^{\beta E_j} > \zeta_j e^{\beta E_j}$ and $k$ be the smallest index larger than $j$ such that $\eta_k e^{\beta E_k} < \zeta_k e^{\beta E_k}$. This picks the segments we shall apply the \PLT\ to. Then:
\begin{equation} \label{eq:jk order}
\eta_j e^{\beta E_j}>\zeta_j e^{\beta E_j}\geq\zeta_k e^{\beta E_k}>\eta_k e^{\beta E_k},
\end{equation}
and note that $\eta_i=\zeta_i$ for $j<i<k$.

We now determine the amount that we need to thermalize by in order to transform the gradient of one of the segments of $\rho$ to that of $\sigma$. Define $\lambda_1$ to be the value of $\lambda$ such that:
\begin{equation}
\left(1-\lambda_1\right)\eta_j e^{\beta E_j} + \frac{\lambda_1\left(\eta_j+\eta_k\right)}{e^{-\beta E_j}+e^{-\beta E_k}}=\zeta_j e^{\beta E_j},
\end{equation}
and $\lambda_2$ to be such that:
\begin{equation}
\left(1-\lambda_2\right)\eta_k e^{\beta E_k} + \frac{\lambda_2\left(\eta_j+\eta_k\right)}{e^{-\beta E_j}+e^{-\beta E_k}}=\zeta_k e^{\beta E_k}.
\end{equation}
Note that:
\begin{equation}
\eta_j e^{\beta E_j}\geq\frac{\eta_j+\eta_k}{e^{-\beta E_j}+e^{-\beta E_k}}\geq\eta_k e^{\beta E_k},
\end{equation}
and hence $\lambda_1,\lambda_2\geq0$. Also, at least one of $\frac{\left(\eta_j+\eta_k\right)}{e^{-\beta E_j}+e^{-\beta E_k}} \leq \zeta_j e^{\beta E_j }$ or $\frac{\left(\eta_j+\eta_k\right)}{e^{-\beta E_j}+e^{-\beta E_k}} \geq \zeta_k e^{\beta E_k }$ holds as $\zeta_j e^{\beta E_j}\geq\zeta_k e^{\beta E_k}$. Hence at least one of $\lambda_1$ and $\lambda_2$ must lie in the interval $\left[0,1\right]$. Let:
\begin{equation}
\lambda=\min\left\{\lambda_1,\lambda_2\right\}.
\end{equation}

Let $\rho'$ be the state formed by applying the 2-level \longPLT\ $\textrm{\PLT}_{\left\{j,k\right\}}\left(\lambda\right)$ to $\rho$. Note that $\rho'$ has the same $\beta$-order as $\rho$ and $\sigma$. To see this, let $\left\{\eta'_i\right\}_{i=1}^{n}$ be the eigenvalues of $\rho'$ listed according to the $\beta$ ordering of $\rho$. Then:
\begin{align*}
\eta'_i&=\eta_i, \quad\textrm{for }1\leq i<j,\\
\eta'_i&=\zeta_i, \quad\textrm{for }j<i<k,\\
\eta'_i&=\eta_i, \quad\textrm{for }k< i<n,
\end{align*}
as the \longPLT\ does not change the occupation probabilities associated with $i\notin\left\{j,k\right\}$. Without loss of generality, suppose $\lambda=\lambda_1$ . Then $\eta'_j=\zeta_j$ and using Eq.~\eqref{eq:jk order} where appropriate, it is easy to see that:
\begin{equation}
\eta'_i e^{\beta E_i}\geq \eta'_{i+1} e^{\beta E_{i+1}}, \quad\textrm{for both }i\in\left\{1,\dots,k-2\right\} \textrm{ and }i\in\left\{k,\dots,n\right\}.
\end{equation}
To see that $\eta'_{k-1} e^{\beta E_{k-1}}\geq\eta'_{k} e^{\beta E_{k}}$, note that:
\begin{equation}
\eta'_{k} e^{\beta E_{k}}=\left(1-\lambda_1\right)\eta_k e^{\beta E_k} + \frac{\lambda_1\left(\eta_j+\eta_k\right)}{e^{-\beta E_j}+e^{-\beta E_k}}\leq\zeta_k e^{\beta E_k}\leq\zeta_{k-1} e^{\beta E_{k-1}}=\eta'_{k-1} e^{\beta E_{k-1}}.
\end{equation}
Hence the $\beta$-order of $\rho'$ is the same as $\rho$.

As \longPLT\ is a thermal operation, $\rho$ thermo-majorizes $\rho'$. Similarly, $\rho'$ thermo-majorizes $\sigma$. To see this, it suffices to show that Eq. \eqref{eq:majorization} still holds if we replace $\rho$ with $\rho'$. As $\eta_j+\eta_k=\eta'_j+\eta'_k$, this obviously holds for $m<j$ and $m\geq k$. By observing that $\eta'_j\geq\zeta_j$ the remaining cases follow.

Applying the procedure once, sets at least one of the occupation probabilities to that of $\sigma$. Hence, by repeating the procedure at most $n-1$ times, starting each iteration with the output of the previous \longPLT, we obtain $\sigma$. 
\end{proof}




\newpage

\section{States with different $\beta$-order} \label{sec:DBO}

In this appendix, we show that using the full set of \longCOs\ enables one to perform all transformations to block-diagonal states allowed under thermal operations.

Combining \longPLTs\ and \longLTs\ with the ability to append a single ancillary, qubit system with known Hamiltonian in the Gibbs state, makes them more powerful. Indeed, they can be used to perform any transition between block-diagonal states allowed under thermal operations without the need to expend any work. This is captured and proven in the following theorem:
\begin{theorem} \label{th:crude protocol with infty}
Suppose that $\rho$ and $\sigma$ are 
 states of an $n$-level system with Hamiltonian $H_S=\sum_{i=1}^{n} E_i \ketbra{i}{i}$ such that:
\begin{enumerate}
\item $\sigma$ is block-diagonal in the energy eigenbasis.
\item $\rho\stackrel{\textrm{TO}}{\longrightarrow}\sigma$.
\end{enumerate}
Then $\rho$ can be converted into $\sigma$ using \longCOs\ without expending any work.
\end{theorem}
\begin{proof}
To prove this we give a protocol consisting only of: adding (and eventually discarding) an ancilla qubit, Exact \longPF\ protocols (as introduced in Definition \ref{def:EPF}) and \longPLTs. None of these operations cost work. By first decohering $\rho$ in the energy eigenbasis if necessary and using an energy conserving unitary to rotate $\sigma$, we can assume that both $\rho$ and $\sigma$ are diagonal in the energy eigenbasis.

Let $\left(\tau_A,H_A\right)$ denote the known ancilla qubit allowed under Operation 1 of \longCOs. The protocol then runs as follows:
\begin{align*}
\left(\rho,H_S\right)&\longrightarrow\left(\rho\otimes\tau_A,H_S+H_A\right)\\
&\stackrel{\textrm{E}\PF}{\longrightarrow}\left(\rho',H'_{SA}\right)\\
&\stackrel{\PLT}{\longrightarrow}\left(\sigma',H'_{SA}\right)\\
&\stackrel{\textrm{E}\PF}{\longrightarrow}\left(\sigma\otimes\tau_A,H_S+H_A\right)\\
&\longrightarrow\left(\sigma,H_S\right).
\end{align*}
Here $\left(\rho',H'_{SA}\right)$ is a system with the same thermo-majorization curve as $\left(\rho\otimes\tau_A,H_S+H_A\right)$. However, the non-elbow points have been moved (potentially while sending energy levels to infinity) so that on the thermo-majorization diagram, they are vertically in line with the elbows (including the point $(Z,1)$) of $\left(\sigma\otimes\tau_A,H_S+H_A\right)$. This transformation can be performed using E\PFs\ and \ITRs. 

Similarly, $\left(\sigma',H'_{SA}\right)$ is a system with the same thermo-majorization curve as $\left(\sigma\otimes\tau_A,H_S+H_A\right)$ 
but with the non-elbow points moved to lie vertically 
inline with the elbows of $\left(\rho\otimes\tau_A,H_S+H_A\right)$ and at $(Z,1)$. 
Again, this transformation can be performed (and reversed) using the \longPF\ protocol.

Note that by construction, $\left(\rho',H'_{SA}\right)$ has the same $\beta$-ordering as $\left(\sigma',H'_{SA}\right)$ and as $\rho$ thermo-majorizes $\sigma$, $\left(\rho',H'_{SA}\right)$ thermo-majorizes $\left(\sigma',H'_{SA}\right)$. Hence, by Theorem \ref{th:same beta order} it is possible to transform $\left(\rho',H'_{SA}\right)$ into $\left(\sigma',H'_{SA}\right)$ using \longPLTs.

The overall protocol is illustrated in Figure \ref{fig:DBO}.
\end{proof}
\begin{figure}
\centering
\begin{minipage}{.5\columnwidth}
  \centering
  \includegraphics[width=0.9\textwidth]{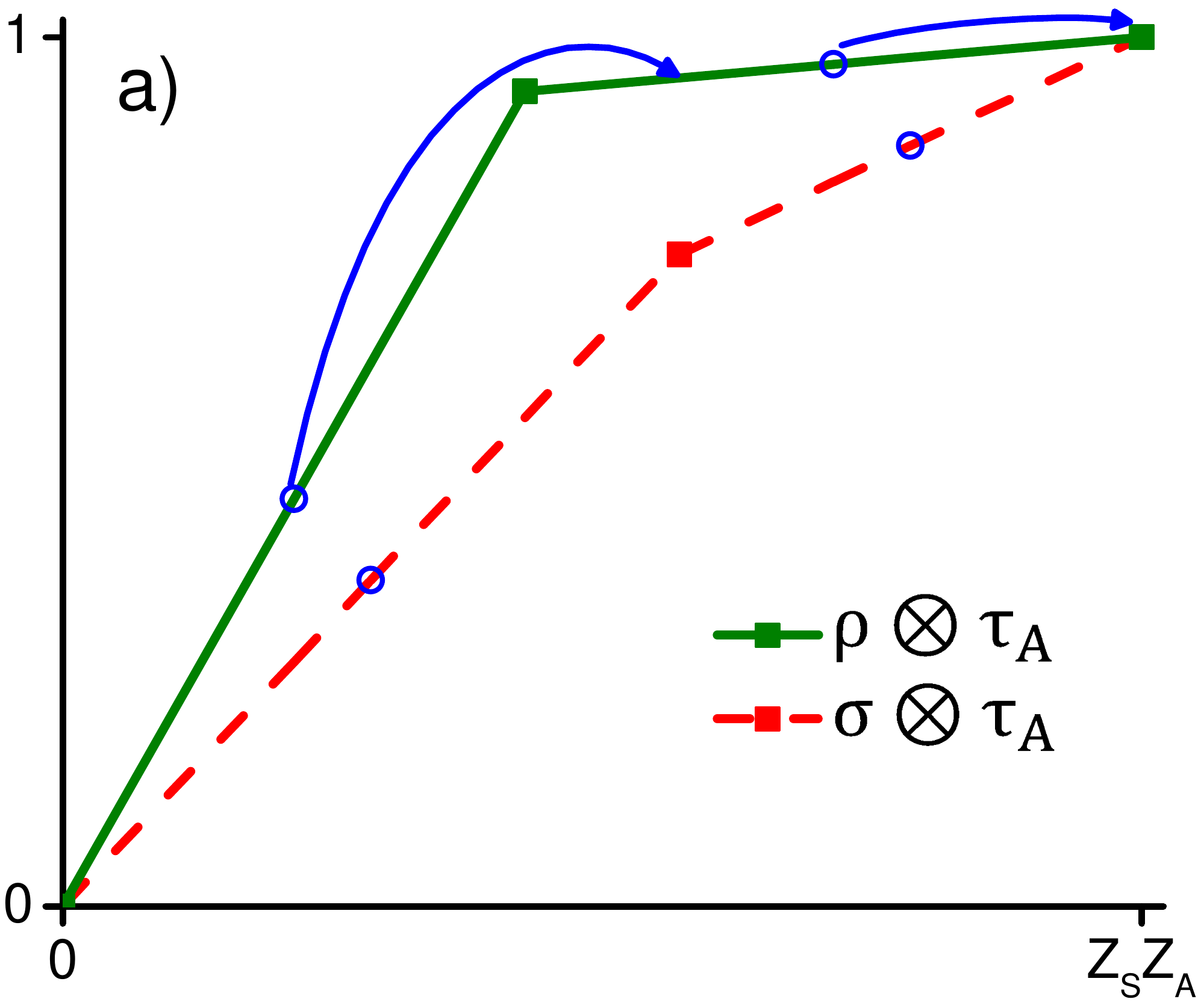}
  \label{fig:DBO1}
	\centerline{(a) \ITRs\ and \PFs}
\end{minipage}%
\begin{minipage}{.5\columnwidth}
  \centering
  \includegraphics[width=0.9\textwidth]{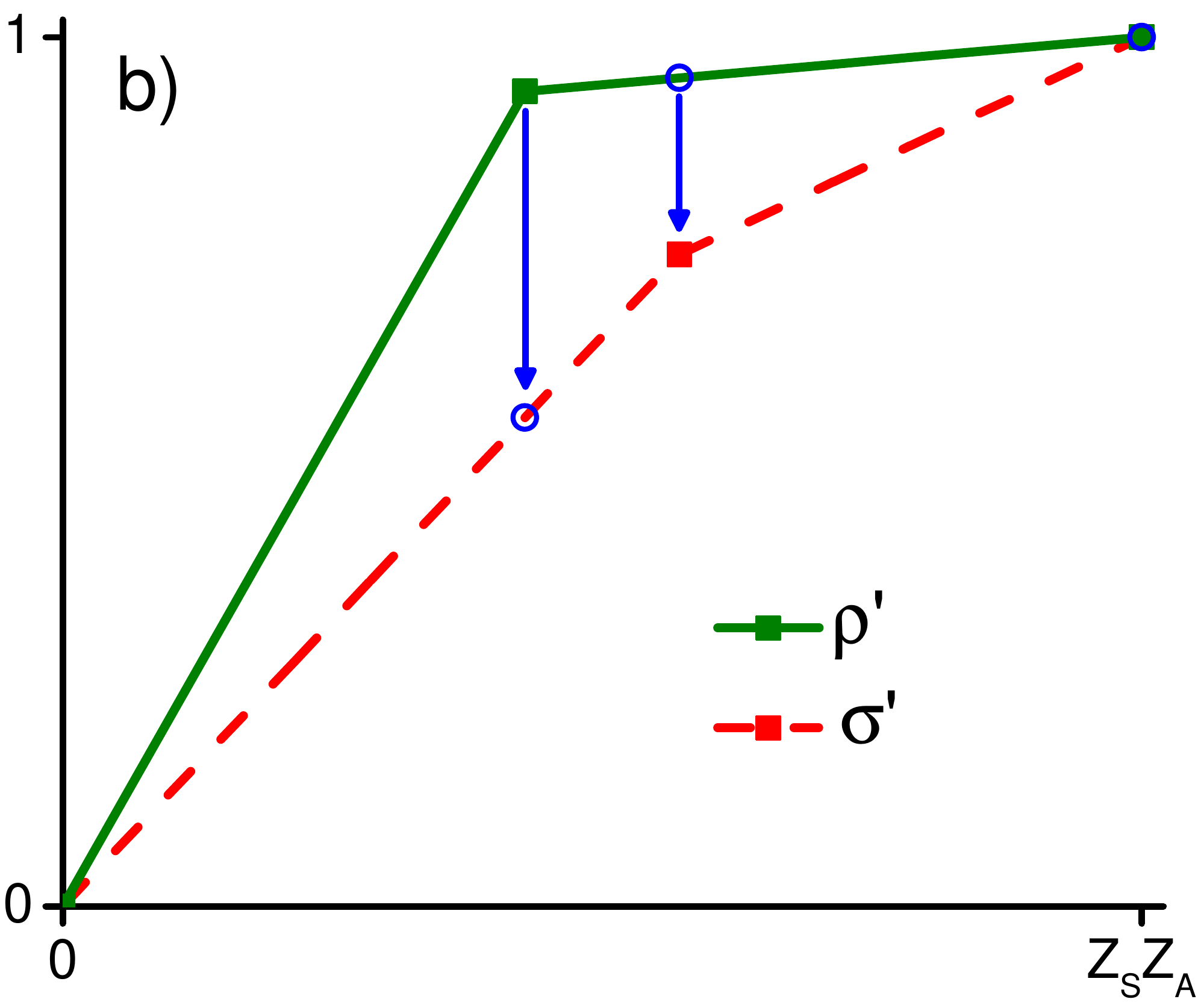}
  \label{fig:DBO2}
	\centerline{(b) Same $\beta$-order protocol applied}
\end{minipage}\\
\vspace{.5cm}
\begin{minipage}{.5\columnwidth}
  \centering
  \includegraphics[width=0.9\textwidth]{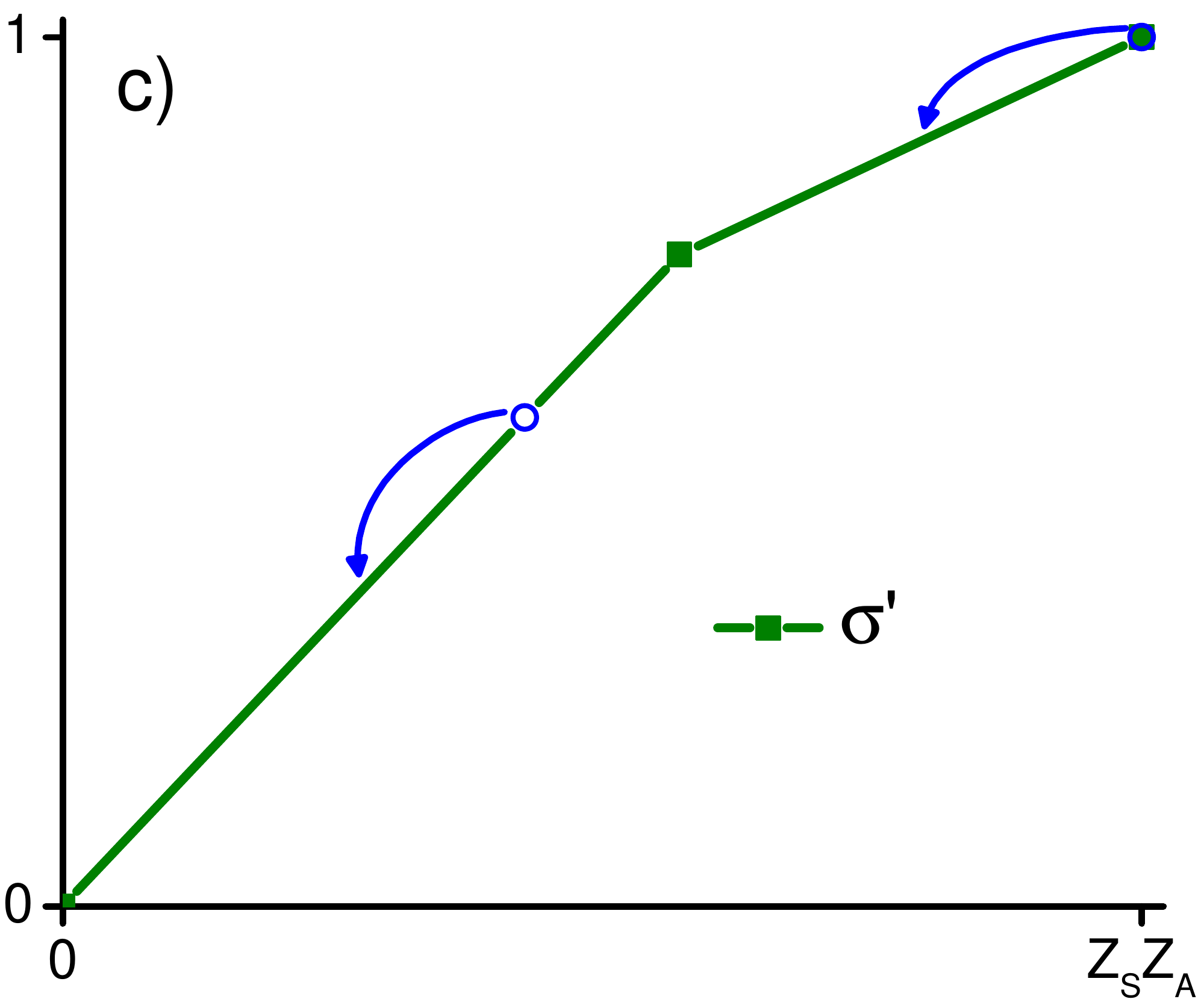}
  \label{fig:DBO3}
	\centerline{(c) \ITRs\ and \PFs}
\end{minipage}%
\begin{minipage}{.5\columnwidth}
  \centering
  \includegraphics[width=.9\textwidth]{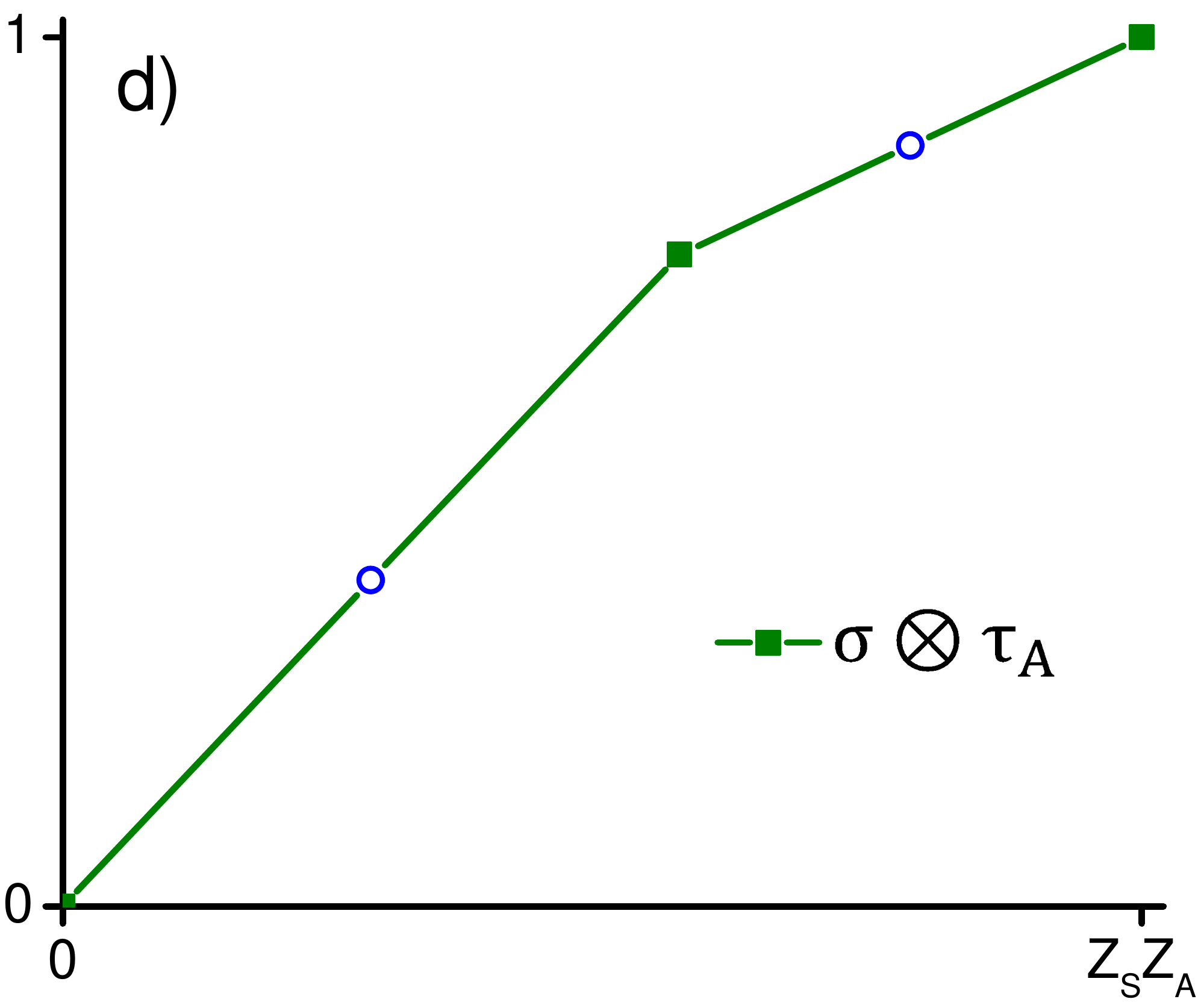}
  \label{fig:DBO4}
	\centerline{(d) Upon discarding $\tau_A$, the final state is $\sigma$}
\end{minipage}
\caption{\emph{\longCOs\ protocol for transforming between states with different $\beta$-orders.} If $\rho$ thermo-majorizes $\sigma$, then it is possible to transform $\rho$ into $\sigma$ using \longCOs. First, a thermal qubit, $\tau_A$, with known Hamiltonian is appended, Figure (a). Using \longITRs\, the blue circles on the thermo-majorization curve of $\rho\otimes\tau_A$ can be moved to be vertically aligned with the `elbows' on the curve for $\sigma\otimes\tau_A$. This forms the state $\rho'$, Figure (b), that has a thermo-majorization curve overlapping that of $\rho\otimes\tau_A$. Using the previously defined protocol for states with the same $\beta$-order, $\rho'$ can be converted into $\sigma'$, Figure (c), that has a thermo-majorization curve overlapping that of $\sigma\otimes\tau_A$. A final round of \ITRs\ converts $\sigma'$ into $\sigma\otimes\tau_A$, Figure (d), and upon discarding $\tau_A$ we obtain $\sigma$.}
\label{fig:DBO}
\end{figure}

The protocol described in the above theorem potentially requires that an energy level be raised to infinite energy. While this can be done at no work cost (provided it is performed infinitely slowly during the \longPF\ protocol), note that such a transition is not required if the thermo-majorization curves of $\rho$ and $\sigma$ do not touch on the interval $\left(0, Z\right)$.
\begin{theorem}
Suppose that $\rho$ and $\sigma$ are 
 states of an $n$-level system with Hamiltonian $H_S=\sum_{i=1}^{n} E_i \ketbra{i}{i}$ such that:
\begin{enumerate}
\item $\sigma$ is block-diagonal in the energy eigenbasis.
\item $\rho\stackrel{\textrm{TO}}{\longrightarrow}\sigma$.
\item The thermo-majorization curves of $\rho$ and $\sigma$ meet only at $\left(0,0\right)$ and $\left(Z_S,1\right)$.
\end{enumerate}
Then $\rho$ can be converted into $\sigma$ using \longCOs, without expending any work and without the need to raise an energy level to infinity.
\end{theorem}
\begin{proof}
Here we sketch how to modify the protocol given in Theorem \ref{th:crude protocol with infty} to avoid needing to raise an energy level to infinity. Again we can assume that $\rho$ and $\sigma$ are diagonal in the energy eigenbasis. The new protocol runs as follows:
\begin{align*}
\left(\rho,H_S\right)\longrightarrow&\left(\rho\otimes\tau_A,H_S+H_A\right)\\
\stackrel{\stackrel{\textrm{A}\PF}{\PLT}}{\longrightarrow}&\left(\tilde{\rho},H_S+H_A\right)\\
\stackrel{\PLT}{\longrightarrow}&\left(\sigma\otimes\tau_A,H_S+H_A\right)\\
\longrightarrow&\left(\sigma,H_S\right).
\end{align*}
Here $\left(\tilde{\rho},H_S+H_A\right)$ is a system with a thermo-majorization curve such that each one of its points (both elbows and non-elbows) are vertically aligned with the points of $\left(\sigma\otimes\tau_A,H_S+H_A\right)$.

To create $\left(\tilde{\rho},H_S+H_A\right)$, we use the following process, illustrated in Figure     \ref{fig:ABO}:
\begin{enumerate}
\item Using Approximate \longPFs, adjust the points of $\rho\otimes\tau_A$ to form $\rho'$ which has non-elbow points vertically aligned with the elbows of $\sigma\otimes\tau_A$. There are $n-1$ such points. As the thermo-majorization curves of $\rho\otimes\tau_A$ and $\sigma\otimes\tau_A$ touch only at $(0,0)$ and at $\left(Z_S Z_A,1\right)$, the A\PF\ can be chosen such that $\rho'$ has the desired alignment, thermo-majorizes $\sigma\otimes\tau_A$ and such that the thermo-majorization curves of $\rho'$ inherits these properties.
\item For each vertically aligned point $i\in\left\{1,\dots n-1\right\}$ on $\rho'$, consider the number of points (both elbows and non-elbows) to the left of it on its thermo-majorization curve. Call this number $r_i$. Compare this quantity to the number of points to the left of the associated vertically aligned point on the thermo-majorization curve of $\sigma\otimes\tau_A$. Call this number $s_i$. If:
\begin{enumerate}
\item $r_i<s_i$: Move the point slightly to the right of its aligned location using a \ITR.
\item $r_i>s_i$: Move the point slightly to the left of its aligned location using a \ITR.
\item $r_i=s_i$: Leave the point where it is.
\end{enumerate}
These \ITRs\ result in a state $\rho''$ with the same thermo-majorization curve as $\rho'$.
\item Defining $i=0$ to be the point $(0,0)$ and $i=n$ to be the point $\left(Z_S Z_A,1\right)$, for each $i\in\left\{1,\dots n\right\}$ thermalize $\rho''$ over the the interval between points $i-1$ and $i$ using \PLTs.
This results in a state $\rho'''$ which has elbows almost vertically aligned with the elbows of $\sigma\otimes\tau_A$.
Provided the movements due to \ITRs\ in Step 2 were chosen to be sufficiently small, as $\rho''$ thermo-majorizes $\sigma\otimes\tau_A$ and their thermo-majorization curves touch only at $(0,0)$ and $\left(Z_S Z_A,1\right)$, $\rho'''$ inherits the same properties.
\item Using Approximate \longPFs, adjust the points of $\rho'''$ to form $\tilde{\rho}$ as defined above. The last time an A\PF\ is applied to an elbow, it should be done in such a way that after the operation, the elbow is precisely vertically aligned with that of $\sigma\otimes\tau_A$. The displacements applied in Step 2 enable this to take place. As $\rho'''$ thermo-majorizes $\sigma\otimes\tau_A$ and their thermo-majorization curves touch only at (0,0) and $\left(Z_S Z_A,1\right)$, $\tilde{\rho}$ again inherits these properties. 
\end{enumerate}

Due to the fact that the protocol uses Approximate \longPFs\ rather than Exact \longPFs, there is no need to raise an energy level to infinity.

As $\left(\tilde{\rho},H_S+H_A\right)$ thermo-majorizes $\sigma\otimes\tau_A$ and they have the same $\beta$-ordering, the transformation can now be completed using \longPLTs\ as described in Theorem \ref{th:same beta order}.
\end{proof}

Note that there are scenarios in which the restriction that the curves touch only at $\left(0,0\right)$ and $\left(Z_S,1\right)$ in the above theorem can be relaxed slightly to demanding that the curves touch only at $\left(0,0\right)$ and on the line $y=1$. For example, this is the case if $\left|W_{\rho\rightarrow\tau_S}\right|>\left|W_{\sigma\rightarrow\tau_S}\right|$, where $\tau_S$ is the Gibbs state of the system's Hamiltonian, i.e. it is possible to extract strictly more work deterministically from $\rho$ than from $\sigma$. In general, allowing the curves to touch at $y=1$ makes the theorem more relevant for situations where we wish to model a change of Hamiltonian or include a work storage system as per Appendix \ref{ssec:wits&switch}.

\begin{figure}[!htbp]
\centering
\setcounter{figure}{18}
\begin{minipage}{.5\textwidth}
  \centering
  \includegraphics[width=.9\linewidth]{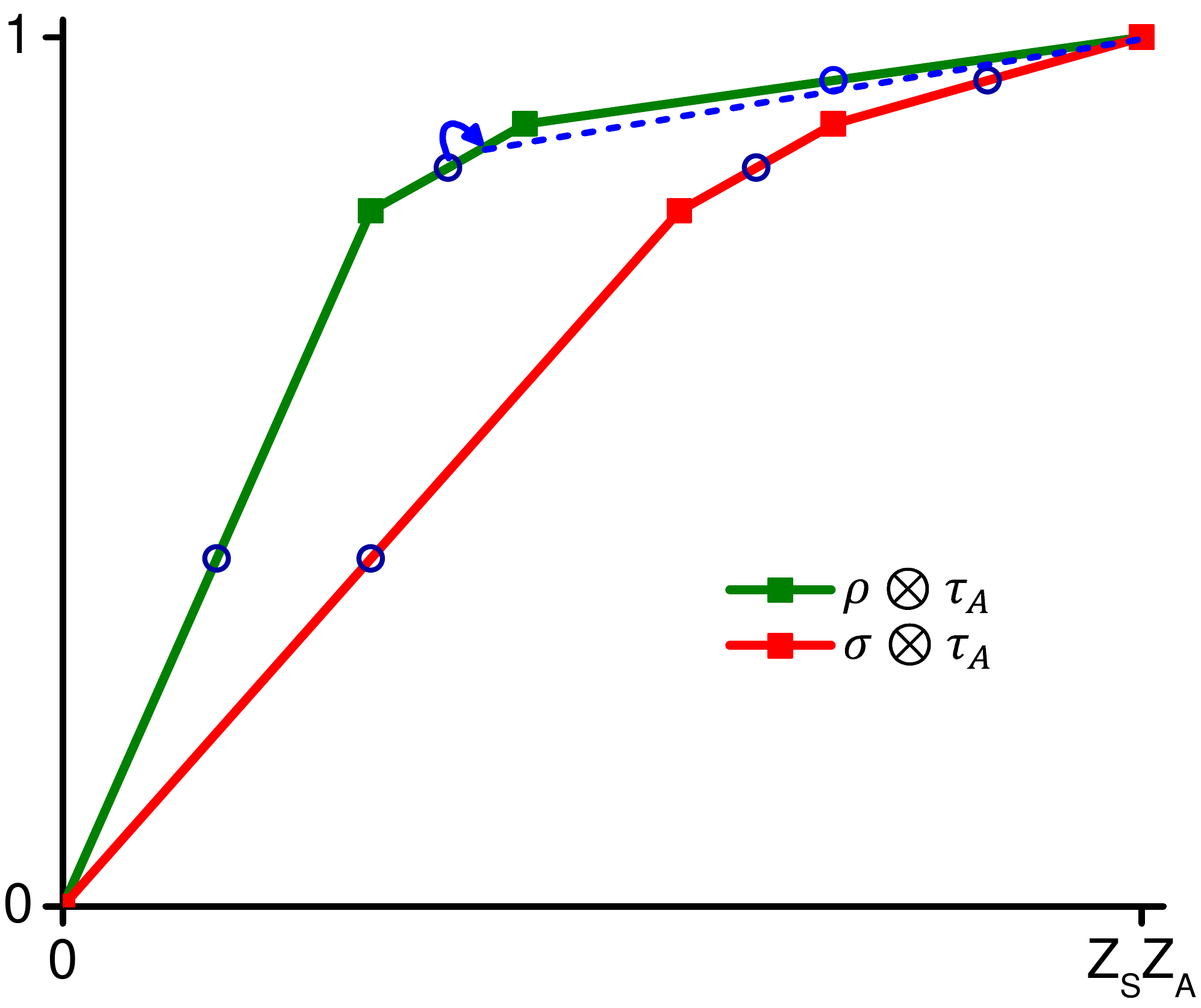}
  \centerline{(a) Initial state. Step 1: A\PFs.}
\end{minipage}%
\begin{minipage}{.5\textwidth}
  \centering
  \includegraphics[width=.9\linewidth]{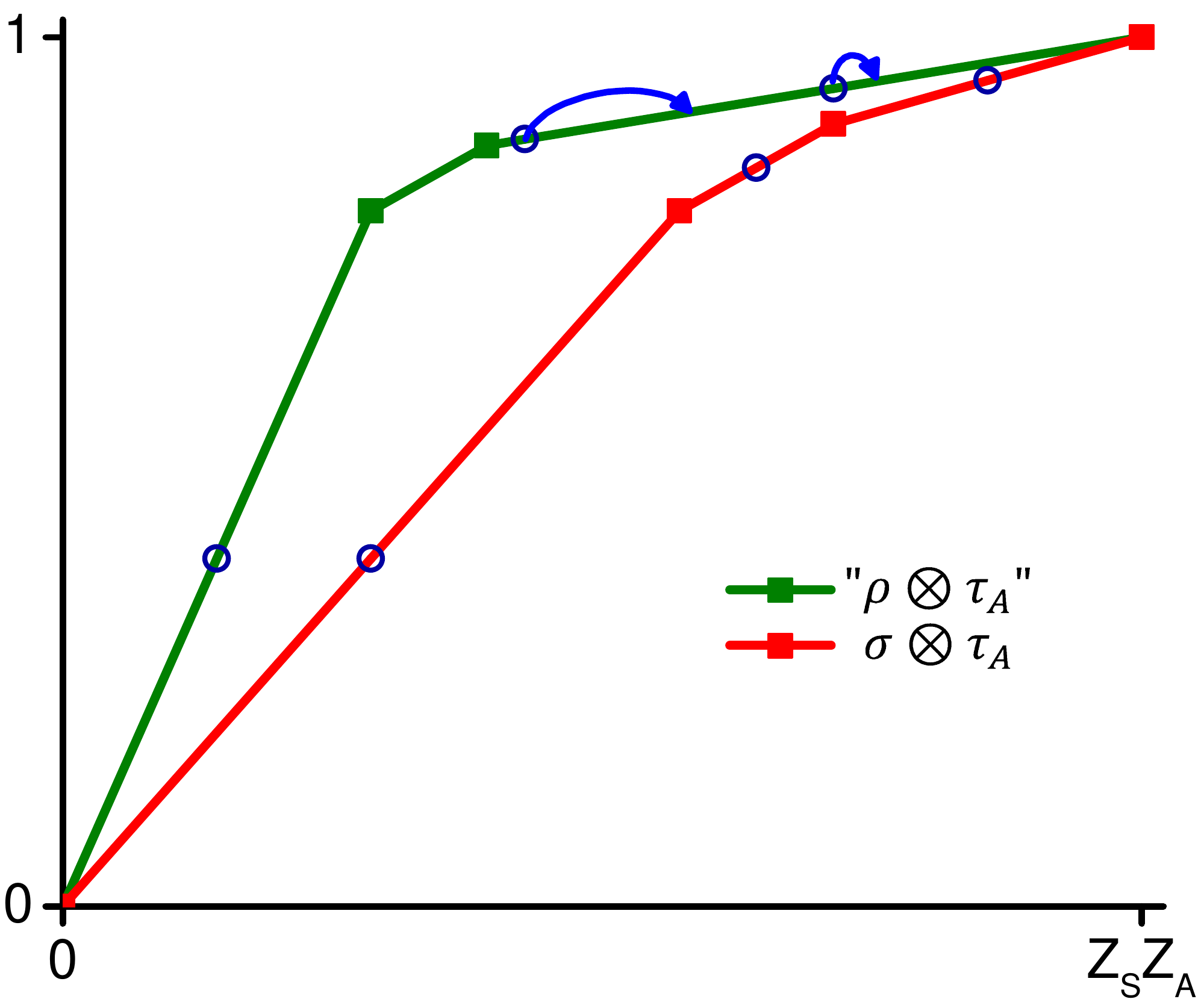}
  \centerline{(b) Step 2: \ITRs.}
\end{minipage}\\
\vspace{1.7cm}
\begin{minipage}{.5\textwidth}
  \centering
  \includegraphics[width=.9\linewidth]{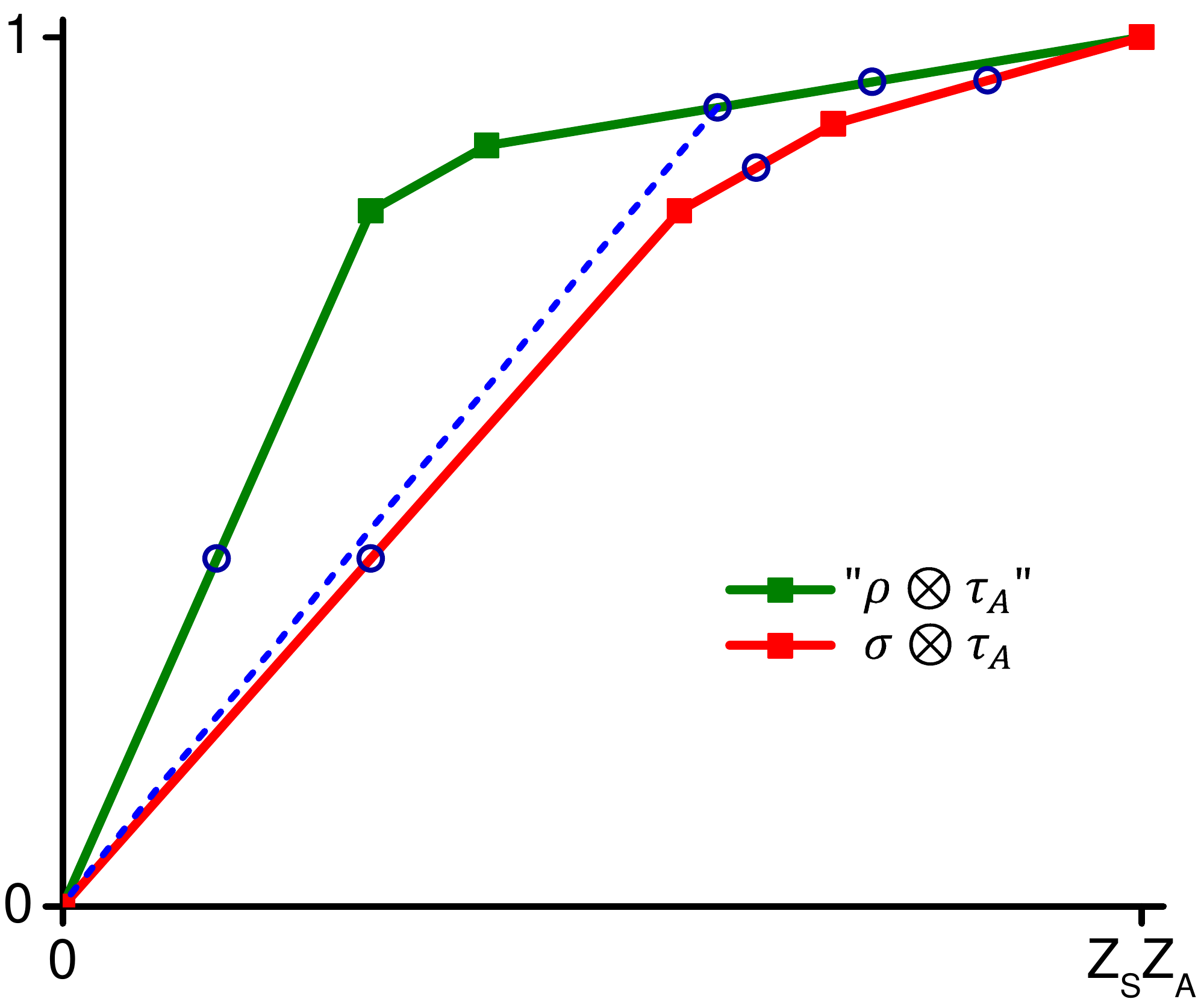}
  \centerline{(c) Step 3: \PLTs.}
\end{minipage}%
\begin{minipage}{.5\textwidth}
  \centering
  \includegraphics[width=.9\linewidth]{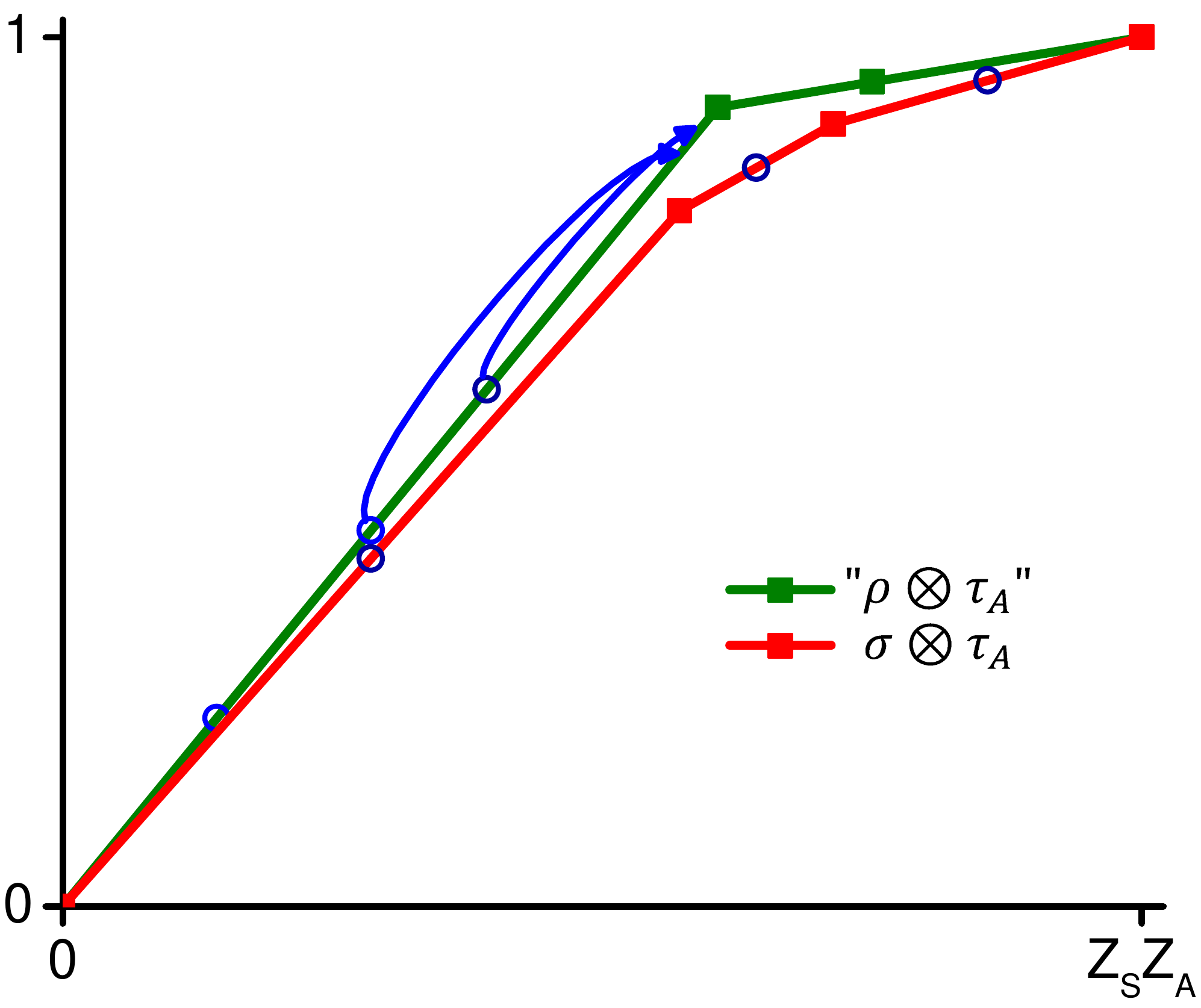}
  \centerline{(d) Step 4: A\PFs.}
\end{minipage}\\
\vspace{1.7cm}
\begin{minipage}{.5\textwidth}
  \centering
  \includegraphics[width=.9\linewidth]{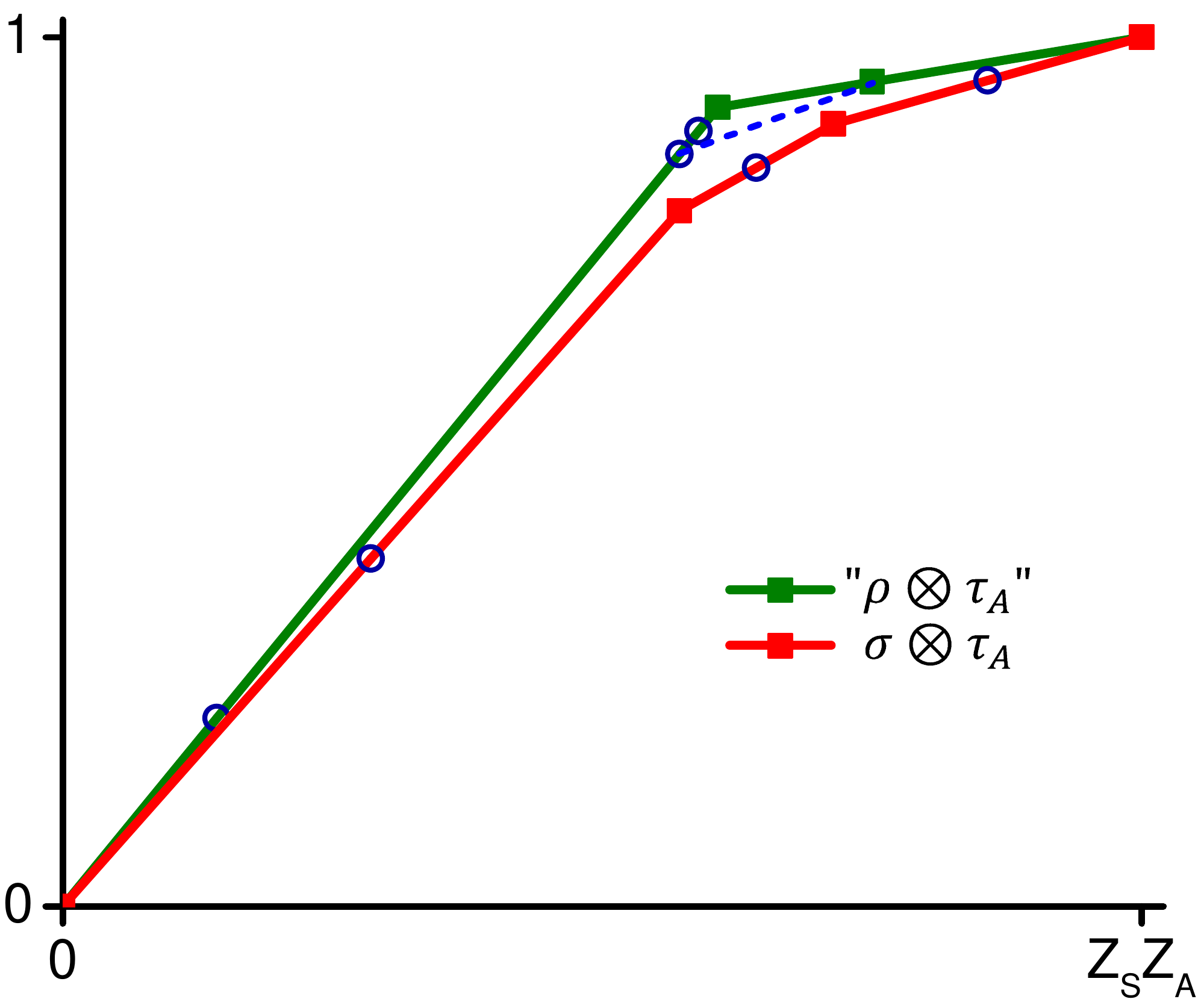}
  \centerline{(e) Step 4: A\PFs\ continued.}
\end{minipage}%
\begin{minipage}{.5\textwidth}
  \centering
  \includegraphics[width=.9\linewidth]{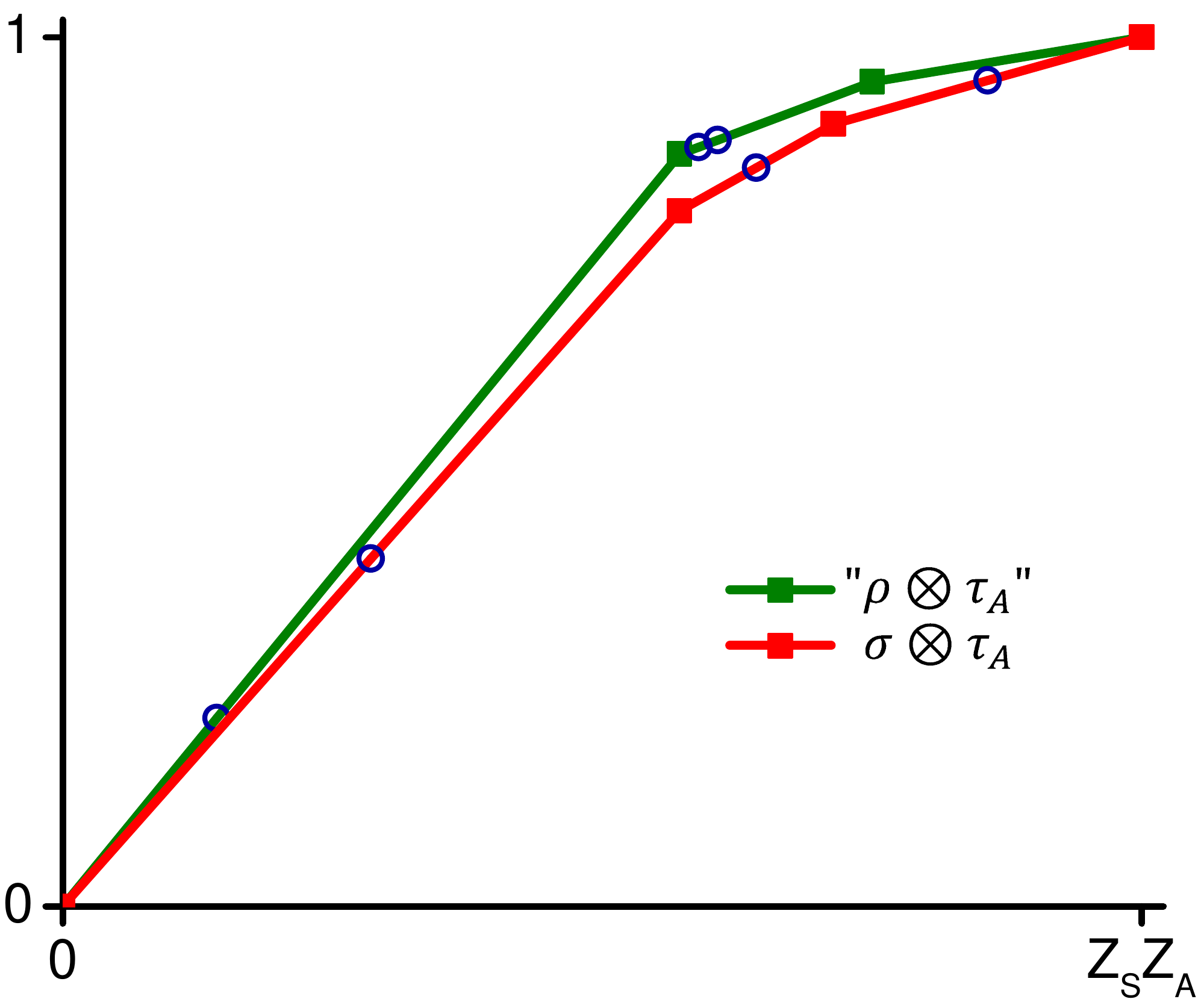}
  \centerline{(f) Step 4: A\PFs\ continued.}
\end{minipage}
\end{figure}

\begin{figure}[!htbp]
\ContinuedFloat
\begin{flushleft}
\begin{minipage}{.5\textwidth}
  \centering
  \includegraphics[width=.9\linewidth]{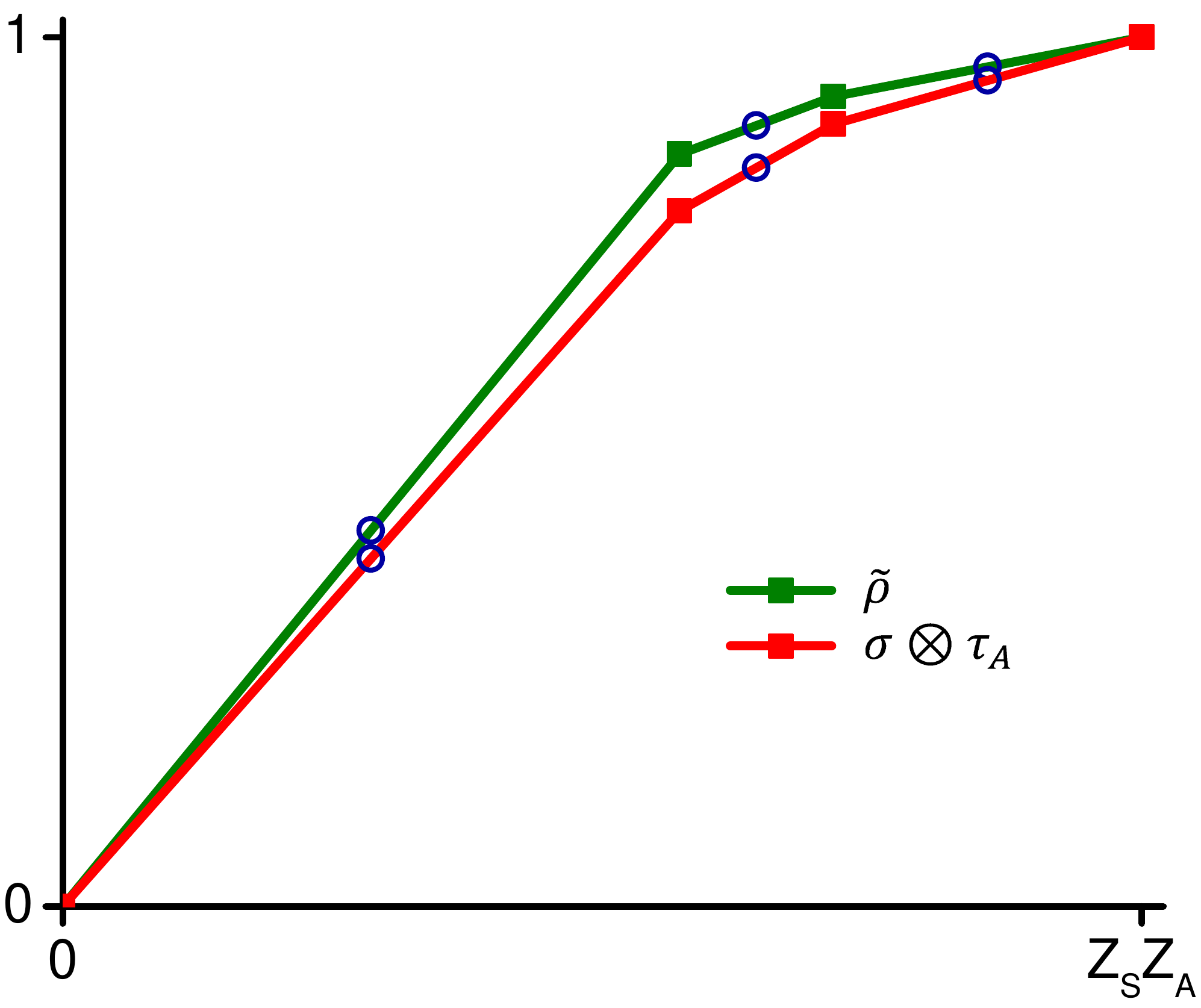}
  \centerline{(g) Final state with the same $\beta$-ordering as $\sigma\otimes\tau_A$.}
\end{minipage}
\end{flushleft}
\caption[$\beta$-order change with Approximate \longPFs.]{\emph{$\beta$-order change with Approximate \longPFs.} In Step 1 of the protocol, A\PFs\ are performed so that the non-elbows of $\rho\otimes\tau_A$ can be horizontally aligned with the elbows of $\sigma\otimes\tau_A$. This transforms Figure (a) into Figure (b). In Step 2, \ITRs\ are used to slightly misaligned the points in anticipation of Step 4. In this example they are misaligned to the right resulting in Figure (c). Next, \PLTs\ are used in Step 3 to generate elbows (almost) horizontally aligned with those of $\sigma\otimes\tau_A$. This leads to Figure (d). Finally, in Step 4 A\PFs\ are applied to exactly match the elbows together with the non-elbow points. This is detailed in Figures (e) and (f). The result of the protocol is shown in Figure (g).}
\label{fig:ABO}
\end{figure}

\newpage

\section{Quantifying worst-case work costs with \longCOs}\label{sec:worstwork}

Instead of using the wit construction summarized in Section \ref{ssec:wits&switch} to analyze the work value of a transformation, under \longCOs\ one can consider the work value of the \longLTs\ used during a process as defined in Eq.~\eqref{eq:LT cost}. In this section, we shall show that using this approach reproduces the results regarding worst-case work obtained under thermal operations in \cite{horodecki2013fundamental, aaberg2013truly}.

\subsection{Extractable work} \label{ss:extractwork}


By setting the final state to be $\tau_S$, the thermal state of the system, in Eq. \eqref{eq:work def}, one can define the single-shot \emph{distillable work} - the amount of work that can be obtained from a state. One can also consider the \emph{smoothed} distillable work, where one allows the possibility of failing to distil positive work with some probability $\epsilon$. Denoting this quantity for a given state $\rho$ by $W^{\epsilon}_{\textrm{distil}}\left(\rho\right)$, in \cite{horodecki2013fundamental, aaberg2013truly} it was shown to be given by:
\begin{align}
\begin{split} \label{eq:distill work}
W^{\epsilon}_{\textrm{distil}}\left(\rho\right)&=F_{\textrm{min}}^{\epsilon}\left(\rho\right)+\frac{1}{\beta}\ln Z\\
&=-\frac{1}{\beta} \left[\ln\left(\tilde{L}_{1-\epsilon}\left(\rho\right)\right)-\ln Z\right],
\end{split}
\end{align}
where $\tilde{L}_{y}$ denotes the horizontal distance between a state's thermo-majorization curve and the $y$-axis at $y$, a quantity discussed more fully in \cite{alhambra2015probability}.

\begin{figure}
\centering
\begin{minipage}{.5\columnwidth}
  \centering
  \includegraphics[width=0.9\textwidth]{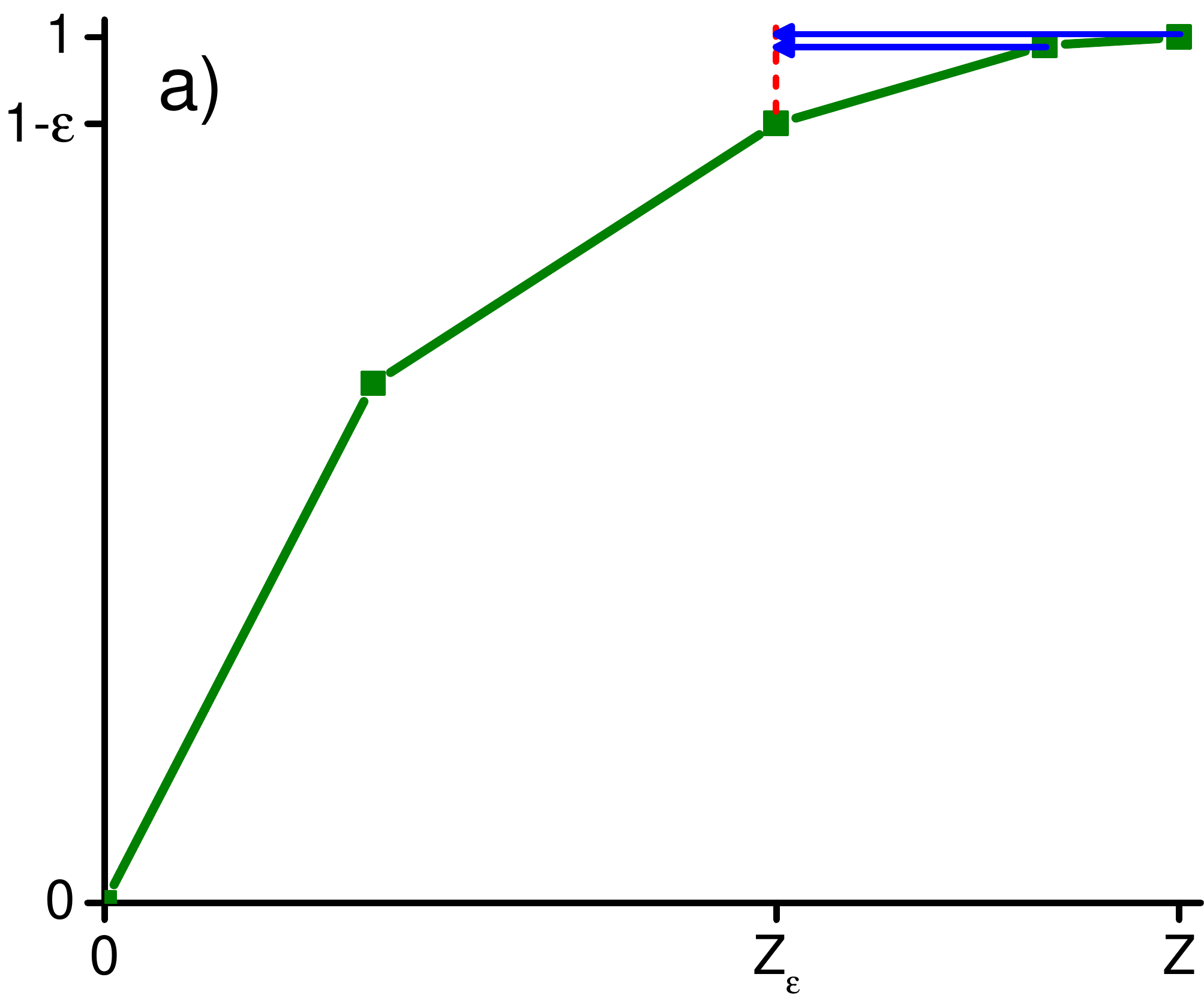}
  \label{fig:ExtractB1}
	\centerline{(a) \LTs}
\end{minipage}%
\begin{minipage}{.5\columnwidth}
  \centering
  \includegraphics[width=0.9\textwidth]{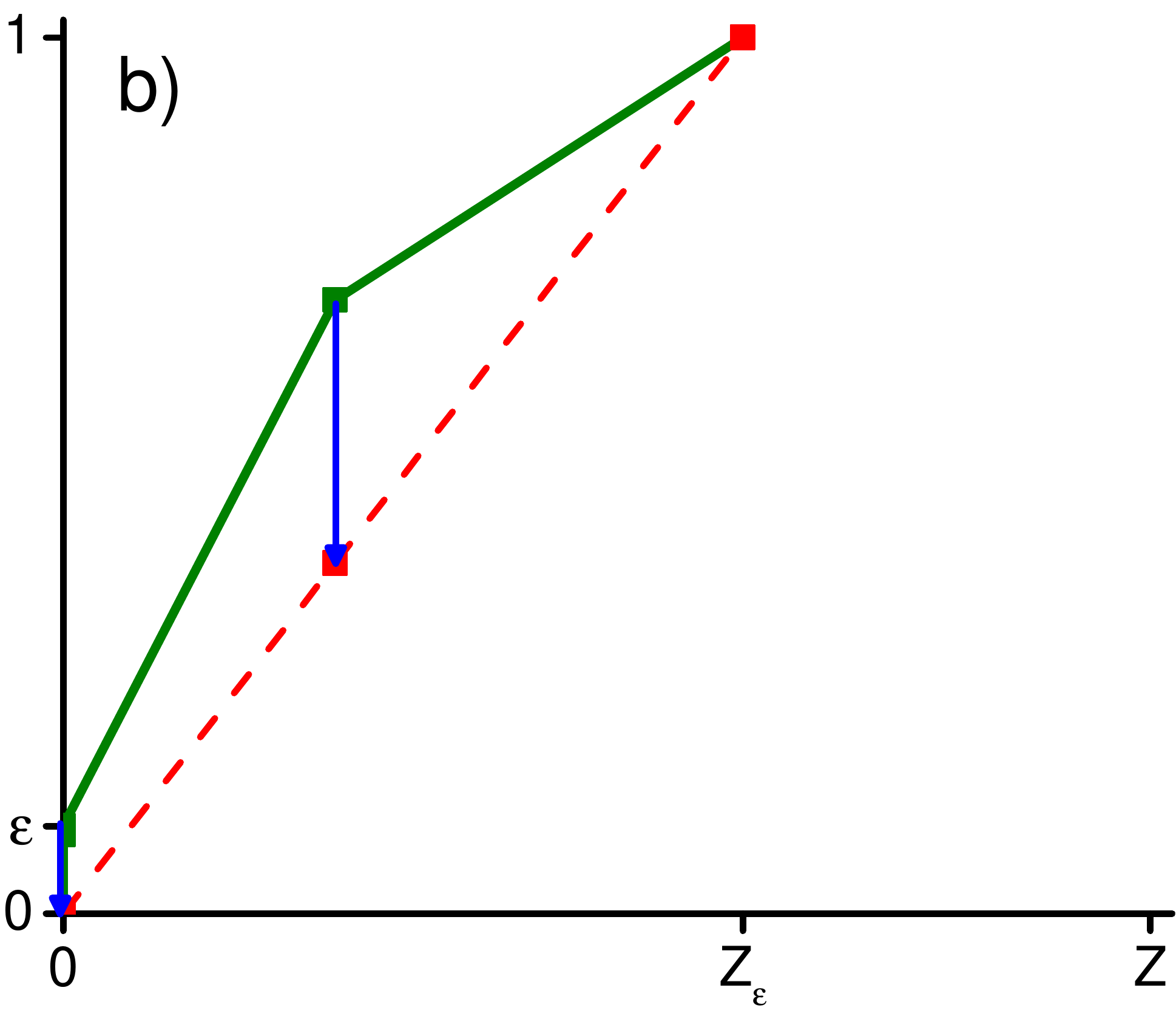}
  \label{fig:ExtractB2}
	\centerline{(b) \PLTs}
\end{minipage}\\
\vspace{.5cm}
\begin{minipage}{.5\columnwidth}
  \centering
  \includegraphics[width=0.9\textwidth]{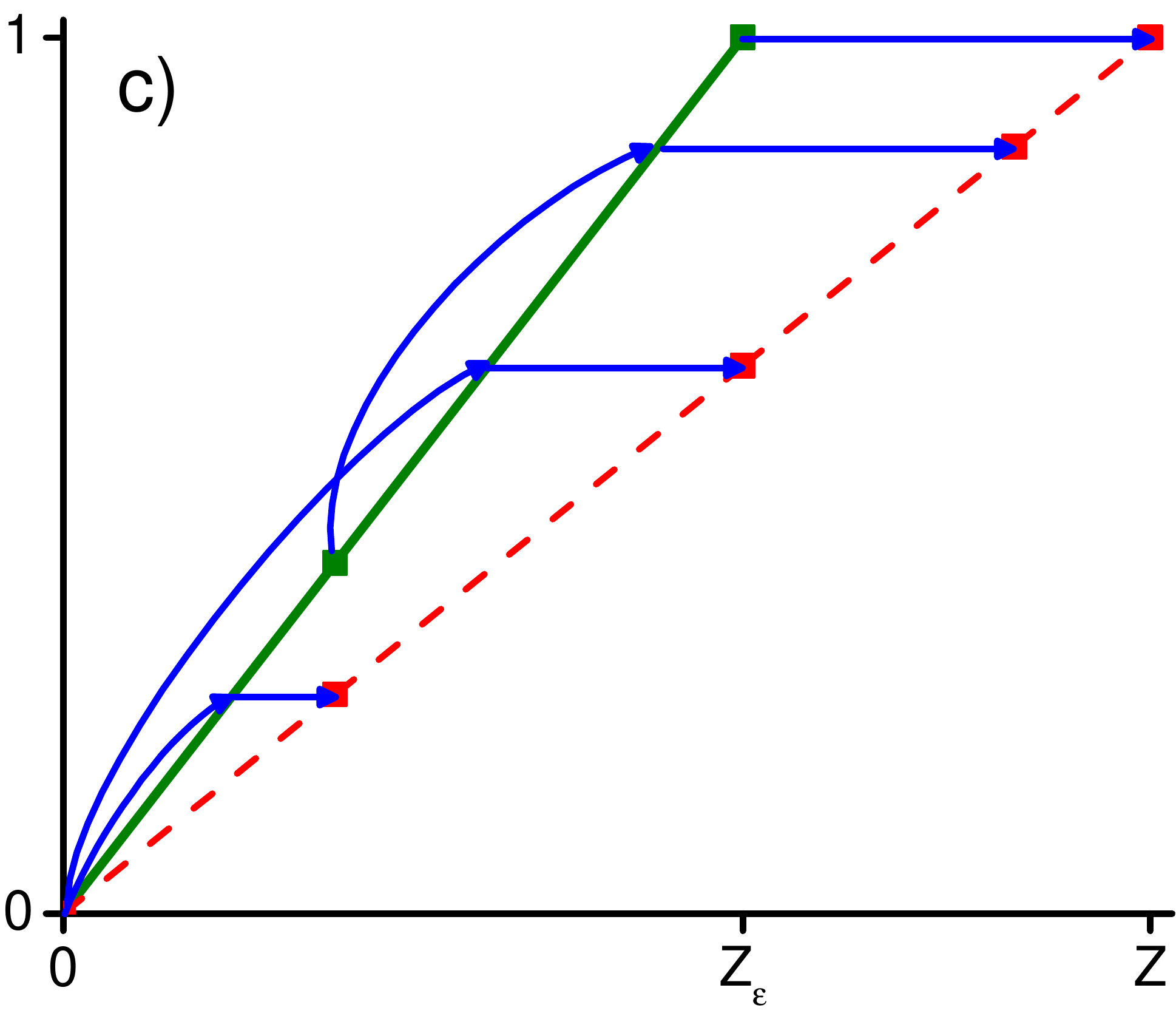}
  \label{fig:ExtractB3}
	\centerline{(c) \ITRs\ and \LTs}
\end{minipage}%
\begin{minipage}{.5\columnwidth}
  \centering
  \includegraphics[width=.9\textwidth]{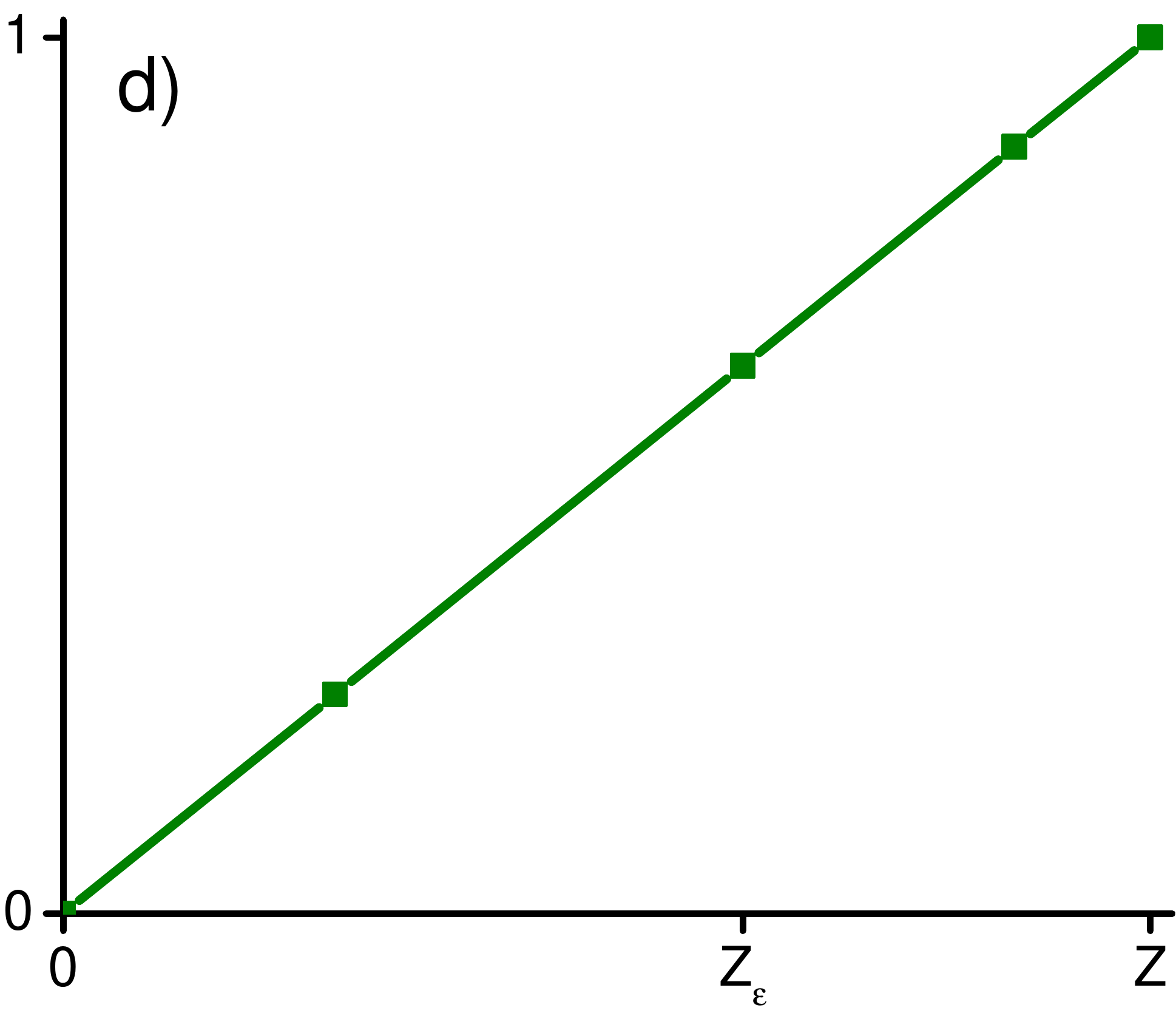}
  \label{fig:ExtractB4}
	\centerline{(d) Final state, $\tau$}
\end{minipage}
\caption{\emph{\longCOs\ protocol for $\epsilon$-deterministic work extraction.} Here we illustrate the protocol for $\epsilon$-deterministic work extraction from \cite{aaberg2013truly} in terms of thermo-majorization diagrams. First, the energy levels to the right of $Z_\epsilon$ are raised to infinity, transforming Figure (a) into Figure (b). Next, the system is fully thermalized, resulting in Figure (c). Finally, by applying \ITRs\ and a deterministic \LT, the system is transformed into the thermal state of the original Hamiltonian, Figure (d). Note that the first step, transforming Figure (a) into Figure (b), involves raising energy levels to infinity. This costs an infinitely large amount of work with probability $\epsilon$ and no work with probability $1-\epsilon$.} \label{fig:Abergs protocol}
\end{figure}

In Figure \ref{fig:Abergs protocol} we represent using thermo-majorization diagrams the protocol from \cite{aaberg2013truly} that distills the amount of work given in Eq.~\eqref{eq:distill work}. This protocol consists solely of \longCOs. Note that as decohering is a \longCO\ which commutes with all other thermal operations \cite{brandao2011resource}, we can first decohere $\rho$ in the energy eigenbasis without altering the amount of work we can extract. Let $Z_\epsilon=\tilde{L}_{1-\epsilon}\left(\rho\right)$ denote the point on the $x$-axis such that those energy levels to the left of it on the thermo-majorization curve of $\rho$ have cumulative population given by $1-\epsilon$ and those energy levels to the right have total weight $\epsilon$. Without loss of generality, we can assume that $\left(Z_\epsilon,1-\epsilon\right)$ is either an elbow or non-elbow on the thermo-majorization curve of $\rho$ as we can always make it so. To do this, we append a thermal qubit ancilla, $\tau_A$, consider the thermo-majorization curve of $\rho\otimes\tau_A$ and precede the extraction protocol with a \longITR\ to move a non-elbow to the desired location.

The protocol then runs as follows:
\begin{enumerate}
\item Raise the energy levels to the right of $Z_\epsilon$ to infinity using \longLTs.
\item Fully thermalize the system.
\item Perform \longITRs\ to horizontally align the points on the system's thermo-majorization curve with those of the target thermal state.
\item Perform a \longLT\ to transform the system into the thermal state of the initial Hamiltonian.
\end{enumerate}
Two stages of this protocol have non-zero work value. The \longLT\ that takes Figure \ref{fig:Abergs protocol}(a) to Figure \ref{fig:Abergs protocol}(b) costs an infinitely large amount of work with probability $\epsilon$ and no work with probability $1-\epsilon$. The second \longLT\ that takes Figure \ref{fig:Abergs protocol}(c) to Figure \ref{fig:Abergs protocol}(d), has deterministic work yield $\frac{1}{\beta}\ln\left(\frac{Z}{Z_\epsilon}\right)$. Hence the overall protocol achieves the work value given in Eq.~\eqref{eq:distill work}.
 
\begin{figure}
\centering
\begin{minipage}{.5\columnwidth}
  \centering
  \includegraphics[width=0.9\textwidth]{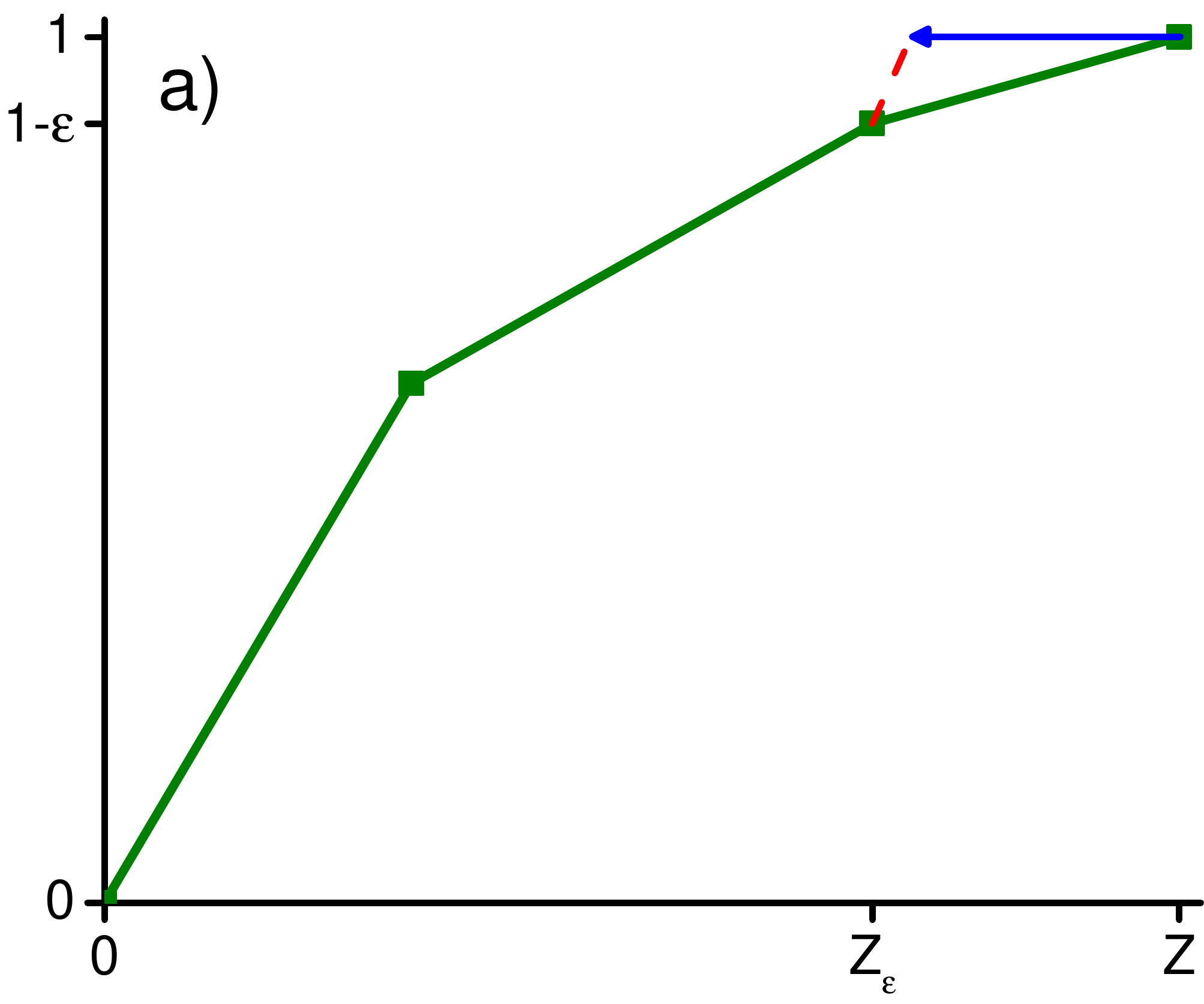}
  \label{fig:ExtractV1}
	\centerline{(a) \LTs}
\end{minipage}%
\begin{minipage}{.5\columnwidth}
  \centering
  \includegraphics[width=0.9\textwidth]{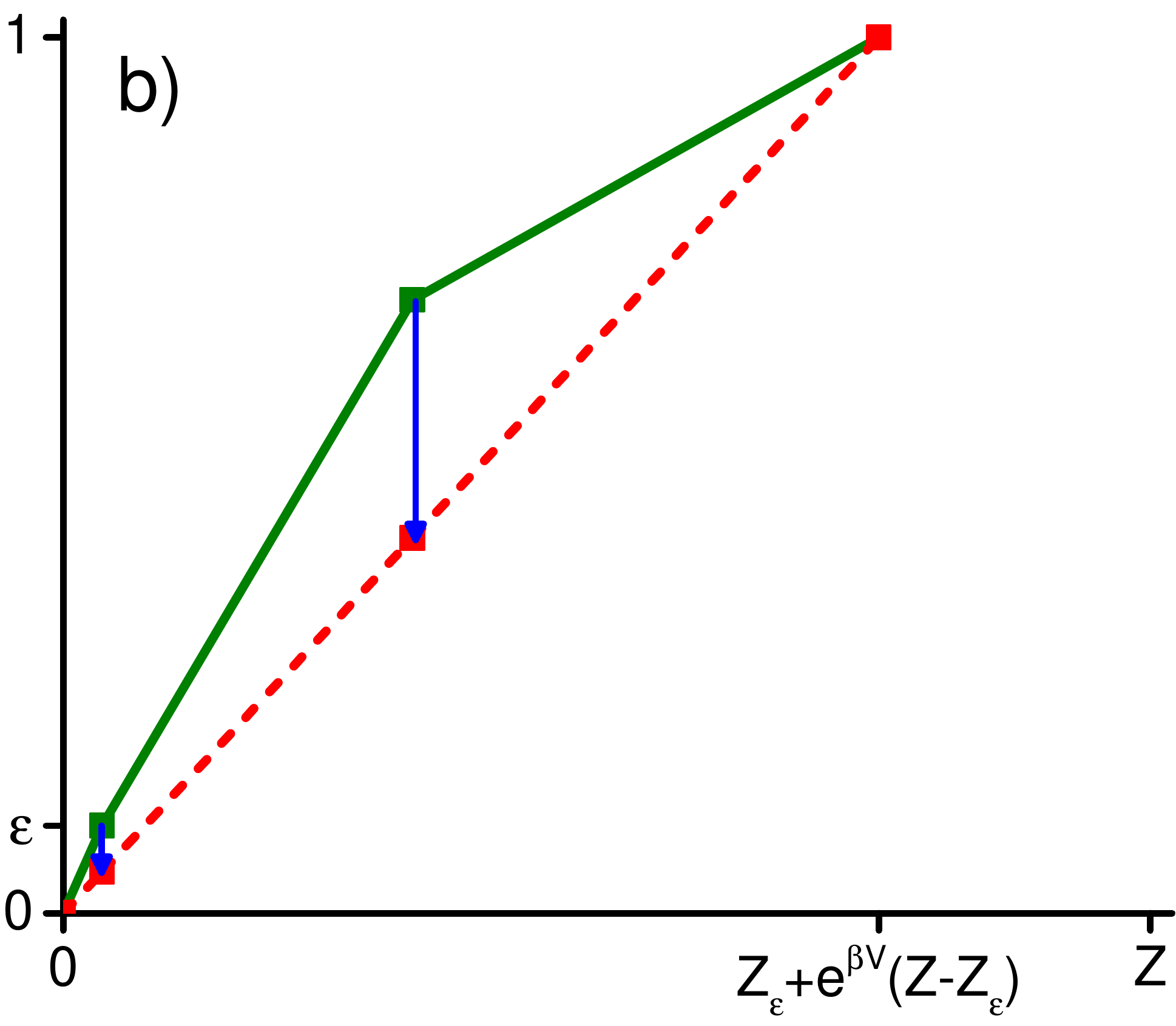}
  \label{fig:ExtractV2}
	\centerline{(b) \PLTs}
\end{minipage}\\
\vspace{.5cm}
\begin{minipage}{.5\columnwidth}
  \centering
  \includegraphics[width=0.9\textwidth]{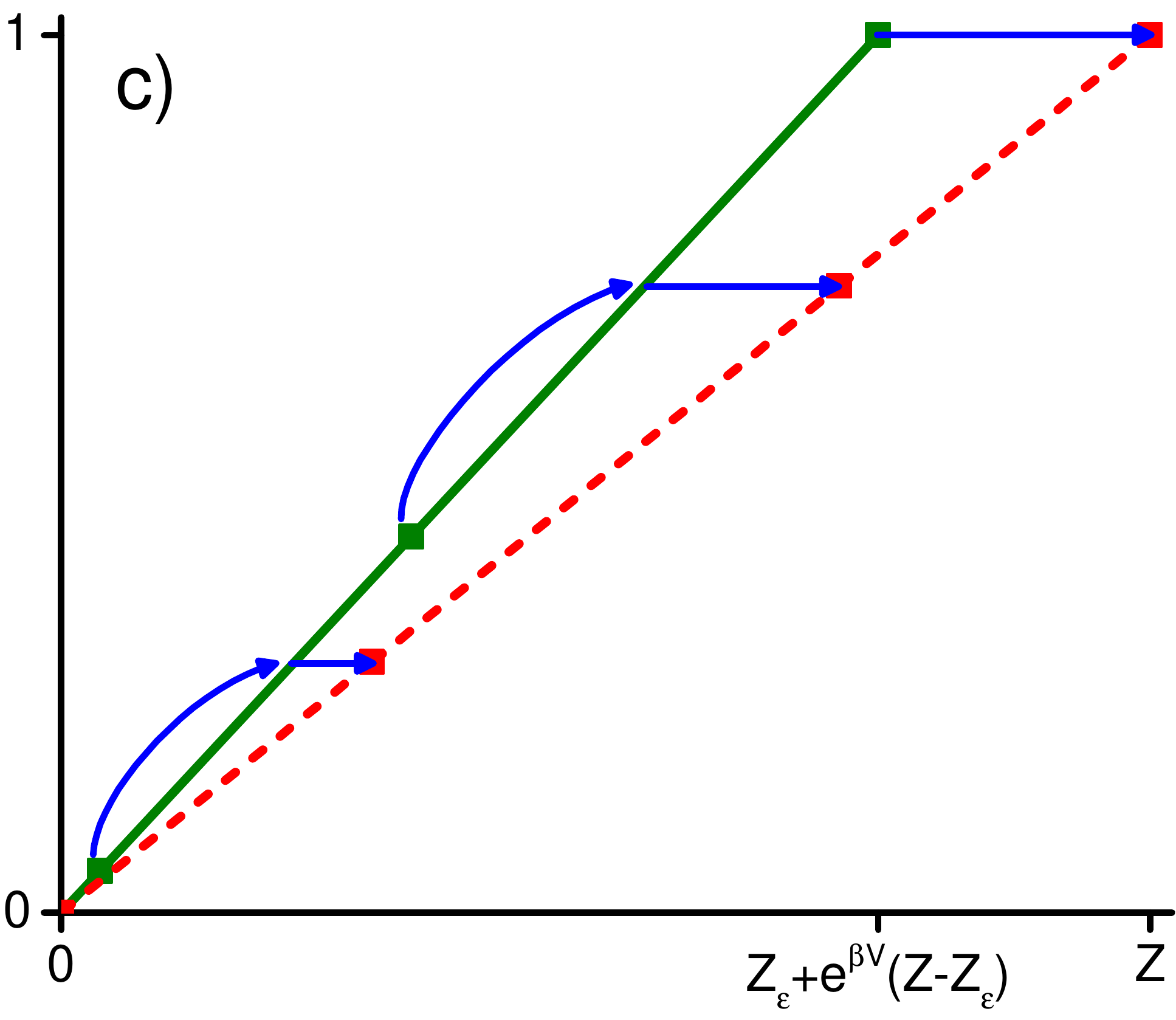}
  \label{fig:ExtractV3}
	\centerline{(c) \ITRs\ and \LTs}
\end{minipage}%
\begin{minipage}{.5\columnwidth}
  \centering
  \includegraphics[width=.9\textwidth]{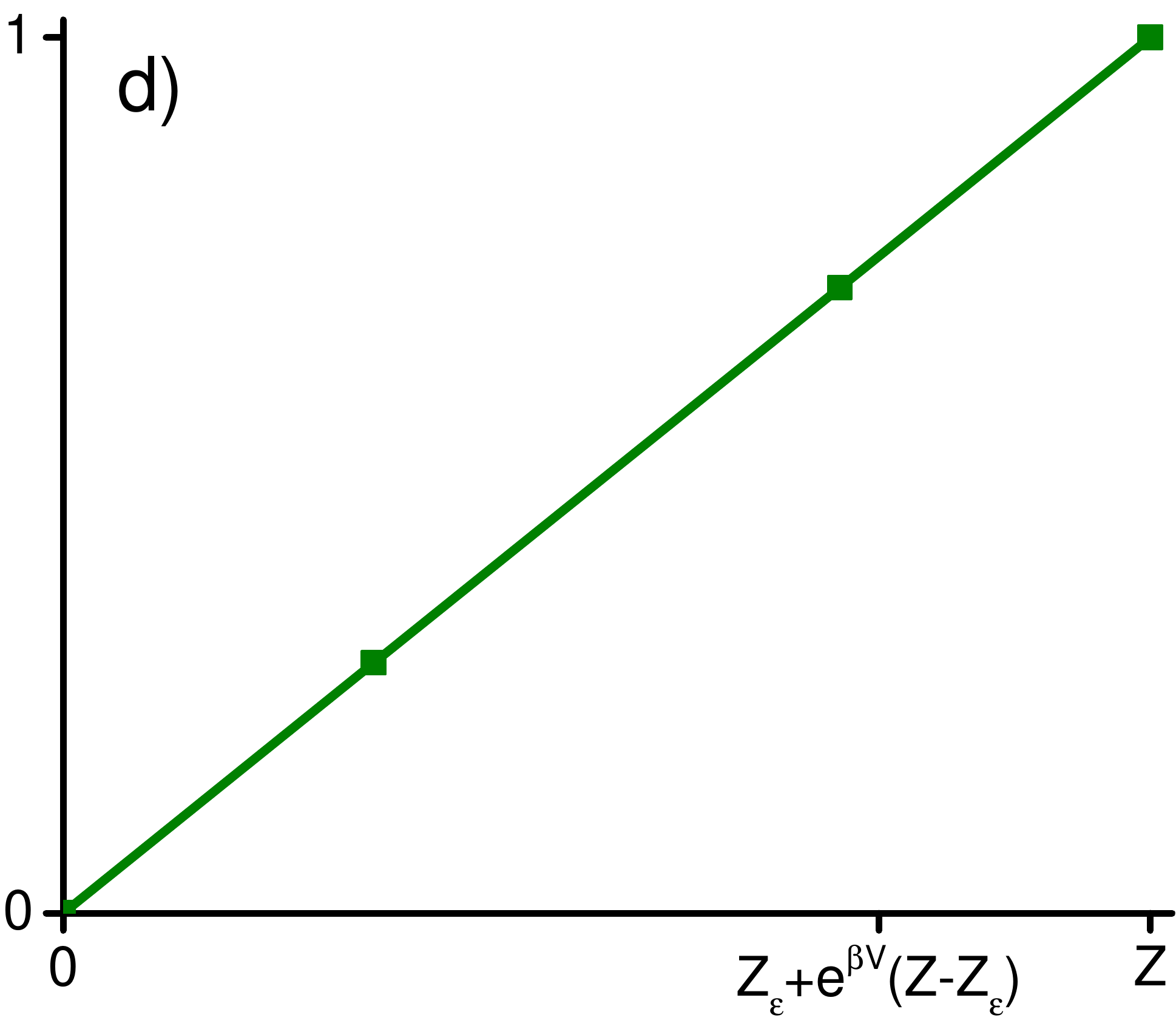}
  \label{fig:ExtractV4}
	\centerline{(d) Final state, $\tau$}
\end{minipage}
\caption{\emph{\longCOs\ protocol for $\epsilon$-deterministic work extraction without infinite cost upon failure.} Here we illustrate the protocol for $\epsilon$-deterministic work extraction without infinite work cost upon failure. First the energy levels to the right of $Z_\epsilon$ are raised by an amount $V$. The dotted red line in Figure (a) illustrates the effect of the \longLT\ on the thermo-majorization diagram and the higher the energy levels are raised, the steeper this line will be. In Figure (b), the resultant curve has been $\beta$-ordered, and this segment now lies between $y=0$ and $y=\epsilon$. Next the system is fully thermalized, resulting in Figure (c). Finally, by applying \ITRs\ and a deterministic \LT, the system is transformed into the thermal state of the original Hamiltonian, Figure (d). Note that this protocol produces less work when it succeeds than that described in Figure \ref{fig:Abergs protocol}.} \label{fig:finite extract protocol}
\end{figure}

However, when this protocol fails, it fails spectacularly and costs an infinitely large amount of work. As discussed in \cite{aaberg2013truly}, this can be avoided if, rather than raising energy levels to infinity during the first \longLT, we instead raise them to a large but finite amount. However, this reduces the amount of work that is extracted when the protocol succeeds. The tradeoff between the cost of failure and the benefit of success is analyzed through the following protocol, illustrated in Figure \ref{fig:finite extract protocol}.

Let $V$ denote the amount that we are willing to raise the energy levels to the right of $Z_\epsilon$ by (as we are raising energy levels, this will cost work). The protocol then runs as follows:
\begin{enumerate}
\item Raise the unoccupied energy levels to the right of $Z_\epsilon$ to infinity using \longLTs.
\item Raise the occupied energy levels to the right of $Z_\epsilon$ by the amount $V$ using \longLTs.
\item Fully thermalize the entire system.
\item Perform \longITRs\ to horizontally align the points on the system's thermo-majorization curve with those of the target thermal state.
\item Perform a \longLT\ to transform the system into the thermal state of the initial Hamiltonian.
\end{enumerate}
Again, two stages of this protocol have non-zero work value. The \longLT\ that takes Figure \ref{fig:finite extract protocol}(a) to Figure \ref{fig:finite extract protocol}(b) has work value $-V$ with probability $\epsilon$ and zero with probability $1-\epsilon$. The second \longLT\ that takes Figure \ref{fig:finite extract protocol}(c) to Figure \ref{fig:finite extract protocol}(d), has deterministic work yield $\frac{1}{\beta}\ln\left(\frac{Z}{Z_\epsilon+e^{-\beta V}\left(\tilde{L}_{1}\left(\rho\right)-Z_\epsilon\right)}\right)$. Hence, with probability $1-\epsilon$ the protocol produces a work value of:
\begin{equation}
\frac{1}{\beta}\ln\left(\frac{Z}{Z_\epsilon+e^{-\beta V}\left(\tilde{L}_{1}\left(\rho\right)-Z_\epsilon\right)}\right),
\end{equation}
while with probability $\epsilon$ the work value is:
\begin{equation}
\frac{1}{\beta}\ln\left(\frac{Z}{Z_\epsilon+e^{\beta V}\left(\tilde{L}_{1}\left(\rho\right)-Z_\epsilon\right)}\right)-V.
\end{equation}
As $V$ tends to infinity, we recover Eq.~\eqref{eq:distill work} while when $V=0$, the amount of work extracted is equal to $W^{0}_{\textrm{distil}}\left(\rho\right)$.

\subsection{Work of formation} \label{ss:formwork}

By setting the initial state to be $\tau_S$, the thermal state of the system, in Eq. \eqref{eq:work def}, one can define the single-shot \emph{work of formation} - the amount of work required to form a state without coherences. Denoting the work of formation for block-diagonal $\rho$ by $W_{\textrm{form}}\left(\rho\right)$, in \cite{horodecki2013fundamental} it was shown to be given by:
\begin{align}
\begin{split} \label{eq:work form}
W_{\textrm{form}}\left(\rho\right)&=-F_{\textrm{max}}\left(\rho\right)-\frac{1}{\beta}\ln Z_S,\\
&=-\frac{1}{\beta}\left[\ln\left(\eta^{\left(\rho\right)}_1 e^{\beta E_1^{\left(\rho\right)}}\right)+\ln Z_S\right],
\end{split}
\end{align}
where the superscript $\left(\rho\right)$ again denotes that the occupation probabilities of $\rho$ and the associated energy levels of $H_S$ have been $\beta$-ordered.

\begin{figure}
\centering
\begin{minipage}{.5\columnwidth}
  \centering
  \includegraphics[width=0.9\textwidth]{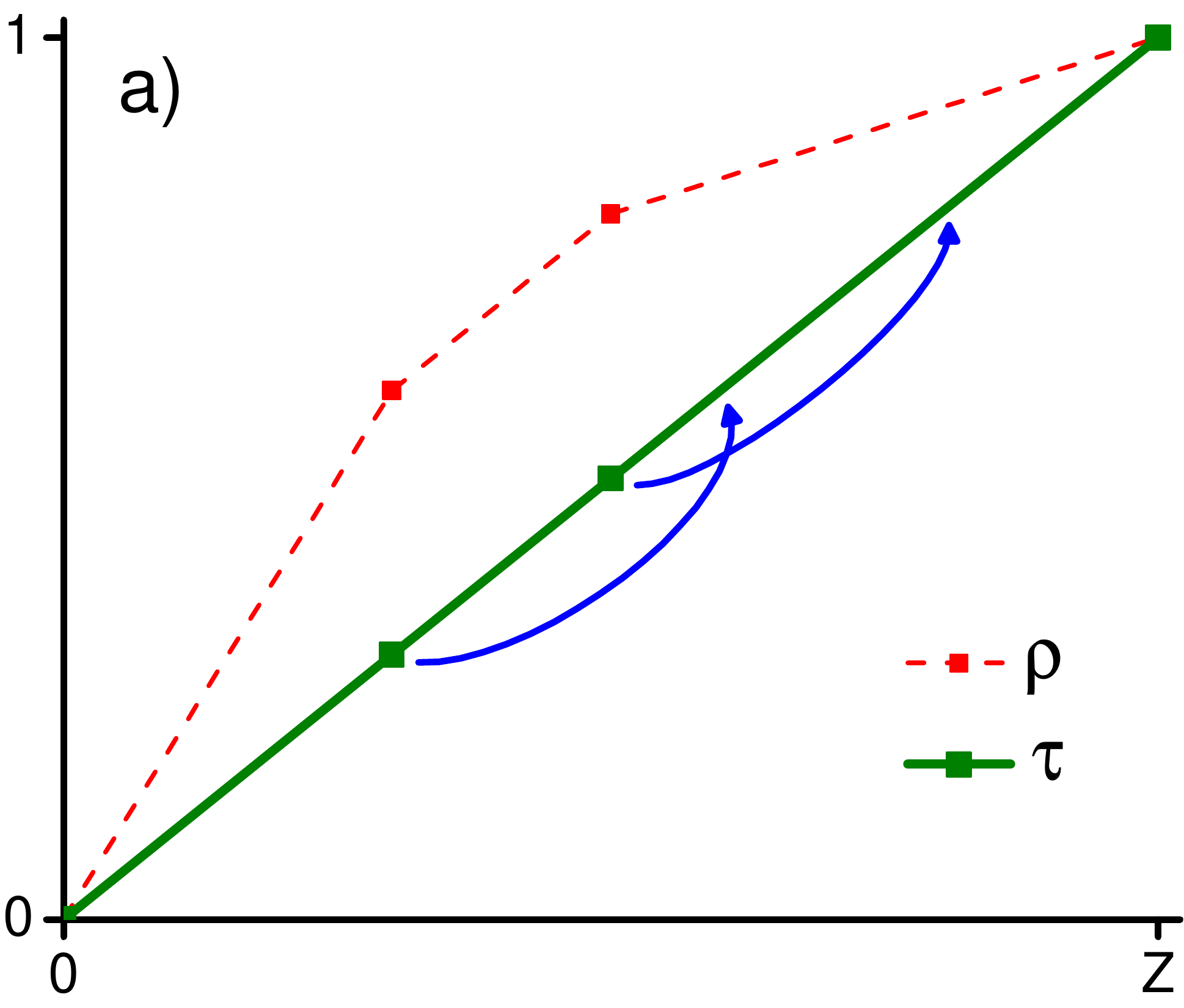}
  \label{fig:Form1}
	\centerline{(a) \ITRs}
\end{minipage}%
\begin{minipage}{.5\columnwidth}
  \centering
  \includegraphics[width=0.9\textwidth]{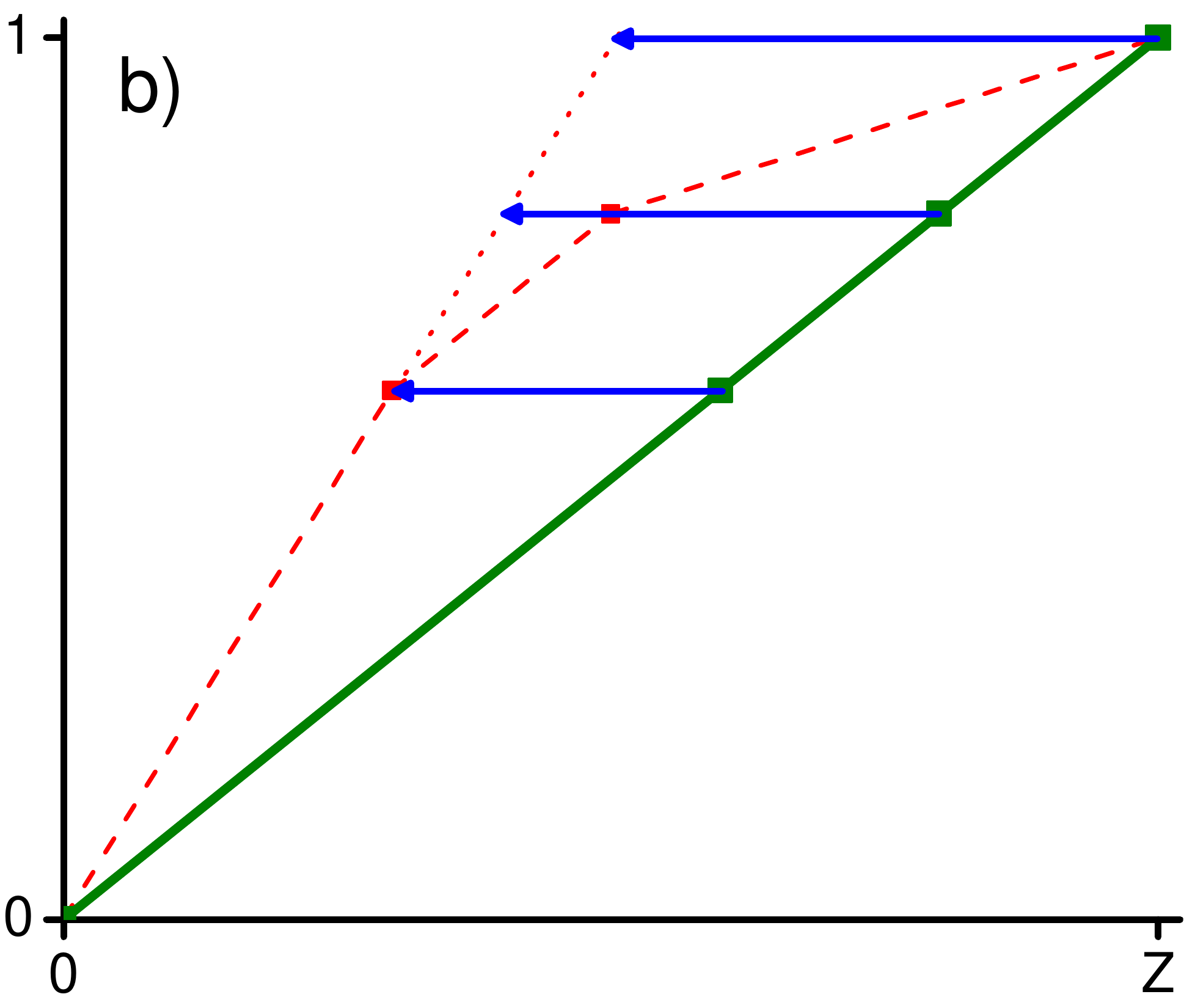}
  \label{fig:Form2}
	\centerline{(b) \LTs}
\end{minipage}\\
\vspace{.5cm}
\begin{minipage}{.5\columnwidth}
  \centering
  \includegraphics[width=0.9\textwidth]{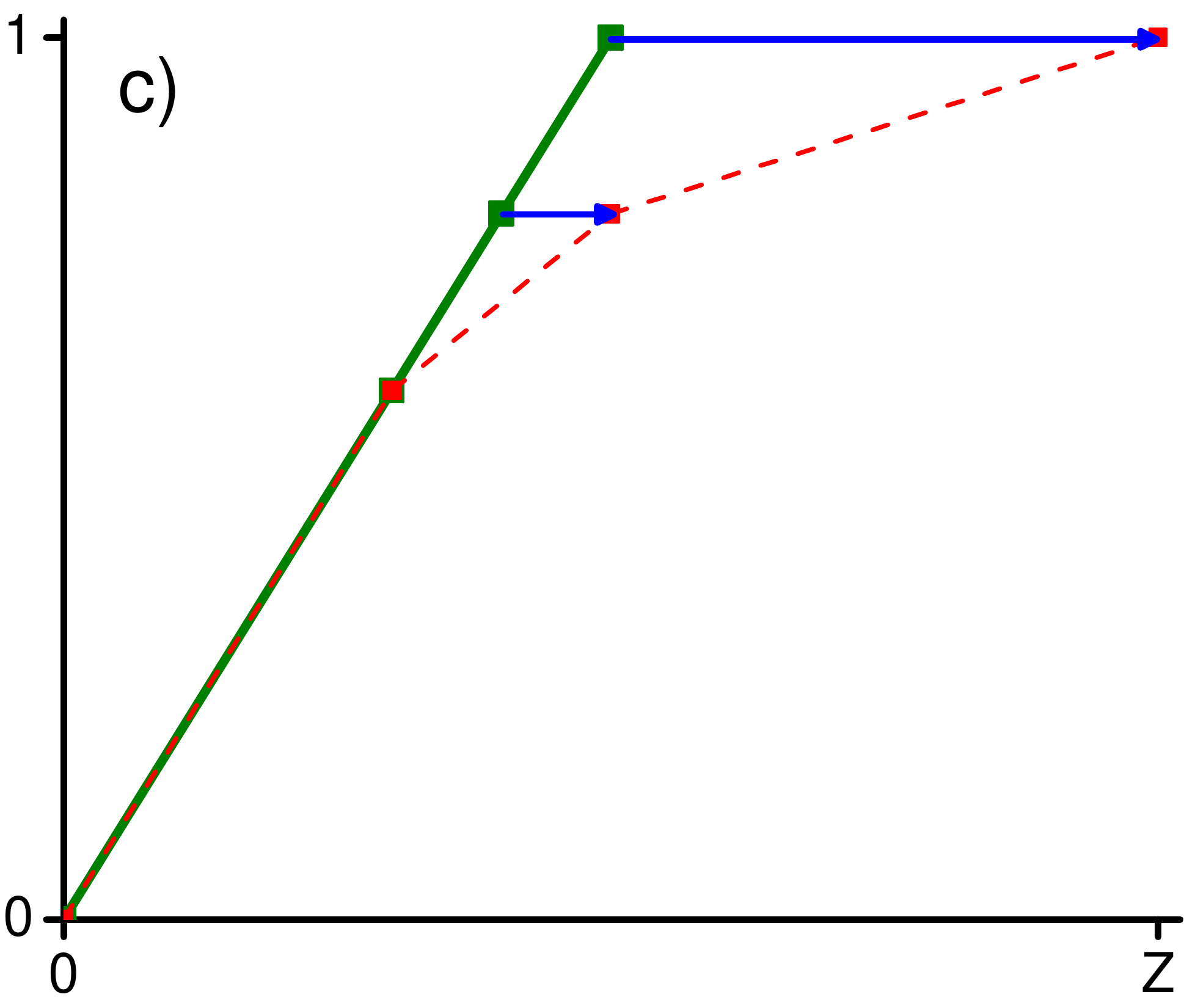}
  \label{fig:Form3}
	\centerline{(c) \LTs}
\end{minipage}%
\begin{minipage}{.5\columnwidth}
  \centering
  \includegraphics[width=.9\textwidth]{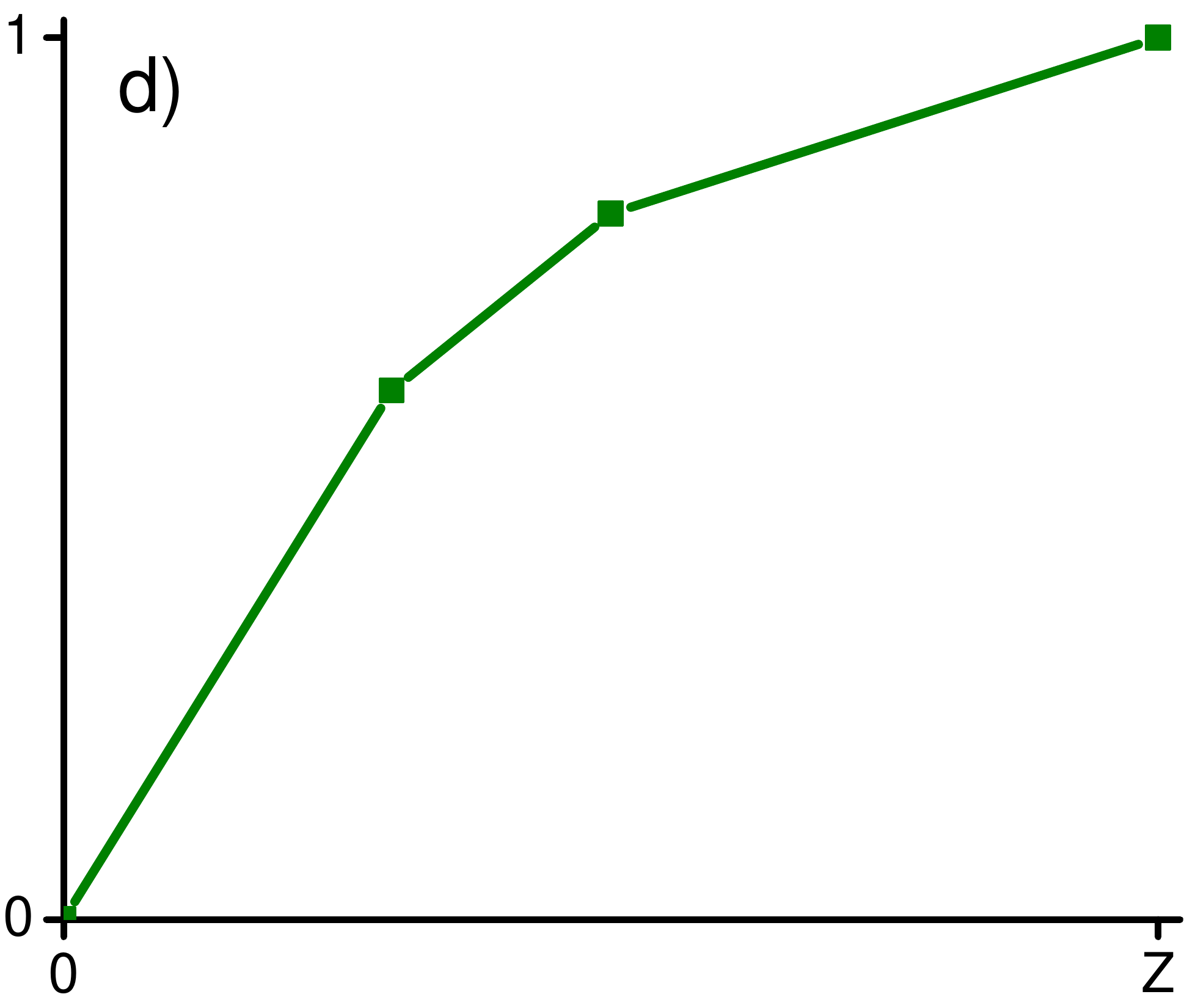}
  \label{fig:Form4}
	\centerline{(d) Final state, $\rho$}
\end{minipage}
\caption{\emph{\longCOs\ protocol for state formation.} Here we illustrate the protocol for state formation. First \ITRs\ are applied to horizontally align the points of the initial state with those of $\rho$. This results in Figure (b). Next a \LT\ is performed to raise the system's energy levels and produce a curve that just thermo-majorizes the curve of $\rho$ as shown in Figure (c). This step costs work. Finally, energy levels are lowered where necessary using a \LT\ to produce the final state, Figure (d). As energy levels are lowered, this final step does not cost work.} \label{fig:work of formation}
\end{figure}

The protocol to form $\rho$ is illustrated in Figure~\ref{fig:work of formation} and runs as follows:
\begin{enumerate}
\item \longITRs\ are performed to transform $(\tau_S,H_S)$ into $(\rho,H')$ where $H'$ is such that $Z'=Z_S$ and $\tau'=\rho$.
\item All energy levels are raised by the same amount using a \longLT\ to form a curve that just thermo-majorizes that of the target state.
\item A second \longLT\ is performed to lower energy levels so as to match the energy levels of $H_S$.
\end{enumerate}
As the first step consists of \longITRs, by Lemma \ref{le:ITR cost} the work value of this step can be taken to be zero. Let the first \longLT\ be parametrized by $\mathcal{E}=\left\{h_i\right\}_{i=1}^{n}$ and the second \longLT\ by $\mathcal{E}'=\left\{h'_i\right\}_{i=1}^{n}$. Here:
\begin{align}
h_i&=\frac{1}{\beta}\left[\ln\left(\eta^{\left(\rho\right)}_1 e^{\beta E_1^{\left(\rho\right)}}\right)+\ln Z_S\right], \quad \forall i,\\
h'_i&\leq0, \quad\forall i.
\end{align}
Hence, using Eq.~\eqref{eq:LT cost}:
\begin{align}
W_{\LT_{\mathcal{E}}}&=-\frac{1}{\beta}\left[\ln\left(\eta^{\left(\rho\right)}_1 e^{\beta E_1^{\left(\rho\right)}}\right)+\ln Z_S\right],\\
W_{\LT_{\mathcal{E}'}}&=0,
\end{align}
and the overall formation protocol has work cost given by Eq.~\eqref{eq:work form}.

\subsection{Work of transformation}

Given two states, $\rho$ and $\sigma$, recall from Section \ref{ssec:wits} that $\Wtran{\rho}{\sigma}$ denotes the deterministic work required (if $\Wtran{\rho}{\sigma}$ is negative)  or gained (if $\Wtran{\rho}{\sigma}$ is positive) in converting $\rho$ into $\sigma$ using thermal operations. This quantity can readily be obtained under \longCOs. If one first performs the \longLT, $\LT_{\mathcal{E}}$ with $\mathcal{E}=\left\{h_i=\Wtran{\rho}{\sigma}\right\}_{i=1}^{n}$ to $\left(\rho,H_S\right)$ by definition, one is left with a system that thermo-majorizes $\left(\sigma,H_S\right)$. This system can now be converted into $\left(\sigma,H_S\right)$ using the protocol given in Theorem \ref{th:crude protocol with infty} and the work cost of the \longLT\ used is $\Wtran{\rho}{\sigma}$.

\newpage

\section{\longCOs\ in the spirit of Szilard boxes} \label{sec:Szilard}

\begin{figure}[!htbp]
\centering
\begin{minipage}{.9\columnwidth}
  \centering
  \includegraphics[width=0.5\textwidth]{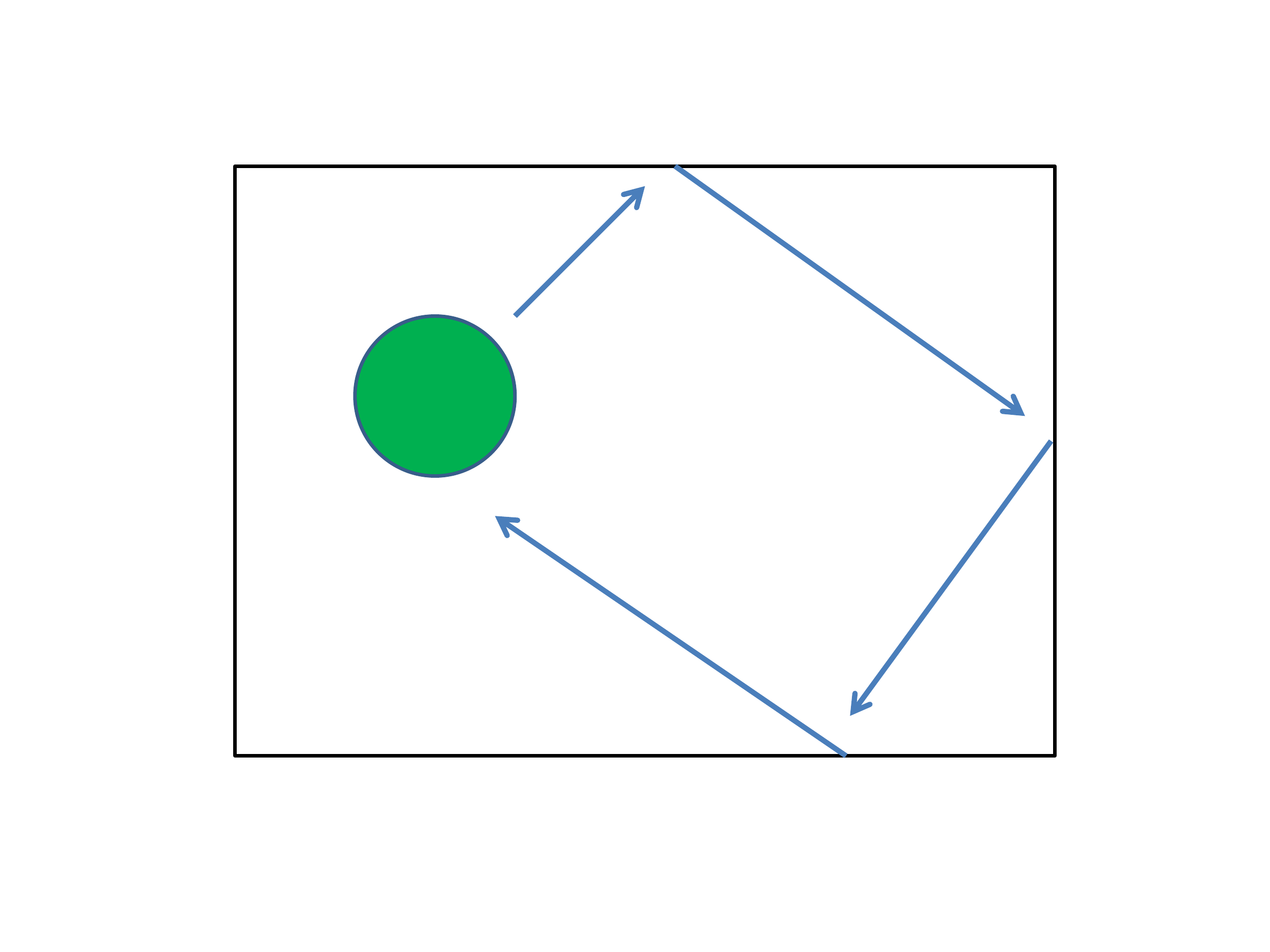}
	\caption{A molecule in a box which is in contact with a thermal reservoir at temperature $T$. When we have many boxes, some probability of finding molecule in each of them is assigned. A piston can be added to change the pressure/volume or a wall can be moved to perform state transitions. The box can be used to describe the Szilard engine and the Maxwell demon thought experiment, where a piston is added to the box so that it is always pushed out by the molecule.}
	\label{fig:box}
\end{minipage}%
\end{figure}

In this section we consider the case where the Hamiltonian is trivial, $H=0$, and we 
show how state-to-state transformations under Thermal Operation can be realized in the spirit of Szilard boxes with partitions (see, Figure \ref{fig:box}). 
This provides a illustrative physical realization of $T$-transforms, while allowing one to quantify the work extracted in any process. This may give some readers a physical
insight into previous sections.
We will consider an arbitrary transformation, as well as work distillation and the work of formation.

The setup is as follows: we consider $d$ boxes (representing a $d$-level system), where each of the boxes has some probability $p_i$ ($i \in \left\{1, \ldots, d\right\}$) of finding a molecule in it. This is a realization of a density matrix $\rho=\sum p_i \ketbra{i}{i}$ and is general, since $H=0$. We can change the parameters of the boxes like volume and pressure, by adding a piston and letting it move in either directions or by temporarily removing and changing the placing of the walls of the boxes. We also assume we are able to shuffle the boxes to arrange the corresponding probabilities at no work cost.

\noindent  
\subsection{Work of distillation}

\begin{figure}[!htbp]
\centering
\begin{minipage}{.9\columnwidth}
  \centering
  \includegraphics[width=0.5\textwidth]{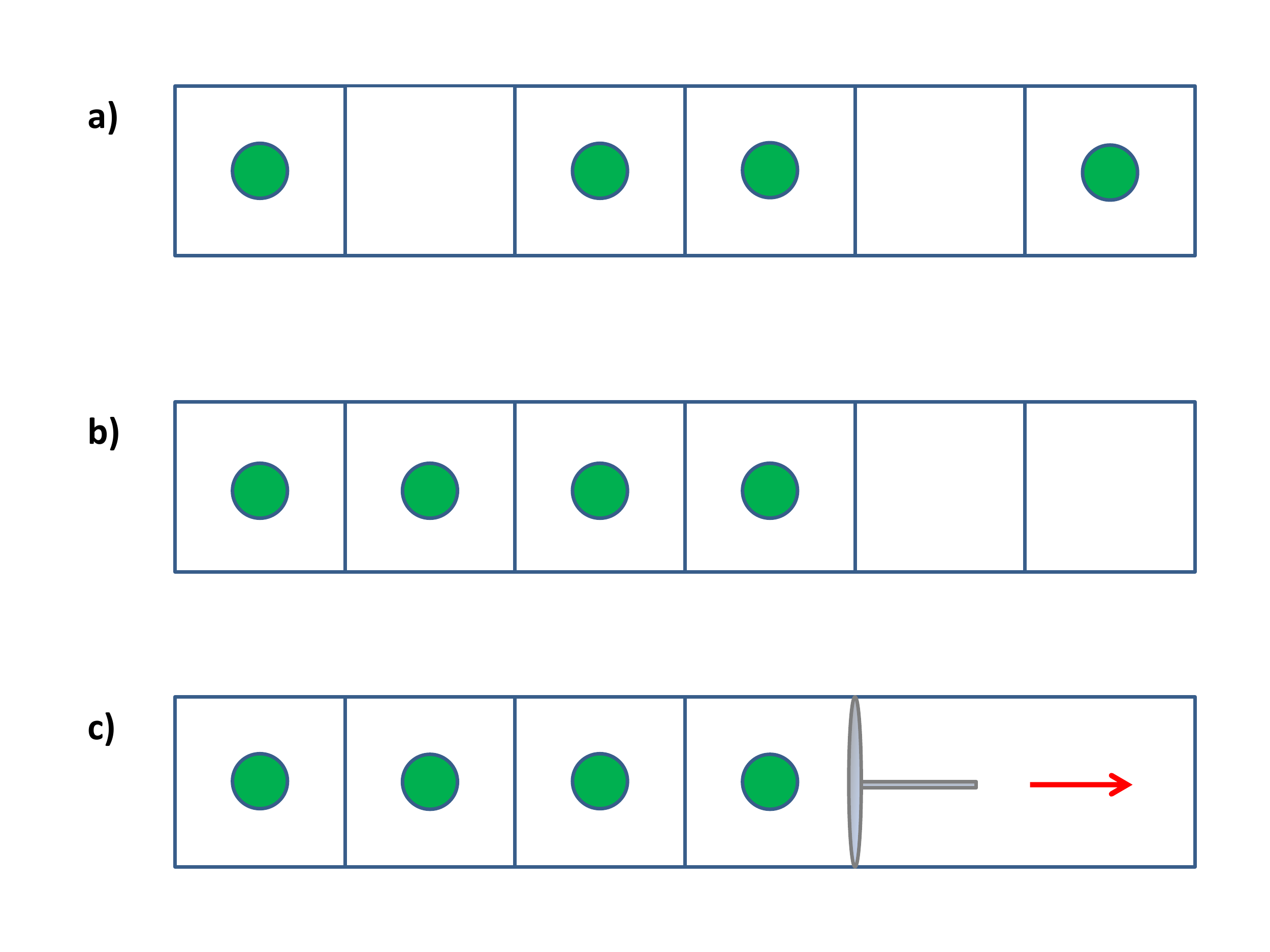}
	\caption{\emph{Distillation with Szilard boxes.} The probability of finding the molecule in one of the boxes marked with a green ball is at least $1-\epsilon$, which the empty boxes have a vanishingly small chance of being occupied. To distil work we first rearrange the boxes so that the unoccupied boxes are on the left, next we put in a piston and let it be pushed out to the right.}
	\label{fig:Dist}
\end{minipage}%
\end{figure}

To distill work with probability at least $1-\epsilon$ as in \cite{dahlsten2011inadequacy}, we first re-order of boxes, so they are arranged according to non-increasing probabilities of finding a molecule in it, $p_i$. Consider a situation when we have $d$ boxes, but most of the time, the molecule is found in one of 
$R$ of them (so that the probability of being in one of these $R$ boxes is at least $1-\epsilon$). The re-ordering ensures that the boxes which only have total probability at most $\epsilon$ of being occupied are on one side (let's imagine it's the right side). We then add a piston and let it be pushed out to the right thought the unoccupied boxes (for those $\sum p_j\leq\epsilon$). The work gain is due to the change of volume and is equal to $W = kT\operatorname{log\frac{V_i}{V_f}} = kT(\operatorname{log}d-\operatorname{log}R)$, where $kT$ is the product of the Boltzmann constant, $k$, and the temperature, $T$, and $V_i$, $V_f$, respectively the initial and final volume. In principle, the expression for work in proportion to the entropy change. We see, that it is non zero iff we have some "unoccupied" boxes. Note that this extracts work with almost certainty (probability greater than $1-\epsilon$). One can obtain probabilistic work by inserting the piston between the first and second box, then the second and third, etc. Since once the boxes are re-ordered, the largest probabilities of being occupied are always on the left, the piston will be more likely to be pushed to the right, than to the left (which represents the loss of one bit of work).

\noindent
\subsection{Work of formation}

\begin{figure}[!htbp]
\centering
\begin{minipage}{.9\columnwidth}
  \centering
  \includegraphics[width=.5\textwidth]{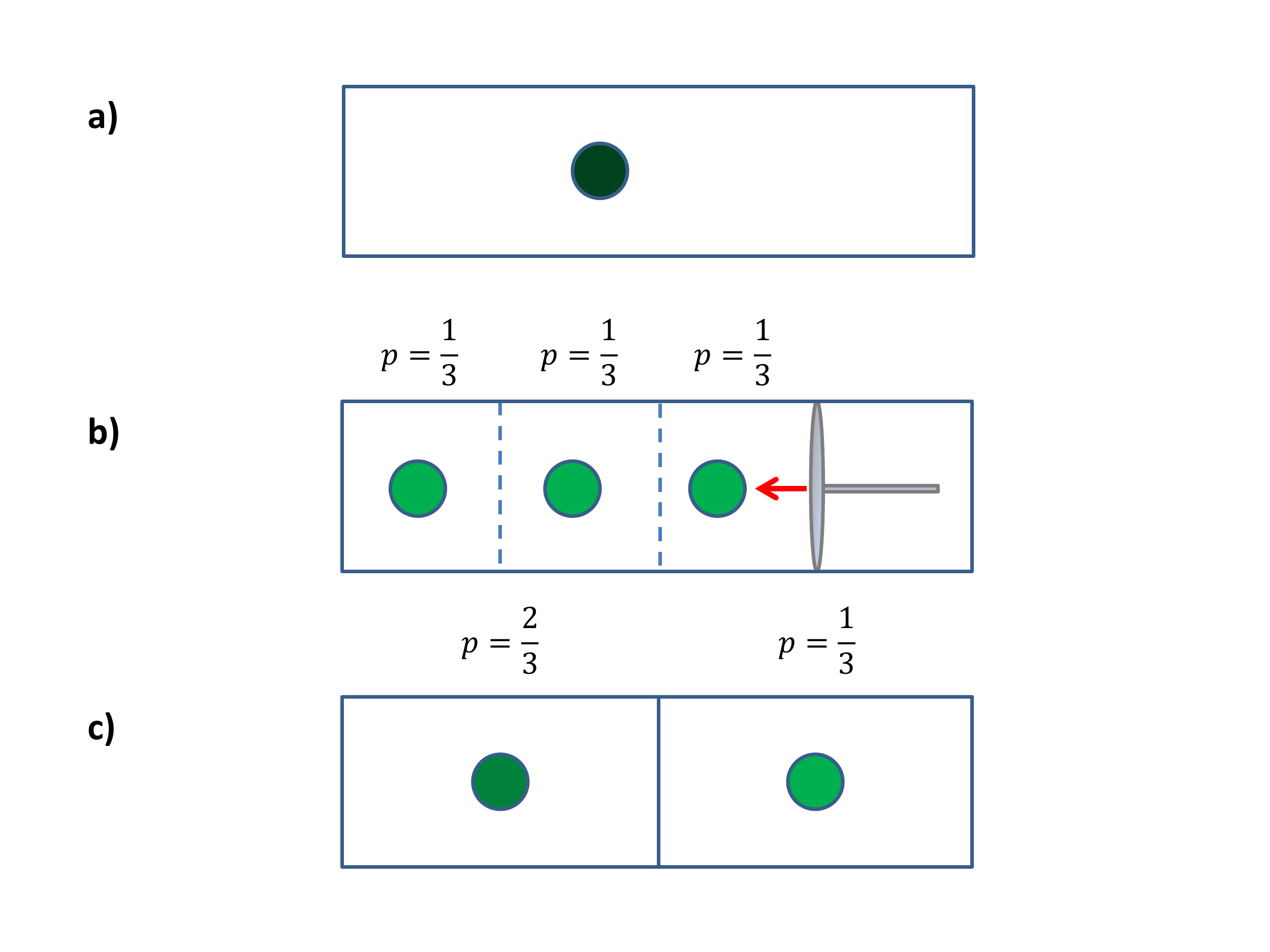}
	\caption{\emph{State formation with Szilard boxes.} In this example, we want to create the state which has probability $2/3$ of the molecule being on the left hand side, and $1/3$ of it being on the right, from a state where the molecule is uniformly distributed (the maximally mixed state). To do so, we push a piston in from the right, so that the volume available in the right hand partition is $1/3$ that of the left hand partition. We then insert a partition into the center and let the piston be pushed back out (which recovers the initial spent work with probability $1/3$.}
	\label{fig:Form}
\end{minipage}
\end{figure}

In Figure \ref{fig:Form} we show how to form a state from the maximally mixed state with $d=2$. The larger the largest eigenvalue $\lambda$, the more the piston has to be pushed in, and it's easy to see that the work required is $W = kT\operatorname{log\frac{V_i}{V_f}} = -kT\operatorname{log}\lambda$. This holds in higher dimension, since once the piston has been pushed in to create the largest eigenvalue, the majorization condition ensures that it can just be pushed back out, extracting work probabilistically in each step.  

Note that we have irreversibility here. If we have no vacancies, i.e. unoccupied boxes, then the work of distillation is equal to 0, while the work of formation can be large, which means that we need to put energy in the system yet cannot get any from it.

\noindent
\subsection{Arbitrary transformations}

An arbitrary transformation can be carried out much in the way of the cascade of T-transforms depicted in Figure \ref{fig:T-transforms}. However, the method of moving probability mass is to insert a piston in place of the partition between two boxes, let the piston be moved out, and then replace the partitions. Namely, we once again
 arrange the boxes in non-increasing order, i.e.
$p_1 \geq p_2 \geq p_3 \geq \ldots \geq p_d$. We want to transform them into boxes with probabilities $q_i$ that are also in non-increasing order $q_1 \geq q_2 \geq q_3 \geq \ldots \geq q_d$, assuming that p majorizes q. Then, initially, we have $p_1 > q_1$. To adjust this probability we take the box with probability $p_1$, and replace it's right partition wall with a piston. We then let the piston be pushed into the direction of the box with $p_2$ lowering the probability from $p_1$ to $q_1$ (if possible). The probability in the second box is now equal to $p_1+p_2-q_1 > q_2$, which can be written as $p_1+p_2 > q_1 + q_2$, but it is already the majorization condition. We can repeat this process to arrange all boxes to transform the initial system into the final one.


%% file: crude.0.7_pc.bbl
\begin{thebibliography}{47}%
\makeatletter
\providecommand \@ifxundefined [1]{%
 \@ifx{#1\undefined}
}%
\providecommand \@ifnum [1]{%
 \ifnum #1\expandafter \@firstoftwo
 \else \expandafter \@secondoftwo
 \fi
}%
\providecommand \@ifx [1]{%
 \ifx #1\expandafter \@firstoftwo
 \else \expandafter \@secondoftwo
 \fi
}%
\providecommand \natexlab [1]{#1}%
\providecommand \enquote  [1]{``#1''}%
\providecommand \bibnamefont  [1]{#1}%
\providecommand \bibfnamefont [1]{#1}%
\providecommand \citenamefont [1]{#1}%
\providecommand \href@noop [0]{\@secondoftwo}%
\providecommand \href [0]{\begingroup \@sanitize@url \@href}%
\providecommand \@href[1]{\@@startlink{#1}\@@href}%
\providecommand \@@href[1]{\endgroup#1\@@endlink}%
\providecommand \@sanitize@url [0]{\catcode `\\12\catcode `\$12\catcode
  `\&12\catcode `\#12\catcode `\^12\catcode `\_12\catcode `\%12\relax}%
\providecommand \@@startlink[1]{}%
\providecommand \@@endlink[0]{}%
\providecommand \url  [0]{\begingroup\@sanitize@url \@url }%
\providecommand \@url [1]{\endgroup\@href {#1}{\urlprefix }}%
\providecommand \urlprefix  [0]{URL }%
\providecommand \Eprint [0]{\href }%
\providecommand \doibase [0]{http://dx.doi.org/}%
\providecommand \selectlanguage [0]{\@gobble}%
\providecommand \bibinfo  [0]{\@secondoftwo}%
\providecommand \bibfield  [0]{\@secondoftwo}%
\providecommand \translation [1]{[#1]}%
\providecommand \BibitemOpen [0]{}%
\providecommand \bibitemStop [0]{}%
\providecommand \bibitemNoStop [0]{.\EOS\space}%
\providecommand \EOS [0]{\spacefactor3000\relax}%
\providecommand \BibitemShut  [1]{\csname bibitem#1\endcsname}%
\let\auto@bib@innerbib\@empty
\bibitem [{\citenamefont {Scovil}\ and\ \citenamefont
  {Schulz-DuBois}(1959)}]{Scovil1959masers}%
  \BibitemOpen
  \bibfield  {author} {\bibinfo {author} {\bibfnamefont {H.~E.~D.}\
  \bibnamefont {Scovil}}\ and\ \bibinfo {author} {\bibfnamefont {E.~O.}\
  \bibnamefont {Schulz-DuBois}},\ }\href {\doibase 10.1103/PhysRevLett.2.262}
  {\bibfield  {journal} {\bibinfo  {journal} {Phys. Rev. Lett.}\ }\textbf
  {\bibinfo {volume} {2}},\ \bibinfo {pages} {262} (\bibinfo {year}
  {1959})}\BibitemShut {NoStop}%
\bibitem [{\citenamefont {Scully}(2002)}]{scully2002afterburner}%
  \BibitemOpen
  \bibfield  {author} {\bibinfo {author} {\bibfnamefont {M.~O.}\ \bibnamefont
  {Scully}},\ }\href {\doibase 10.1103/PhysRevLett.88.050602} {\bibfield
  {journal} {\bibinfo  {journal} {Phys. Rev. Lett.}\ }\textbf {\bibinfo
  {volume} {88}},\ \bibinfo {pages} {050602} (\bibinfo {year}
  {2002})}\BibitemShut {NoStop}%
\bibitem [{\citenamefont {Rousselet}\ \emph {et~al.}(1994)\citenamefont
  {Rousselet}, \citenamefont {Salome}, \citenamefont {Ajdari},\ and\
  \citenamefont {Prost}}]{rousselet1994directional}%
  \BibitemOpen
  \bibfield  {author} {\bibinfo {author} {\bibfnamefont {J.}~\bibnamefont
  {Rousselet}}, \bibinfo {author} {\bibfnamefont {L.}~\bibnamefont {Salome}},
  \bibinfo {author} {\bibfnamefont {A.}~\bibnamefont {Ajdari}}, \ and\ \bibinfo
  {author} {\bibfnamefont {J.}~\bibnamefont {Prost}},\ }\href@noop {}
  {\bibfield  {journal} {\bibinfo  {journal} {Nature}\ }\textbf {\bibinfo
  {volume} {370}},\ \bibinfo {pages} {446} (\bibinfo {year}
  {1994})}\BibitemShut {NoStop}%
\bibitem [{\citenamefont {Faucheux}\ \emph {et~al.}(1995)\citenamefont
  {Faucheux}, \citenamefont {Bourdieu}, \citenamefont {Kaplan},\ and\
  \citenamefont {Libchaber}}]{Faucheux1995ratchet}%
  \BibitemOpen
  \bibfield  {author} {\bibinfo {author} {\bibfnamefont {L.~P.}\ \bibnamefont
  {Faucheux}}, \bibinfo {author} {\bibfnamefont {L.~S.}\ \bibnamefont
  {Bourdieu}}, \bibinfo {author} {\bibfnamefont {P.~D.}\ \bibnamefont
  {Kaplan}}, \ and\ \bibinfo {author} {\bibfnamefont {A.~J.}\ \bibnamefont
  {Libchaber}},\ }\href {\doibase 10.1103/PhysRevLett.74.1504} {\bibfield
  {journal} {\bibinfo  {journal} {Phys. Rev. Lett.}\ }\textbf {\bibinfo
  {volume} {74}},\ \bibinfo {pages} {1504} (\bibinfo {year}
  {1995})}\BibitemShut {NoStop}%
\bibitem [{\citenamefont {Baugh}\ \emph {et~al.}(2005)\citenamefont {Baugh},
  \citenamefont {Moussa}, \citenamefont {Ryan}, \citenamefont {Nayak},\ and\
  \citenamefont {Laflamme}}]{baugh2005experimental}%
  \BibitemOpen
  \bibfield  {author} {\bibinfo {author} {\bibfnamefont {J.}~\bibnamefont
  {Baugh}}, \bibinfo {author} {\bibfnamefont {O.}~\bibnamefont {Moussa}},
  \bibinfo {author} {\bibfnamefont {C.}~\bibnamefont {Ryan}}, \bibinfo {author}
  {\bibfnamefont {A.}~\bibnamefont {Nayak}}, \ and\ \bibinfo {author}
  {\bibfnamefont {R.}~\bibnamefont {Laflamme}},\ }\href@noop {} {\bibfield
  {journal} {\bibinfo  {journal} {Nature}\ }\textbf {\bibinfo {volume} {438}},\
  \bibinfo {pages} {470} (\bibinfo {year} {2005})}\BibitemShut {NoStop}%
\bibitem [{\citenamefont {Ruch}\ and\ \citenamefont
  {Mead}(1976)}]{ruch1976principle}%
  \BibitemOpen
  \bibfield  {author} {\bibinfo {author} {\bibfnamefont {E.}~\bibnamefont
  {Ruch}}\ and\ \bibinfo {author} {\bibfnamefont {A.}~\bibnamefont {Mead}},\
  }\href@noop {} {\bibfield  {journal} {\bibinfo  {journal} {Theoretical
  Chemistry Accounts: Theory, Computation, and Modeling (Theoretica Chimica
  Acta)}\ }\textbf {\bibinfo {volume} {41}},\ \bibinfo {pages} {95} (\bibinfo
  {year} {1976})}\BibitemShut {NoStop}%
\bibitem [{\citenamefont {Horodecki}\ and\ \citenamefont
  {Oppenheim}(2013)}]{horodecki2013fundamental}%
  \BibitemOpen
  \bibfield  {author} {\bibinfo {author} {\bibfnamefont {M.}~\bibnamefont
  {Horodecki}}\ and\ \bibinfo {author} {\bibfnamefont {J.}~\bibnamefont
  {Oppenheim}},\ }\href@noop {} {\bibfield  {journal} {\bibinfo  {journal}
  {Nature communications}\ }\textbf {\bibinfo {volume} {4}} (\bibinfo {year}
  {2013})}\BibitemShut {NoStop}%
\bibitem [{\citenamefont {{Brandao}}\ \emph {et~al.}(2015)\citenamefont
  {{Brandao}}, \citenamefont {{Horodecki}}, \citenamefont {{Ng}}, \citenamefont
  {{Oppenheim}},\ and\ \citenamefont {{Wehner}}}]{brandao2013second}%
  \BibitemOpen
  \bibfield  {author} {\bibinfo {author} {\bibfnamefont {F.~G.~S.~L.}\
  \bibnamefont {{Brandao}}}, \bibinfo {author} {\bibfnamefont {M.}~\bibnamefont
  {{Horodecki}}}, \bibinfo {author} {\bibfnamefont {N.~H.~Y.}\ \bibnamefont
  {{Ng}}}, \bibinfo {author} {\bibfnamefont {J.}~\bibnamefont {{Oppenheim}}}, \
  and\ \bibinfo {author} {\bibfnamefont {S.}~\bibnamefont {{Wehner}}},\
  }\href@noop {} {\bibfield  {journal} {\bibinfo  {journal} {Proc. Natl. Acad.
  Sci.}\ }\textbf {\bibinfo {volume} {112}},\ \bibinfo {pages} {3275} (\bibinfo
  {year} {2015})}\BibitemShut {NoStop}%
\bibitem [{Note1()}]{Note1}%
  \BibitemOpen
  \bibinfo {note} {Note that these generalized free energy constraints can be
  obtained from the thermo-majorization criteria provided one is allowed to use
  an ancillary system which is returned in its initial state at the end of
  protocol. As such, we can restrict ourselves to considering the
  thermo-majorization criteria here.}\BibitemShut {Stop}%
\bibitem [{\citenamefont {Wilming}\ \emph {et~al.}(2014)\citenamefont
  {Wilming}, \citenamefont {Gallego},\ and\ \citenamefont
  {Eisert}}]{wilming2014weak}%
  \BibitemOpen
  \bibfield  {author} {\bibinfo {author} {\bibfnamefont {H.}~\bibnamefont
  {Wilming}}, \bibinfo {author} {\bibfnamefont {R.}~\bibnamefont {Gallego}}, \
  and\ \bibinfo {author} {\bibfnamefont {J.}~\bibnamefont {Eisert}},\
  }\href@noop {} {\bibfield  {journal} {\bibinfo  {journal} {arXiv preprint
  arXiv:1411.3754}\ } (\bibinfo {year} {2014})}\BibitemShut {NoStop}%
\bibitem [{\citenamefont {Janzing}\ \emph {et~al.}(2000)\citenamefont
  {Janzing}, \citenamefont {Wocjan}, \citenamefont {Zeier}, \citenamefont
  {Geiss},\ and\ \citenamefont {Beth}}]{janzing2000thermodynamic}%
  \BibitemOpen
  \bibfield  {author} {\bibinfo {author} {\bibfnamefont {D.}~\bibnamefont
  {Janzing}}, \bibinfo {author} {\bibfnamefont {P.}~\bibnamefont {Wocjan}},
  \bibinfo {author} {\bibfnamefont {R.}~\bibnamefont {Zeier}}, \bibinfo
  {author} {\bibfnamefont {R.}~\bibnamefont {Geiss}}, \ and\ \bibinfo {author}
  {\bibfnamefont {T.}~\bibnamefont {Beth}},\ }\href@noop {} {\bibfield
  {journal} {\bibinfo  {journal} {International Journal of Theoretical
  Physics}\ }\textbf {\bibinfo {volume} {39}},\ \bibinfo {pages} {2717}
  (\bibinfo {year} {2000})}\BibitemShut {NoStop}%
\bibitem [{\citenamefont {Streater}(1995)}]{Streater_dynamics}%
  \BibitemOpen
  \bibfield  {author} {\bibinfo {author} {\bibfnamefont {R.~F.}\ \bibnamefont
  {Streater}},\ }\href@noop {} {\emph {\bibinfo {title} {Statistical Dynamics:
  A Stochastic Approach to nonequilibrium Thermodynamics}}}\ (\bibinfo
  {publisher} {Imperial College Press},\ \bibinfo {address} {London, UK},\
  \bibinfo {year} {1995})\BibitemShut {NoStop}%
\bibitem [{\citenamefont {Brand\~ao}\ \emph {et~al.}(2013)\citenamefont
  {Brand\~ao}, \citenamefont {Horodecki}, \citenamefont {Oppenheim},
  \citenamefont {Renes},\ and\ \citenamefont {Spekkens}}]{brandao2011resource}%
  \BibitemOpen
  \bibfield  {author} {\bibinfo {author} {\bibfnamefont {F.~G. S.~L.}\
  \bibnamefont {Brand\~ao}}, \bibinfo {author} {\bibfnamefont {M.}~\bibnamefont
  {Horodecki}}, \bibinfo {author} {\bibfnamefont {J.}~\bibnamefont
  {Oppenheim}}, \bibinfo {author} {\bibfnamefont {J.~M.}\ \bibnamefont
  {Renes}}, \ and\ \bibinfo {author} {\bibfnamefont {R.~W.}\ \bibnamefont
  {Spekkens}},\ }\href {\doibase 10.1103/PhysRevLett.111.250404} {\bibfield
  {journal} {\bibinfo  {journal} {Phys. Rev. Lett.}\ }\textbf {\bibinfo
  {volume} {111}},\ \bibinfo {pages} {250404} (\bibinfo {year}
  {2013})}\BibitemShut {NoStop}%
\bibitem [{\citenamefont {\AA{}berg}(2014)}]{aberg2014catalytic}%
  \BibitemOpen
  \bibfield  {author} {\bibinfo {author} {\bibfnamefont {J.}~\bibnamefont
  {\AA{}berg}},\ }\href@noop {} {\bibfield  {journal} {\bibinfo  {journal}
  {Phys. Rev. Lett.}\ }\textbf {\bibinfo {volume} {113}},\ \bibinfo {pages}
  {150402} (\bibinfo {year} {2014})}\BibitemShut {NoStop}%
\bibitem [{\citenamefont {{Lostaglio}}\ \emph {et~al.}(2015)\citenamefont
  {{Lostaglio}}, \citenamefont {{Korzekwa}}, \citenamefont {{Jennings}},\ and\
  \citenamefont {{Rudolph}}}]{korzekwa2015extraction}%
  \BibitemOpen
  \bibfield  {author} {\bibinfo {author} {\bibfnamefont {M.}~\bibnamefont
  {{Lostaglio}}}, \bibinfo {author} {\bibfnamefont {K.}~\bibnamefont
  {{Korzekwa}}}, \bibinfo {author} {\bibfnamefont {D.}~\bibnamefont
  {{Jennings}}}, \ and\ \bibinfo {author} {\bibfnamefont {T.}~\bibnamefont
  {{Rudolph}}},\ }\href@noop {} {\bibfield  {journal} {\bibinfo  {journal}
  {Phys. Rev. X}\ }\textbf {\bibinfo {volume} {5}},\ \bibinfo {pages} {021001}
  (\bibinfo {year} {2015})}\BibitemShut {NoStop}%
\bibitem [{\citenamefont {{\'C}wikli{\'n}ski}\ \emph
  {et~al.}(2015)\citenamefont {{\'C}wikli{\'n}ski}, \citenamefont
  {Studzi{\'n}ski}, \citenamefont {Horodecki},\ and\ \citenamefont
  {Oppenheim}}]{cwiklinski2014limitations}%
  \BibitemOpen
  \bibfield  {author} {\bibinfo {author} {\bibfnamefont {P.}~\bibnamefont
  {{\'C}wikli{\'n}ski}}, \bibinfo {author} {\bibfnamefont {M.}~\bibnamefont
  {Studzi{\'n}ski}}, \bibinfo {author} {\bibfnamefont {M.}~\bibnamefont
  {Horodecki}}, \ and\ \bibinfo {author} {\bibfnamefont {J.}~\bibnamefont
  {Oppenheim}},\ }\href@noop {} {\bibfield  {journal} {\bibinfo  {journal}
  {Phys. Rev. Lett.}\ }\textbf {\bibinfo {volume} {115}},\ \bibinfo {pages}
  {210403} (\bibinfo {year} {2015})}\BibitemShut {NoStop}%
\bibitem [{\citenamefont {{\AA}berg}(2013)}]{aaberg2013truly}%
  \BibitemOpen
  \bibfield  {author} {\bibinfo {author} {\bibfnamefont {J.}~\bibnamefont
  {{\AA}berg}},\ }\href@noop {} {\bibfield  {journal} {\bibinfo  {journal}
  {Nature communications}\ }\textbf {\bibinfo {volume} {4}} (\bibinfo {year}
  {2013})}\BibitemShut {NoStop}%
\bibitem [{\citenamefont {Egloff}\ \emph {et~al.}(2015)\citenamefont {Egloff},
  \citenamefont {Dahlsten}, \citenamefont {Renner},\ and\ \citenamefont
  {Vedral}}]{egloff2015measure}%
  \BibitemOpen
  \bibfield  {author} {\bibinfo {author} {\bibfnamefont {D.}~\bibnamefont
  {Egloff}}, \bibinfo {author} {\bibfnamefont {O.}~\bibnamefont {Dahlsten}},
  \bibinfo {author} {\bibfnamefont {R.}~\bibnamefont {Renner}}, \ and\ \bibinfo
  {author} {\bibfnamefont {V.}~\bibnamefont {Vedral}},\ }\href@noop {}
  {\bibfield  {journal} {\bibinfo  {journal} {New Journal of Physics}\ }\textbf
  {\bibinfo {volume} {17}},\ \bibinfo {pages} {073001} (\bibinfo {year}
  {2015})}\BibitemShut {NoStop}%
\bibitem [{\citenamefont {Renes}(2014)}]{renes2014work}%
  \BibitemOpen
  \bibfield  {author} {\bibinfo {author} {\bibfnamefont {J.~M.}\ \bibnamefont
  {Renes}},\ }\href@noop {} {\bibfield  {journal} {\bibinfo  {journal} {The
  European Physical Journal Plus}\ }\textbf {\bibinfo {volume} {129}},\
  \bibinfo {pages} {1} (\bibinfo {year} {2014})}\BibitemShut {NoStop}%
\bibitem [{\citenamefont {{Renes}}(2015)}]{renes2015work}%
  \BibitemOpen
  \bibfield  {author} {\bibinfo {author} {\bibfnamefont {J.~M.}\ \bibnamefont
  {{Renes}}},\ }\href@noop {} {\bibfield  {journal} {\bibinfo  {journal} {ArXiv
  e-prints}\ } (\bibinfo {year} {2015})},\ \Eprint
  {http://arxiv.org/abs/1510.03695} {arXiv:1510.03695} \BibitemShut {NoStop}%
\bibitem [{\citenamefont {Skrzypczyk}\ \emph {et~al.}(2014)\citenamefont
  {Skrzypczyk}, \citenamefont {Short},\ and\ \citenamefont
  {Popescu}}]{Skrzypczyk13}%
  \BibitemOpen
  \bibfield  {author} {\bibinfo {author} {\bibfnamefont {P.}~\bibnamefont
  {Skrzypczyk}}, \bibinfo {author} {\bibfnamefont {A.~J.}\ \bibnamefont
  {Short}}, \ and\ \bibinfo {author} {\bibfnamefont {S.}~\bibnamefont
  {Popescu}},\ }\href@noop {} {\bibfield  {journal} {\bibinfo  {journal} {Nat.
  Commun.}\ }\textbf {\bibinfo {volume} {5}},\ \bibinfo {pages} {4185}
  (\bibinfo {year} {2014})}\BibitemShut {NoStop}%
\bibitem [{\citenamefont {Horodecki}\ \emph {et~al.}(2003)\citenamefont
  {Horodecki}, \citenamefont {Horodecki},\ and\ \citenamefont
  {Oppenheim}}]{uniqueinfo}%
  \BibitemOpen
  \bibfield  {author} {\bibinfo {author} {\bibfnamefont {M.}~\bibnamefont
  {Horodecki}}, \bibinfo {author} {\bibfnamefont {P.}~\bibnamefont
  {Horodecki}}, \ and\ \bibinfo {author} {\bibfnamefont {J.}~\bibnamefont
  {Oppenheim}},\ }\href@noop {} {\bibfield  {journal} {\bibinfo  {journal}
  {Phys. Rev. A}\ }\textbf {\bibinfo {volume} {67}},\ \bibinfo {pages} {062104}
  (\bibinfo {year} {2003})},\ \Eprint {http://arxiv.org/abs/quant-ph/0212019}
  {quant-ph/0212019} \BibitemShut {NoStop}%
\bibitem [{\citenamefont {Muirhead}(1902)}]{muirhead1902some}%
  \BibitemOpen
  \bibfield  {author} {\bibinfo {author} {\bibfnamefont {R.~F.}\ \bibnamefont
  {Muirhead}},\ }\href@noop {} {\bibfield  {journal} {\bibinfo  {journal}
  {Proceedings of the Edinburgh Mathematical Society}\ }\textbf {\bibinfo
  {volume} {21}},\ \bibinfo {pages} {144} (\bibinfo {year} {1902})}\BibitemShut
  {NoStop}%
\bibitem [{\citenamefont {Hardy}\ \emph {et~al.}(1952)\citenamefont {Hardy},
  \citenamefont {Littlewood},\ and\ \citenamefont
  {P{\'o}lya}}]{hardy1952inequalities}%
  \BibitemOpen
  \bibfield  {author} {\bibinfo {author} {\bibfnamefont {G.~H.}\ \bibnamefont
  {Hardy}}, \bibinfo {author} {\bibfnamefont {J.~E.}\ \bibnamefont
  {Littlewood}}, \ and\ \bibinfo {author} {\bibfnamefont {G.}~\bibnamefont
  {P{\'o}lya}},\ }\href@noop {} {\emph {\bibinfo {title} {Inequalities}}}\
  (\bibinfo  {publisher} {Cambridge university press},\ \bibinfo {year}
  {1952})\BibitemShut {NoStop}%
\bibitem [{\citenamefont {{Gour}}\ \emph {et~al.}(2015)\citenamefont {{Gour}},
  \citenamefont {{M{\"u}ller}}, \citenamefont {{Narasimhachar}}, \citenamefont
  {{Spekkens}},\ and\ \citenamefont {{Yunger Halpern}}}]{gour2013resource}%
  \BibitemOpen
  \bibfield  {author} {\bibinfo {author} {\bibfnamefont {G.}~\bibnamefont
  {{Gour}}}, \bibinfo {author} {\bibfnamefont {M.~P.}\ \bibnamefont
  {{M{\"u}ller}}}, \bibinfo {author} {\bibfnamefont {V.}~\bibnamefont
  {{Narasimhachar}}}, \bibinfo {author} {\bibfnamefont {R.~W.}\ \bibnamefont
  {{Spekkens}}}, \ and\ \bibinfo {author} {\bibfnamefont {N.}~\bibnamefont
  {{Yunger Halpern}}},\ }\href@noop {} {\bibfield  {journal} {\bibinfo
  {journal} {Phys. Rep.}\ }\textbf {\bibinfo {volume} {583}},\ \bibinfo {pages}
  {1} (\bibinfo {year} {2015})}\BibitemShut {NoStop}%
\bibitem [{\citenamefont {Geusic}\ \emph {et~al.}(1967)\citenamefont {Geusic},
  \citenamefont {Schulz-DuBios},\ and\ \citenamefont
  {Scovil}}]{Geusic1967quatum}%
  \BibitemOpen
  \bibfield  {author} {\bibinfo {author} {\bibfnamefont {J.~E.}\ \bibnamefont
  {Geusic}}, \bibinfo {author} {\bibfnamefont {E.~O.}\ \bibnamefont
  {Schulz-DuBios}}, \ and\ \bibinfo {author} {\bibfnamefont {H.~E.~D.}\
  \bibnamefont {Scovil}},\ }\href {\doibase 10.1103/PhysRev.156.343} {\bibfield
   {journal} {\bibinfo  {journal} {Phys. Rev.}\ }\textbf {\bibinfo {volume}
  {156}},\ \bibinfo {pages} {343} (\bibinfo {year} {1967})}\BibitemShut
  {NoStop}%
\bibitem [{\citenamefont {Alicki}(1979)}]{Alicki_1979heat}%
  \BibitemOpen
  \bibfield  {author} {\bibinfo {author} {\bibfnamefont {R.}~\bibnamefont
  {Alicki}},\ }\href@noop {} {\bibfield  {journal} {\bibinfo  {journal} {J.
  Phys. A: Math. Theor.}\ }\textbf {\bibinfo {volume} {12}},\ \bibinfo {pages}
  {0305} (\bibinfo {year} {1979})}\BibitemShut {NoStop}%
\bibitem [{\citenamefont {Gelbwaser-Klimovsky}\ \emph
  {et~al.}(2013)\citenamefont {Gelbwaser-Klimovsky}, \citenamefont {Alicki},\
  and\ \citenamefont {Kurizki}}]{Klimovsky_2013thermal}%
  \BibitemOpen
  \bibfield  {author} {\bibinfo {author} {\bibfnamefont {D.}~\bibnamefont
  {Gelbwaser-Klimovsky}}, \bibinfo {author} {\bibfnamefont {R.}~\bibnamefont
  {Alicki}}, \ and\ \bibinfo {author} {\bibfnamefont {G.}~\bibnamefont
  {Kurizki}},\ }\href@noop {} {\bibfield  {journal} {\bibinfo  {journal} {Phys.
  Rev. E}\ }\textbf {\bibinfo {volume} {87}},\ \bibinfo {pages} {012140}
  (\bibinfo {year} {2013})}\BibitemShut {NoStop}%
\bibitem [{\citenamefont {Howard}(1997)}]{howard1997molecular}%
  \BibitemOpen
  \bibfield  {author} {\bibinfo {author} {\bibfnamefont {J.}~\bibnamefont
  {Howard}},\ }\href@noop {} {\bibfield  {journal} {\bibinfo  {journal}
  {Nature}\ }\textbf {\bibinfo {volume} {389}},\ \bibinfo {pages} {561}
  (\bibinfo {year} {1997})}\BibitemShut {NoStop}%
\bibitem [{\citenamefont {Geva}\ and\ \citenamefont
  {Kosloff}(1992)}]{geva1992classical}%
  \BibitemOpen
  \bibfield  {author} {\bibinfo {author} {\bibfnamefont {E.}~\bibnamefont
  {Geva}}\ and\ \bibinfo {author} {\bibfnamefont {R.}~\bibnamefont {Kosloff}},\
  }\href@noop {} {\bibfield  {journal} {\bibinfo  {journal} {The Journal of
  chemical physics}\ }\textbf {\bibinfo {volume} {97}},\ \bibinfo {pages}
  {4398} (\bibinfo {year} {1992})}\BibitemShut {NoStop}%
\bibitem [{\citenamefont {H\"anggi}\ and\ \citenamefont
  {Marchesoni}(2009)}]{Hanggi2009brownian}%
  \BibitemOpen
  \bibfield  {author} {\bibinfo {author} {\bibfnamefont {P.}~\bibnamefont
  {H\"anggi}}\ and\ \bibinfo {author} {\bibfnamefont {F.}~\bibnamefont
  {Marchesoni}},\ }\href {\doibase 10.1103/RevModPhys.81.387} {\bibfield
  {journal} {\bibinfo  {journal} {Rev. Mod. Phys.}\ }\textbf {\bibinfo {volume}
  {81}},\ \bibinfo {pages} {387} (\bibinfo {year} {2009})}\BibitemShut
  {NoStop}%
\bibitem [{\citenamefont {Allahverdyan}\ and\ \citenamefont
  {Nieuwenhuizen}(2000)}]{allahverdyan2000extraction}%
  \BibitemOpen
  \bibfield  {author} {\bibinfo {author} {\bibfnamefont {A.~E.}\ \bibnamefont
  {Allahverdyan}}\ and\ \bibinfo {author} {\bibfnamefont {T.~M.}\ \bibnamefont
  {Nieuwenhuizen}},\ }\href {\doibase 10.1103/PhysRevLett.85.1799} {\bibfield
  {journal} {\bibinfo  {journal} {Phys. Rev. Lett.}\ }\textbf {\bibinfo
  {volume} {85}},\ \bibinfo {pages} {1799} (\bibinfo {year}
  {2000})}\BibitemShut {NoStop}%
\bibitem [{\citenamefont {Feldmann}\ and\ \citenamefont
  {Kosloff}(2006)}]{feldmann2006lubrication}%
  \BibitemOpen
  \bibfield  {author} {\bibinfo {author} {\bibfnamefont {T.}~\bibnamefont
  {Feldmann}}\ and\ \bibinfo {author} {\bibfnamefont {R.}~\bibnamefont
  {Kosloff}},\ }\href {\doibase 10.1103/PhysRevE.73.025107} {\bibfield
  {journal} {\bibinfo  {journal} {Phys. Rev. E}\ }\textbf {\bibinfo {volume}
  {73}},\ \bibinfo {pages} {025107} (\bibinfo {year} {2006})}\BibitemShut
  {NoStop}%
\bibitem [{\citenamefont {Linden}\ \emph {et~al.}(2010)\citenamefont {Linden},
  \citenamefont {Popescu},\ and\ \citenamefont {Skrzypczyk}}]{linden2010small}%
  \BibitemOpen
  \bibfield  {author} {\bibinfo {author} {\bibfnamefont {N.}~\bibnamefont
  {Linden}}, \bibinfo {author} {\bibfnamefont {S.}~\bibnamefont {Popescu}}, \
  and\ \bibinfo {author} {\bibfnamefont {P.}~\bibnamefont {Skrzypczyk}},\
  }\href@noop {} {\bibfield  {journal} {\bibinfo  {journal} {Physical review
  letters}\ }\textbf {\bibinfo {volume} {105}},\ \bibinfo {pages} {130401}
  (\bibinfo {year} {2010})}\BibitemShut {NoStop}%
\bibitem [{\citenamefont {Dahlsten}\ \emph {et~al.}(2011)\citenamefont
  {Dahlsten}, \citenamefont {Renner}, \citenamefont {Rieper},\ and\
  \citenamefont {Vedral}}]{dahlsten2011inadequacy}%
  \BibitemOpen
  \bibfield  {author} {\bibinfo {author} {\bibfnamefont {O.}~\bibnamefont
  {Dahlsten}}, \bibinfo {author} {\bibfnamefont {R.}~\bibnamefont {Renner}},
  \bibinfo {author} {\bibfnamefont {E.}~\bibnamefont {Rieper}}, \ and\ \bibinfo
  {author} {\bibfnamefont {V.}~\bibnamefont {Vedral}},\ }\href@noop {}
  {\bibfield  {journal} {\bibinfo  {journal} {New Journal of Physics}\ }\textbf
  {\bibinfo {volume} {13}},\ \bibinfo {pages} {053015} (\bibinfo {year}
  {2011})}\BibitemShut {NoStop}%
\bibitem [{\citenamefont {Del~Rio}\ \emph {et~al.}(2011)\citenamefont
  {Del~Rio}, \citenamefont {{\AA}berg}, \citenamefont {Renner}, \citenamefont
  {Dahlsten},\ and\ \citenamefont {Vedral}}]{del2011thermodynamic}%
  \BibitemOpen
  \bibfield  {author} {\bibinfo {author} {\bibfnamefont {L.}~\bibnamefont
  {Del~Rio}}, \bibinfo {author} {\bibfnamefont {J.}~\bibnamefont {{\AA}berg}},
  \bibinfo {author} {\bibfnamefont {R.}~\bibnamefont {Renner}}, \bibinfo
  {author} {\bibfnamefont {O.}~\bibnamefont {Dahlsten}}, \ and\ \bibinfo
  {author} {\bibfnamefont {V.}~\bibnamefont {Vedral}},\ }\href@noop {}
  {\bibfield  {journal} {\bibinfo  {journal} {Nature}\ }\textbf {\bibinfo
  {volume} {474}},\ \bibinfo {pages} {61} (\bibinfo {year} {2011})}\BibitemShut
  {NoStop}%
\bibitem [{\citenamefont {Faist}\ \emph {et~al.}(2012)\citenamefont {Faist},
  \citenamefont {Dupuis}, \citenamefont {Oppenheim},\ and\ \citenamefont
  {Renner}}]{faist2012quantitative}%
  \BibitemOpen
  \bibfield  {author} {\bibinfo {author} {\bibfnamefont {P.}~\bibnamefont
  {Faist}}, \bibinfo {author} {\bibfnamefont {F.}~\bibnamefont {Dupuis}},
  \bibinfo {author} {\bibfnamefont {J.}~\bibnamefont {Oppenheim}}, \ and\
  \bibinfo {author} {\bibfnamefont {R.}~\bibnamefont {Renner}},\ }\href@noop {}
  {\bibfield  {journal} {\bibinfo  {journal} {arXiv preprint arXiv:1211.1037}\
  } (\bibinfo {year} {2012})}\BibitemShut {NoStop}%
\bibitem [{\citenamefont {Halpern}\ and\ \citenamefont
  {Renes}(2014)}]{halpern2014beyond}%
  \BibitemOpen
  \bibfield  {author} {\bibinfo {author} {\bibfnamefont {N.~Y.}\ \bibnamefont
  {Halpern}}\ and\ \bibinfo {author} {\bibfnamefont {J.~M.}\ \bibnamefont
  {Renes}},\ }\href@noop {} {\bibfield  {journal} {\bibinfo  {journal} {arXiv
  preprint arXiv:1409.3998}\ } (\bibinfo {year} {2014})}\BibitemShut {NoStop}%
\bibitem [{\citenamefont {Anders}\ and\ \citenamefont
  {Giovannetti}(2013)}]{anders2013thermodynamics}%
  \BibitemOpen
  \bibfield  {author} {\bibinfo {author} {\bibfnamefont {J.}~\bibnamefont
  {Anders}}\ and\ \bibinfo {author} {\bibfnamefont {V.}~\bibnamefont
  {Giovannetti}},\ }\href@noop {} {\bibfield  {journal} {\bibinfo  {journal}
  {New Journal of Physics}\ }\textbf {\bibinfo {volume} {15}},\ \bibinfo
  {pages} {033022} (\bibinfo {year} {2013})}\BibitemShut {NoStop}%
\bibitem [{\citenamefont {Mueller}\ and\ \citenamefont
  {Pastena}(2015)}]{mueller2014work}%
  \BibitemOpen
  \bibfield  {author} {\bibinfo {author} {\bibfnamefont {M.~P.}\ \bibnamefont
  {Mueller}}\ and\ \bibinfo {author} {\bibfnamefont {M.}~\bibnamefont
  {Pastena}},\ }\href@noop {} {\bibfield  {journal} {\bibinfo  {journal} {Phys.
  Rev. Lett.}\ }\textbf {\bibinfo {volume} {115}},\ \bibinfo {pages} {150402}
  (\bibinfo {year} {2015})}\BibitemShut {NoStop}%
\bibitem [{\citenamefont {Huber}\ \emph {et~al.}(2015)\citenamefont {Huber},
  \citenamefont {Perarnau-Llobet}, \citenamefont {Hovhannisyan}, \citenamefont
  {Skrzypczyk}, \citenamefont {Klockl}, \citenamefont {Brunner},\ and\
  \citenamefont {Acin}}]{huber2014thermodynamic}%
  \BibitemOpen
  \bibfield  {author} {\bibinfo {author} {\bibfnamefont {M.}~\bibnamefont
  {Huber}}, \bibinfo {author} {\bibfnamefont {M.}~\bibnamefont
  {Perarnau-Llobet}}, \bibinfo {author} {\bibfnamefont {K.~V.}\ \bibnamefont
  {Hovhannisyan}}, \bibinfo {author} {\bibfnamefont {P.}~\bibnamefont
  {Skrzypczyk}}, \bibinfo {author} {\bibfnamefont {C.}~\bibnamefont {Klockl}},
  \bibinfo {author} {\bibfnamefont {N.}~\bibnamefont {Brunner}}, \ and\
  \bibinfo {author} {\bibfnamefont {A.}~\bibnamefont {Acin}},\ }\href@noop {}
  {\bibfield  {journal} {\bibinfo  {journal} {New Journal of Physics}\ }\textbf
  {\bibinfo {volume} {17}},\ \bibinfo {pages} {065008} (\bibinfo {year}
  {2015})}\BibitemShut {NoStop}%
\bibitem [{\citenamefont {{Narasimhachar}}\ and\ \citenamefont
  {{Gour}}(2015)}]{narasimhachar2014low}%
  \BibitemOpen
  \bibfield  {author} {\bibinfo {author} {\bibfnamefont {V.}~\bibnamefont
  {{Narasimhachar}}}\ and\ \bibinfo {author} {\bibfnamefont {G.}~\bibnamefont
  {{Gour}}},\ }\href@noop {} {\bibfield  {journal} {\bibinfo  {journal} {Nature
  Communications}\ }\textbf {\bibinfo {volume} {6}},\ \bibinfo {pages} {7689}
  (\bibinfo {year} {2015})}\BibitemShut {NoStop}%
\bibitem [{\citenamefont {Alhambra}\ \emph {et~al.}(2015)\citenamefont
  {Alhambra}, \citenamefont {Oppenheim},\ and\ \citenamefont
  {Perry}}]{alhambra2015probability}%
  \BibitemOpen
  \bibfield  {author} {\bibinfo {author} {\bibfnamefont {{\'A}.~M.}\
  \bibnamefont {Alhambra}}, \bibinfo {author} {\bibfnamefont {J.}~\bibnamefont
  {Oppenheim}}, \ and\ \bibinfo {author} {\bibfnamefont {C.}~\bibnamefont
  {Perry}},\ }\href@noop {} {\bibfield  {journal} {\bibinfo  {journal} {arXiv
  preprint arXiv:1504.00020}\ } (\bibinfo {year} {2015})}\BibitemShut {NoStop}%
\bibitem [{\citenamefont {{Woods}}\ \emph {et~al.}(2015)\citenamefont
  {{Woods}}, \citenamefont {{Ng}},\ and\ \citenamefont
  {{Wehner}}}]{woods2015maximum}%
  \BibitemOpen
  \bibfield  {author} {\bibinfo {author} {\bibfnamefont {M.~P.}\ \bibnamefont
  {{Woods}}}, \bibinfo {author} {\bibfnamefont {N.}~\bibnamefont {{Ng}}}, \
  and\ \bibinfo {author} {\bibfnamefont {S.}~\bibnamefont {{Wehner}}},\
  }\href@noop {} {\bibfield  {journal} {\bibinfo  {journal} {ArXiv e-prints}\ }
  (\bibinfo {year} {2015})},\ \Eprint {http://arxiv.org/abs/1506.02322}
  {arXiv:1506.02322} \BibitemShut {NoStop}%
\bibitem [{\citenamefont {Lostaglio}\ \emph {et~al.}(2015)\citenamefont
  {Lostaglio}, \citenamefont {Jennings},\ and\ \citenamefont
  {Rudolph}}]{lostaglio2015description}%
  \BibitemOpen
  \bibfield  {author} {\bibinfo {author} {\bibfnamefont {M.}~\bibnamefont
  {Lostaglio}}, \bibinfo {author} {\bibfnamefont {D.}~\bibnamefont {Jennings}},
  \ and\ \bibinfo {author} {\bibfnamefont {T.}~\bibnamefont {Rudolph}},\
  }\href@noop {} {\bibfield  {journal} {\bibinfo  {journal} {Nat. Commun.}\
  }\textbf {\bibinfo {volume} {6}},\ \bibinfo {pages} {6383} (\bibinfo {year}
  {2015})}\BibitemShut {NoStop}%
\bibitem [{\citenamefont {Oppenheim}(2013)}]{talk_qip2013}%
  \BibitemOpen
  \bibfield  {author} {\bibinfo {author} {\bibfnamefont {J.}~\bibnamefont
  {Oppenheim}},\ }\href
  {http://conference.iiis.tsinghua.edu.cn/QIP2013/wp-content/uploads/2012/01/Jonathan-Oppenheim.pdf}
  {\enquote {\bibinfo {title} {Fundamental limitations for quantum and nano
  thermodynamics},}\ } (\bibinfo {year} {2013}),\ \bibinfo {note} {{QIP}
  talk}\BibitemShut {NoStop}%
\bibitem [{\citenamefont {Marshall}\ \emph {et~al.}(2010)\citenamefont
  {Marshall}, \citenamefont {Olkin},\ and\ \citenamefont
  {Arnold}}]{marshall2010inequalities}%
  \BibitemOpen
  \bibfield  {author} {\bibinfo {author} {\bibfnamefont {A.~W.}\ \bibnamefont
  {Marshall}}, \bibinfo {author} {\bibfnamefont {I.}~\bibnamefont {Olkin}}, \
  and\ \bibinfo {author} {\bibfnamefont {B.}~\bibnamefont {Arnold}},\
  }\href@noop {} {\emph {\bibinfo {title} {Inequalities: theory of majorization
  and its applications}}}\ (\bibinfo  {publisher} {Springer Science \& Business
  Media},\ \bibinfo {year} {2010})\BibitemShut {NoStop}%
\end{thebibliography}%
